\documentclass[preprint,12pt]{elsarticle}

\usepackage{color}
\usepackage{amssymb}
\usepackage{amsmath}
\usepackage{setspace}
\usepackage{graphicx}
\usepackage{subfigure}
\usepackage{booktabs}
\usepackage{makecell}
\usepackage[ruled,linesnumbered]{algorithm2e}

\journal{Computer Methods in Applied Mechanics and Engineering}

\begin{document}
	
	\begin{frontmatter}
		
		
		
		\title{A Lagrangian free-stream boundary condition for weakly compressible smoothed particle hydrodynamics}
		
		
		\author{Shuoguo Zhang}
		\author{Wenbin Zhang}
		\author{Chi Zhang}
		\author{Xiangyu Hu\corref{cor1}}
		\ead{xiangyu.hu@tum.de}
		
		\cortext[cor1]{Corresponding author}

		\address{School of Engineering and Design, 
			Technical University of Munich, Garching, 85747, Germany}
		
		
		\begin{abstract}
			
			This paper proposes the first free-stream boundary condition 
			in a purely Lagrangian framework for weakly-compressible smoothed particle hydrodynamics (WCSPH). The boundary condition is implemented 
			based on several numerical techniques, including a spatio-temporal method and an ensuring technique for surface particle identification, 
			far-field corrections of density and velocity near the free-stream boundary, and improved in- and outlet conditions which guarantee
			smooth inflow profile and free outflow particles without the issue of the truncated kernel. The accuracy and effectiveness of the method have been validated by the reference solutions of flow past a flat plate and circular cylinder, respectively. The method has also been tested with more complex flows to demonstrate its applicability, such as fluid-structure interaction (FSI) and 3D flow past a sphere. 
		\end{abstract}
		\begin{graphicalabstract}
		\end{graphicalabstract}
		
		\begin{highlights}
			\item The first free-stream boundary condition in a purely Lagrangian framework is proposed for the weakly-compressible smoothed particle hydrodynamics (WCSPH).
			\item The proposed spatio-temporal identification method based on particle position divergence avoids the misjudgement of surface particles due to uneven particle distribution or low-pressure regions. 
			\item A simpler inlet/outlet boundary condition guarantees a smooth flow profile.
		\end{highlights}
		
		
	\end{frontmatter}
	
	
	\section{Introduction}
	\label{}
	Smoothed particle hydrodynamics (SPH) is a meshless numerical method initially proposed by Lucy \cite{Lucy1977}, Gingold, and Monaghan \cite{Gingold1977} for modelling astrophysical phenomena. Since its inception, it has been successfully applied for modeling fluid \cite{Monaghan1994, Morris1997}, solid \cite{Benz1995, Gray2001} and fluid-structure interaction (FSI) \cite{Rafiee2009, Zhang2021}. SPH method is particularly suitable for the flows with a free surface, open boundary \cite{Federico2012, Tan2015}, wave impacting \cite{Staroszczyk2014, Marrone2011}, and other hydraulic engineering problems \cite{Chen, Vacondio, Matthieu}. This is due to its fully Lagrangian nature and a straightforward free-surface condition, which can be imposed in the weakly-compressible SPH (WCSPH) method to accurately depict free-surface waves and splashes. 
	
    However, SPH still encounters difficulties with some traditional fluid dynamics problems, which hinders the applicability of SPH as a general numerical method for computational fluid dynamics (CFD). One typical case is the aerodynamic problem in which a free-stream boundary condition is often applied to focus on the flow near the geometry other than the far field. Although this boundary condition is quite straightforward for the established Eulerian mesh-based methods, its implementation in SPH remains an open question. Up to now, a fixed boundary, in which static particles are set with the free-stream velocity, is usually employed as a substitute for the far-field in SPH simulations \cite{Tafuni2018, shuoguo}. In order to eliminate the boundary effect, the computational domain has to be enlarged greatly with the compensation of computational efficiency. Very often, these methods also require the implementation of in- and out-flow boundary conditions in the stream direction  
    \cite{Martin2009, Molteni2013, Braun2015, Tafuni2018}. In order to avoid the issue of truncated kernel support near the in- and outlets, two different boundary regions are added, respectively. While the former generates new particles with a free-stream state into the flow domain, the latter deletes particles after they leave. A relevant issue of the inlet is that spurious pressure fluctuation can be produced near the boundary-fluid interface \cite{Ferrand2017}, which requires quite elaborate treatment \cite{Colagrossi2013, Wang2019}. In order to address a similar issue at the outlet, another method called ``do-nothing" freezes the fluid state within the outlet boundary region \cite{Khorasanizade2015, Alvarado-Rodriguez2017, Negi2020} to avoid spurious pressure fluctuation.

    In the SPH method, before imposing the free-stream boundary conditions, the surface particles, i.e. those with truncated kernel support, have to be identified, either in an Eulerian manner such as setting up in- or outlet regions or with Lagrangian particle identification algorithms such as those used in the SPH simulation of free-surface flow. Generally, there are two main, namely algebraic and geometric, Lagrangian identification approaches. In the algebraic approach, \cite{Lee2008}, the free-surface particles are identified based on the SPH approximation of position divergence because the value near-surface is considerably smaller than the inner values. However, as shown later, this approach often leads to misjudgment. One typical case is that surface particle may be missed when particle distribution is non-uniform, especially in splashing flows. Another case is that the particles in the low-pressure region can be wrongly identified as surface particles. In contrast, the geometric approach proposed by Dilts \cite{Dilts, Haque, Lin2019} is more accurate. The difficulty of the geometric approach is the complicated and expensive implementation, which leads to a further attempt of hybridization with the algebraic identification \cite{Marrone2010}. Note that while the geometric and hybrid approaches can effectively alleviate the inaccuracy due to non-uniform particle distribution, they can still produce misjudgment caused by the low-pressure region.
	
	In this paper, other than using fixed static particles, we propose a Lagrangian free-stream boundary condition for the WCSPH method to increase computational efficiency and decrease boundary effects. First, a spatio-temporal identification method and an ensuring technique are developed to detect the surface particles and avoid the misjudgment caused by both non-uniform particle distribution and low-pressure regions. Then, density and velocity corrections are implemented on the surface particles to impose the far-field condition. Note that the present far-field corrections also lead to an improved implementation of inflow and outflow boundary conditions, in which spurious pressure fluctuation is suppressed. The remainder of this paper is organized as follows. First, a brief overview of the WCSPH method is given in Section 2. In Section 3, the proposed free-stream boundary algorithm is detailed. The present method's accuracy, efficiency, and robustness are comprehensively validated with several benchmark tests in Section 4, including the flows over a flat plate and a circular cylinder, flow-induced oscillation of an elastic beam attached to a cylinder, and a 3D flow past a sphere. Finally, concluding remarks are given in Section 5. The code accompanying this work is implemented in the open-source SPH library (SPHinXsys) \cite{Zhang2021CPC}, and is available at https://www.sphinxsys.org.
	
	\section{SPH method}
	Before the details of the Lagrangian free-stream boundary condition, the governing equations and their SPH discretizations are briefly summarized in this section. More details can be referred to Refs. \cite{Zhang2021CPC,Hu2006,Zhang2020,Zhang2017}.
	
	\label{}
	\subsection{Governing equations}
	\label{section2-1}
	The WCSPH is employed to model the flow in the present work. In the Lagrangian form, conservation equations of mass and momentum are respectively written as
	\begin{equation} \label{eq1} 
		\frac{d\rho}{dt} = -\rho\nabla\cdot\mathbf{v},   
	\end{equation}
	\begin{equation} \label{eq2} 
		\frac{d\mathbf{v}}{dt} = -\frac{1}{\rho}\nabla p+\nu\nabla^{2}\mathbf{v},  
	\end{equation}
	where $ \rho $ is the density, $ \mathbf{v} $ the velocity, $ p $ the pressure, $ \nu $ the kinematic viscosity, and $t $ the time. Under the weakly-compressible assumption, the above equations Eq.(\ref{eq1}) and Eq.(\ref{eq2}) are closed by the following artificial isothermal equation of state (EoS)
	\begin{equation} \label{eq3}
		p=c_0^2(\rho-\rho_{0}), 
	\end{equation}
	where $ c_{0} $ is the artificial sound speed, and $ \rho_{0} $ the initial reference density. In order to control the density variation around 1$\%$, 
	corresponding to the Mach number $ M\approx 0.1 $ suggested by Monaghan \cite{Monaghan1994}, $ c_{0} $ is set to be $ c_{0} =10U_{max} $, where $ U_{max} $ is the anticipated maximum flow speed. 
	
	\subsection{SPH discretization}
	Following Ref. \cite{Zhang2021CPC}, the continuity equation Eq.(\ref{eq1}) is discretized, in respect to particle $i$, by using a Riemann-based WCSPH method as
	\begin{equation} \label{eq4}
		\frac{d\rho_{i}}{dt}=2\rho_{i}\sum_{j}\frac{m_{j}}{\rho_{j}}(\mathbf{v}_{i}-\mathbf{v}^{*})\cdot\nabla_{i} W_{ij} , 
	\end{equation}
	where the subscript $ j $ denotes the neighbor particles, $ m $ the particle mass. $ \nabla_{i} W_{ij} $ represents the gradient of the kernel function $ W(|\mathbf{r}_{ij}|,h) $, where  $ \mathbf{r}_{ij}=\mathbf{r}_{i}-\mathbf{r}_{j} $, and $ h $ the smooth length.
	
	Here, the low-dissipation Riemann solver is adopted to solve the inter-particle Riemann problem constructed along the unit vector $\mathbf{e}_{ij}=-\mathbf{r}_{ij}/r_{ij} $ pointing from particle $ i $ to $ j $ \cite{article2017Chi},
	$ \mathbf{v}^{*} = U^{*}\mathbf{e}_{ij}+(\mathbf{\overline{v}}_{ij}-\overline{U}\mathbf{e}_{ij})$, where $ \mathbf{\overline{v}}_{ij} =(\mathbf{v}_{i}+\mathbf{v}_{j})/2 $ is the average velocity between particle $ i $ and $ j $. Then, associated with the initial left and right states ($ L $ and $ R $, respectively) on particles $i$ and $j$, the solution of the Riemann problem $ U^{*} $ can be written as in Ref. \cite{article2017Chi}
	\begin{equation} 
		\begin{cases} \label{eq5}
			U^{*}=\overline{U}+\dfrac{P_{L}-P_{R}}{2\overline{\rho}c_{0}}\\[3mm]
			(\rho_{L},U_{L},P_{L})=(\rho_{i}, \mathbf{v}_{i}\cdot \mathbf{e}_{ij}, p_{i})\\[3mm]
			(\rho_{R},U_{R},P_{R})=(\rho_{j}, \mathbf{v}_{j}\cdot \mathbf{e}_{ij}, p_{j})
		\end{cases}.
	\end{equation}
	
	Here, $ \overline{U}=(U_{L}+U_{R})/2 $ and $\overline{\rho}=(\rho_{L}+\rho_{R})/2 $ are inter-particle averages. Since the density error becomes more significant 
	as the flow Reynolds number increases, the particle density is re-initialized before updated in continuity equation Eq.(\ref{eq4}) at each time step with \cite{Zhang2021CPC}
	\begin{equation} \label{eq6}
		\rho_{i}=\rho_{0}\dfrac{\sum W_{ij}}{\sum W^{0}_{ij}}.
	\end{equation}
	Here, the superscript $ 0 $ represents the initial reference value at the beginning of the simulation.
	
	For flow with moderate Reynolds number as considered in this work, the conservation equation of momentum Eq.(\ref{eq2}) is discretized without Reimann solvers to decrease the numerical dissipation \cite{Zhang2021CPC,Zhang2020}
	\begin{equation} \label{eq7}
		\frac{d\mathbf{v}_{i}}{dt}=-2\sum_{j}\dfrac{m_{j}}{\rho_{i}\rho_{j}}(\frac{p_{i}\rho_{j}+p_{j}\rho_{i}}{\rho_{i}+\rho_{j}})\frac{\partial W_{ij}}{\partial r_{ij}}\mathbf{e}_{ij}+2\sum_{j}m_{j}\frac{\eta\mathbf{v}_{ij} }{\rho_{i}\rho_{j}r_{ij}}\frac{\partial W_{ij}}{\partial r_{ij}},
	\end{equation}
	where $ \eta $ is the dynamic viscosity, $ \mathbf{v}_{ij}=\mathbf{v}_{i}-\mathbf{v}_{j} $ the relative velocity. 
	
	Here, the dual-criteria time stepping with position Verlet method is applied for time integration to increase computational efficiency and conservation properties. The details are referred to Refs.  \cite{Zhang2021, Zhang2021CPC,Zhang2020}. 
	\subsection{Transport velocity formulation}
	\label{section2-3}
	The tensile instability associated with negative pressure may affect particle distribution and lead to particle clumping or void regions. Therefore, transport-velocity formulation, which is applied to regularize the particle distribution \cite{Adami2013, Zhang2017}, can be written as follows
	\begin{equation} \label{eq8}
		\widetilde{\mathbf{v}}_{i}(t + \delta t)=\mathbf{v}_{i}(t)+\delta t(\frac{\widetilde{d}\mathbf{v}_{i}}{dt}-p_{max}\sum_{j} \frac{2m_{j}}{\rho_{i}\rho_{j}} \frac{\partial W_{ij}}{\partial r_{ij}}\mathbf{e_{ij}}).
	\end{equation}
	Here, the global background $ p_{max} $ is chosen as $p_{max}=\alpha\rho_{0}\mathbf{v}_{max}^{2},$, where $ \mathbf{v}_{max} $ is the maximum particle velocity at each time step. Note that the empirical coefficient $ \alpha $  generally ranges from 5.0 to 10.0 and herein it is chosen as constant $ \alpha = 7.0$. 
	\section{Free-stream boundary condition}
	This section introduces the free-stream boundary condition, where the dynamic conditions are directly imposed on the detected surface particles. To achieve this purpose, several sub-steps, including the spatio-temporal identification method, surface particles ensuring, far-field corrections, and in- and outflow boundary conditions, are proposed.
	
	\subsection{Spatio-temporal identification}
	\label{section3-1}
	\begin{figure}[htb!]
		\centering    
		\subfigure[Particle categories in a splashing flow: green-inner fluid particles, red-outmost surface particles, and blue-inner surface particles around low-pressure regions. Left: the standard position-divergence method identifies surface particles even those near the inner low regions. Right: the present method accurately screens out true surface particles without the misjudgment due to low-pressure regions.]
		{
			\begin{minipage}[b]{0.45\linewidth}
				\includegraphics[width=1\textwidth]{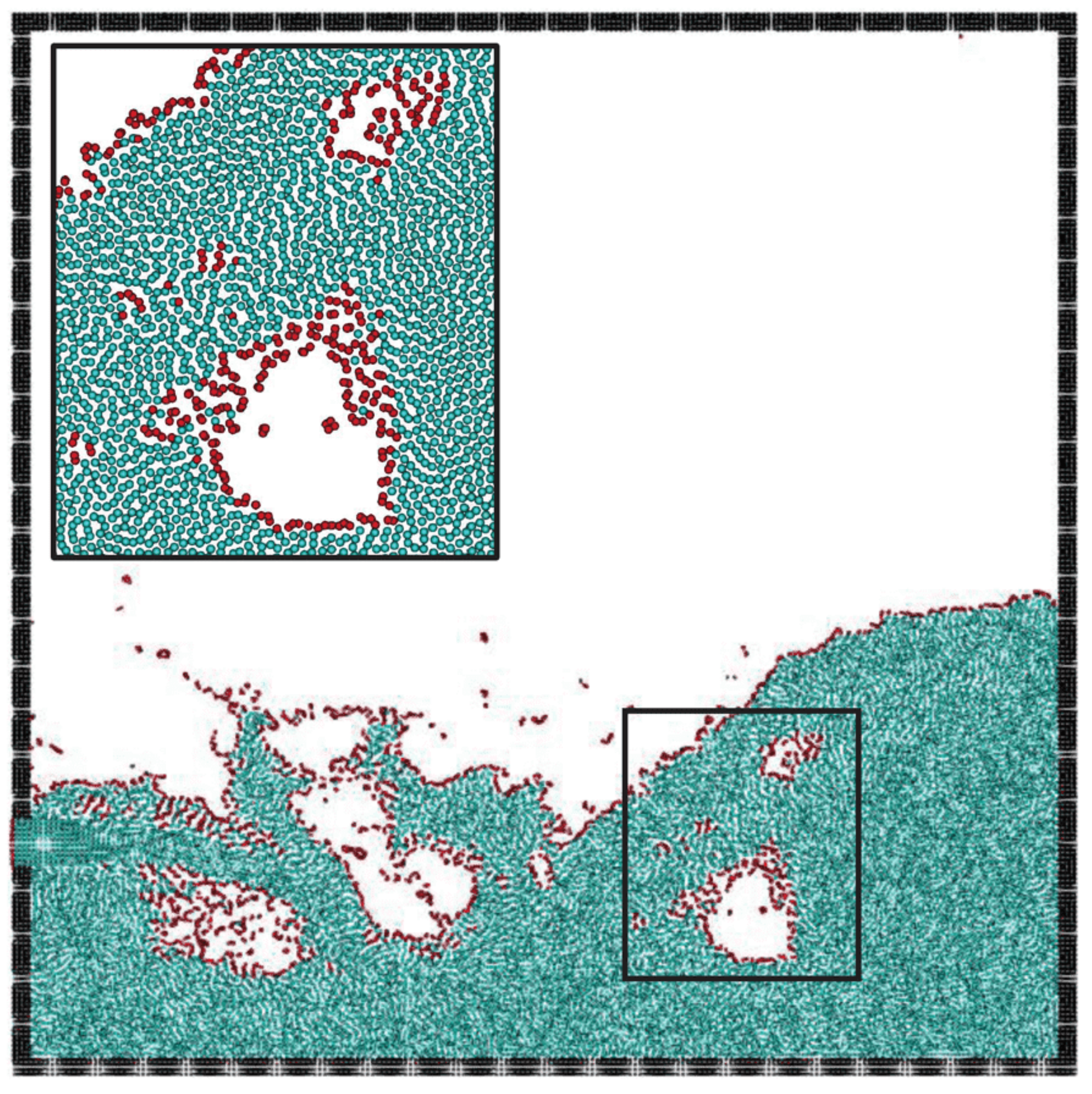}
			\end{minipage}
			\begin{minipage}[b]{0.45\linewidth}
				\includegraphics[width=1\textwidth]{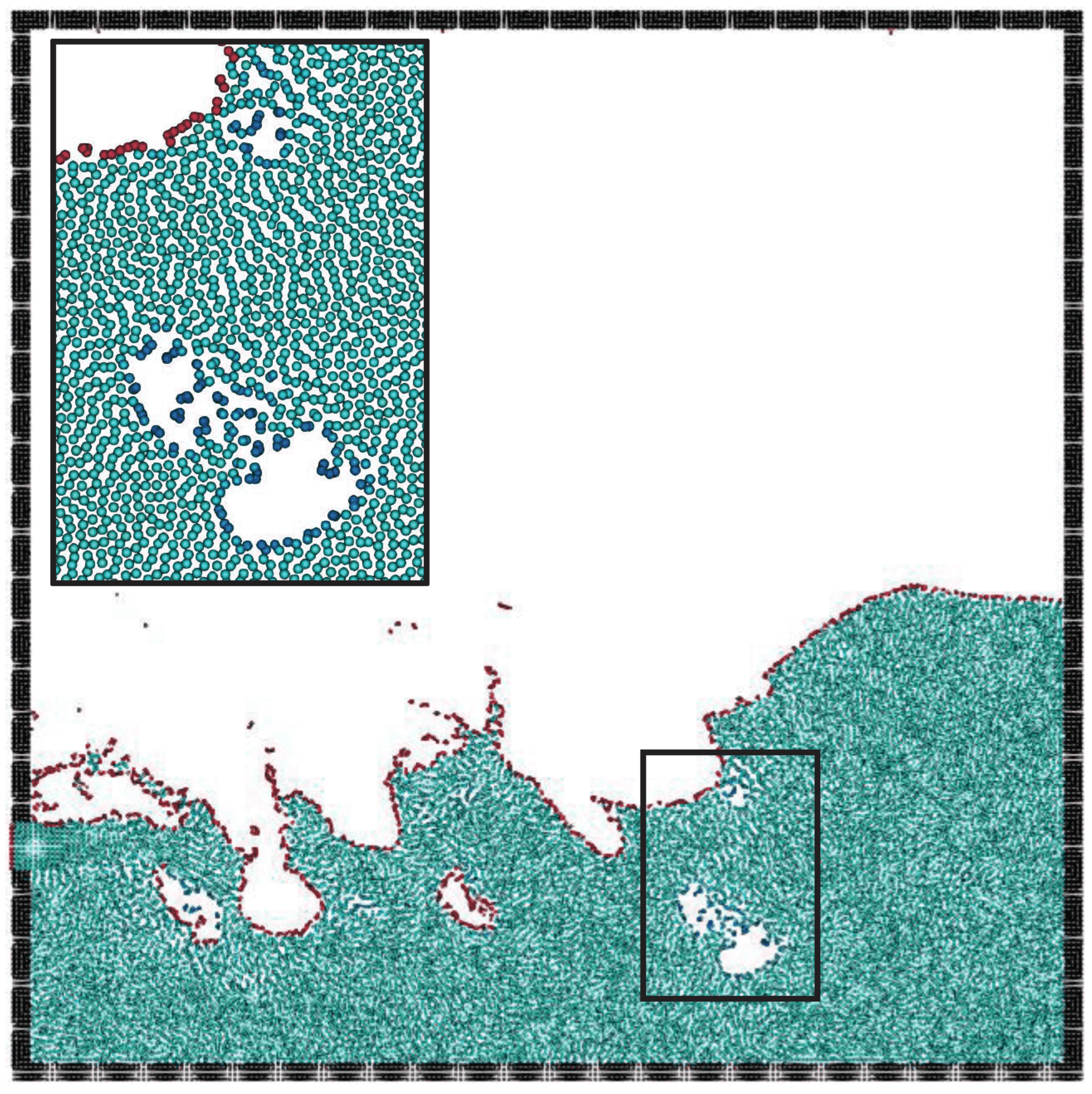}
			\end{minipage}
			\label{fig1a}
		}	 
		
		\subfigure[Particle categories in a right-going flow past a cylinder: 
		blue-inner fluid particles, green-solid particles of the cylinder, and grey-inner surface particles around low-pressure regions. Left: the misjudgement of the standard position-divergence method leads to non-physical voids behind the cylinder. Right: the present method avoids possible misjudgments, and the non-physical voids do not appear.]
		{
			\begin{minipage}[b]{0.45\linewidth}
				\includegraphics[width=1\textwidth]{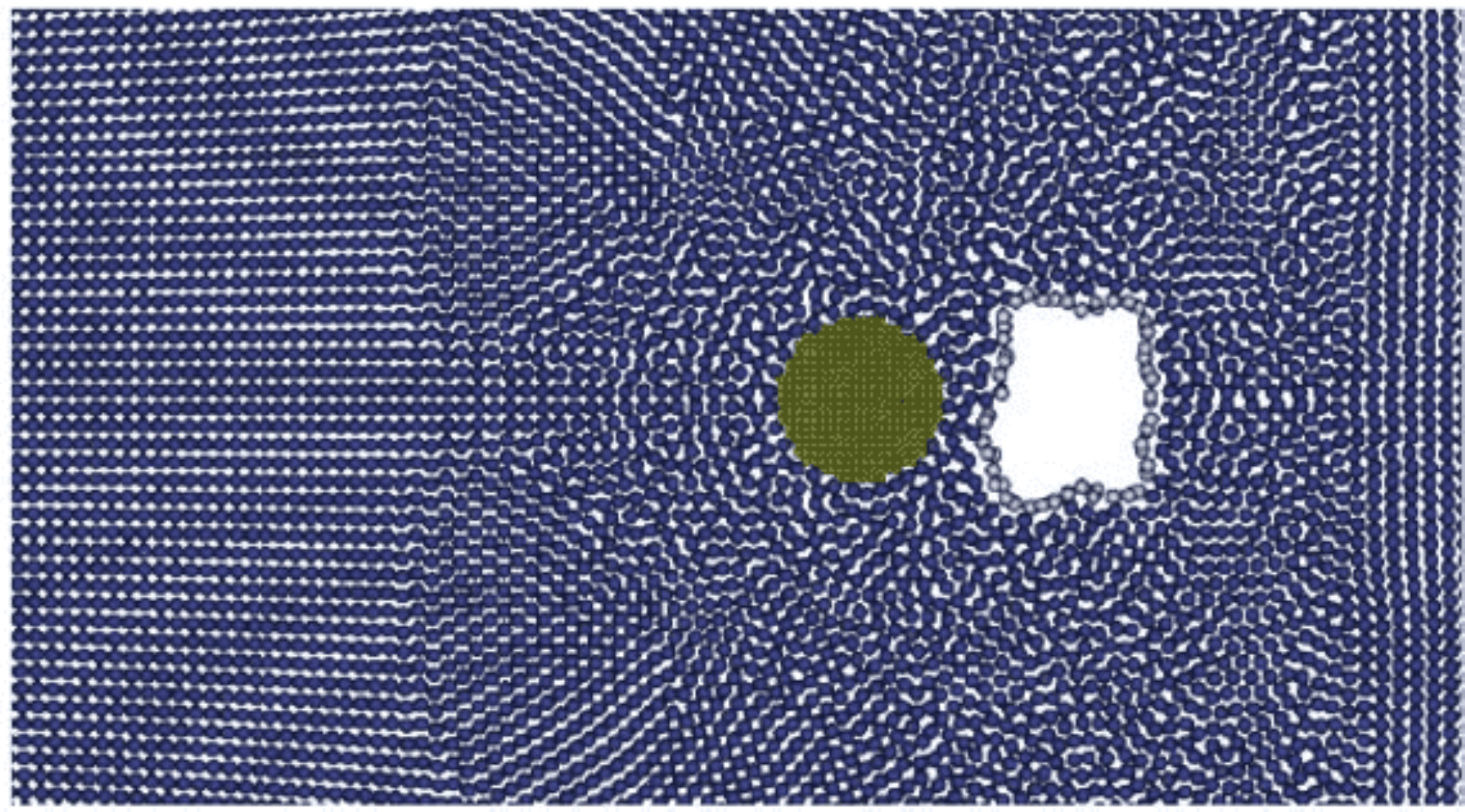}
			\end{minipage}
			\begin{minipage}[b]{0.45\linewidth}
				\includegraphics[width=1\textwidth]{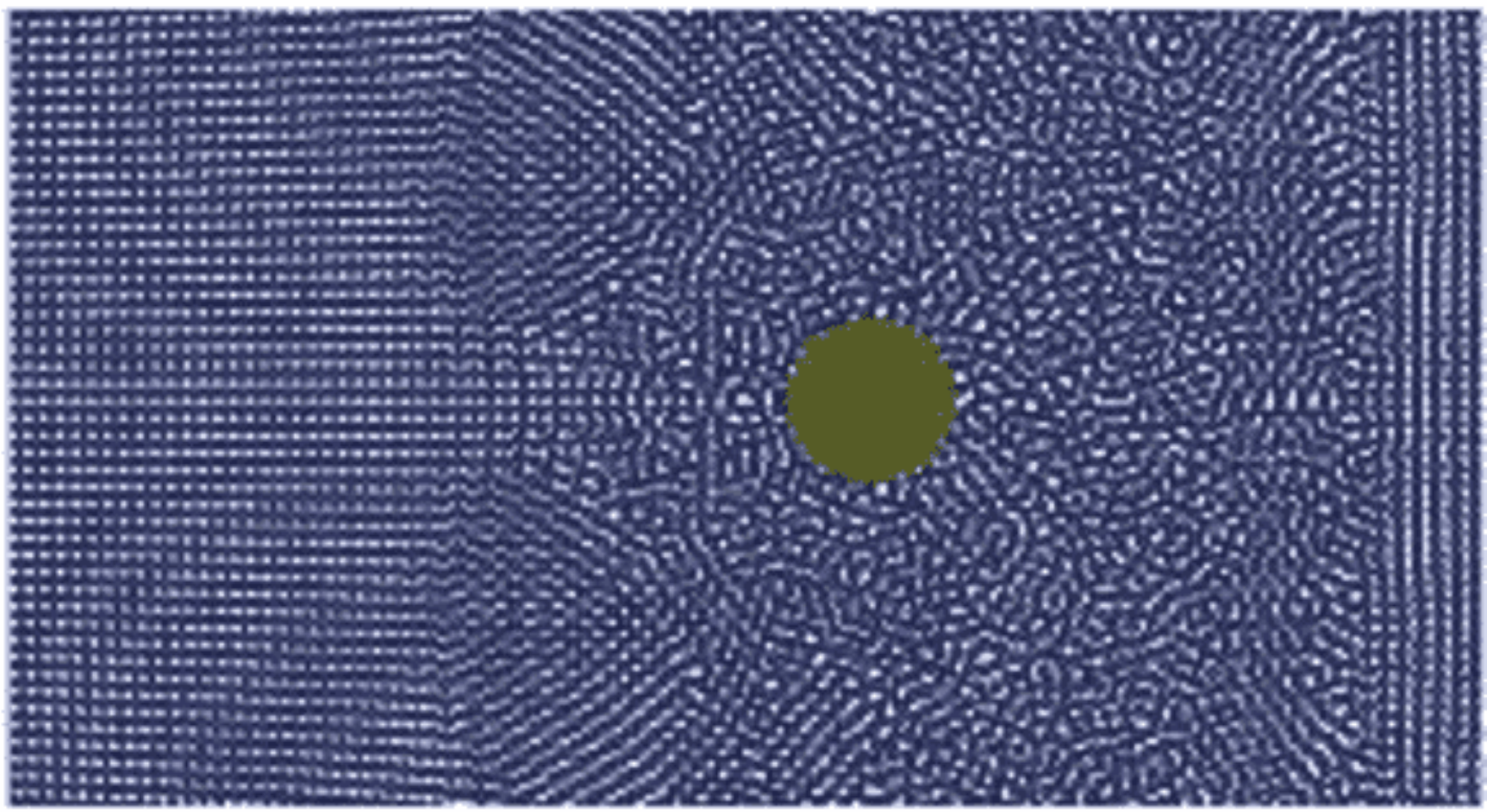}
			\end{minipage}
			\label{fig1b}
		}	
		\caption{Qualitative comparison of the standard position-divergence method \cite{Lee2008}  (left) and the present spatio-temporal identification method (right) on the identification of surface particles.}
		\label{fig1}
	\end{figure}
	The divergence of particle position represented by $ \nabla \cdot \mathbf{r}_{i} $ can be discretized as follows \cite{Lee2008}
	\begin{equation} \label{eq9}
		\nabla \cdot \mathbf{r}_{i} =\sum_{j}\frac{m_{j}}{\rho_{j}} \mathbf{r_{ij}} \cdot \nabla W_{ij}. 
	\end{equation} 
	It is known that an inner fluid particle with full kernel supported has a value $ \nabla \cdot \mathbf{r}_{i} \approx 2.0 $ for 2D simulations. In contrast, a particle adjacent to the fluid surface has a considerably smaller value since it has fewer neighbour particles in its support domain. Considering this difference, Lee et al. \cite{Lee2008} chose a threshold of $ \nabla \cdot \mathbf{r}_{i}=1.5 $ to distinguish the inner and surface particles.
	
	Although very often effective, this method sometimes misjudges surface particles. One issue is that it misjudges some inner fluid particles located at the low-pressure region in Fig.\ref{fig1a} can also be misjudged as surface ones. This issue is because the particle distribution can be slightly sparser in the low-pressure region due to the applied weakly compressible assumption, and the correspondingly decreased position divergence leads to misjudgment. Another issue is that it can miss, as demonstrated in the example indicated in Fig.\ref{fig1a}, some surface particles in the very non-regular particle distribution, especially splashing regions.  
	
	Since a Lagrangian fluid particle moves along the streamline, a surface particle at the present step should also be a surface particle or at least a neighbour of a surface particle in the previous time step. Considering this spatio-temporal characteristic, we propose a new identification method based on the position-divergence method to identify true surface particles without suffering from the first issue in low-pressure regions, as shown in Fig.\ref{fig1a}. The detailed procedures are explained and summarized in Algo.\ref{algorithm1}. Together with the transport velocity formulation in Section \ref{section2-3} to further regularize the particle distribution, this algorithm efficiently eliminates non-physical voids in the low-pressure region as indicated in Fig.\ref{fig1b}.   
	
	Note that the spatio-temporal identification method is still not able to address the issue of very non-regular particles, as also shown in the insert of Fig.\ref{fig1a}, which will be handled by a surface particle ensuring technique in the next section.

	{\center 
		\begin{algorithm}[htbp]
			\label{algorithm1}
			\SetAlgoLined
			\KwData{Particle category in previous time step $ \theta_{i}$,  particle category in current time step $ \beta_{i}$, position divergence $ \nabla \cdot \mathbf{r}_{i} $, number of fluid paritcles $ n $, number of neighbour fluid particles $ k $. }
			\KwResult{surface particles: $  \theta_{i} = 1 $, $  \beta_{i} = 1 $. inner particles: $ \theta_{i} = 0 $, $ \beta_{i} = 0 $. } 
			
			\textbf{Procedure} Initialization\;
			\For{particle $i=1$ \KwTo $n$}
			{
				$ \theta_{i} = 1 $\;
			}
			\textbf{Procedure} Solver\;
			\For{particle $i=1$ \KwTo $n$}{
				$ \nabla \cdot \mathbf{r}_{i} = \sum_{j}\frac{m_{j}}{\rho_{j}} \mathbf{r_{ij}} \cdot \nabla W_{ij} $\;
				$ \beta_{i} = 0 $\;
				\If{$ \nabla \cdot \mathbf{r}_{i}<1.5 $}{
					\eIf{$ \theta_{i} = 1 $}
					{
						$ \beta_{i} = 1 $
					}
					{
						\For{neighbour particle $j = 1$ \KwTo $k$}
						{
							\If{$ \theta_{ij} = 1 $}
							{
								$ \beta_{i} = 1 $
				}}}}
			}
			\textbf{Procedure} Update\;
			\For{particle $i=1$ \KwTo $n$}
			{
				$ \theta_{i} = \beta_{i} $
			}
			\caption{The spatio-temporal identification method. The main procedures of the method are: initialization (lines 1 to 4),  solver (lines 5 to 20) and update (lines 21 to 24).}
		\end{algorithm} 
	}
	\subsection{Surface particle ensuring} 
	\label{section3-2}
\begin{figure}[htbp]
    	\centering 
    	\includegraphics[width =0.75\textwidth]{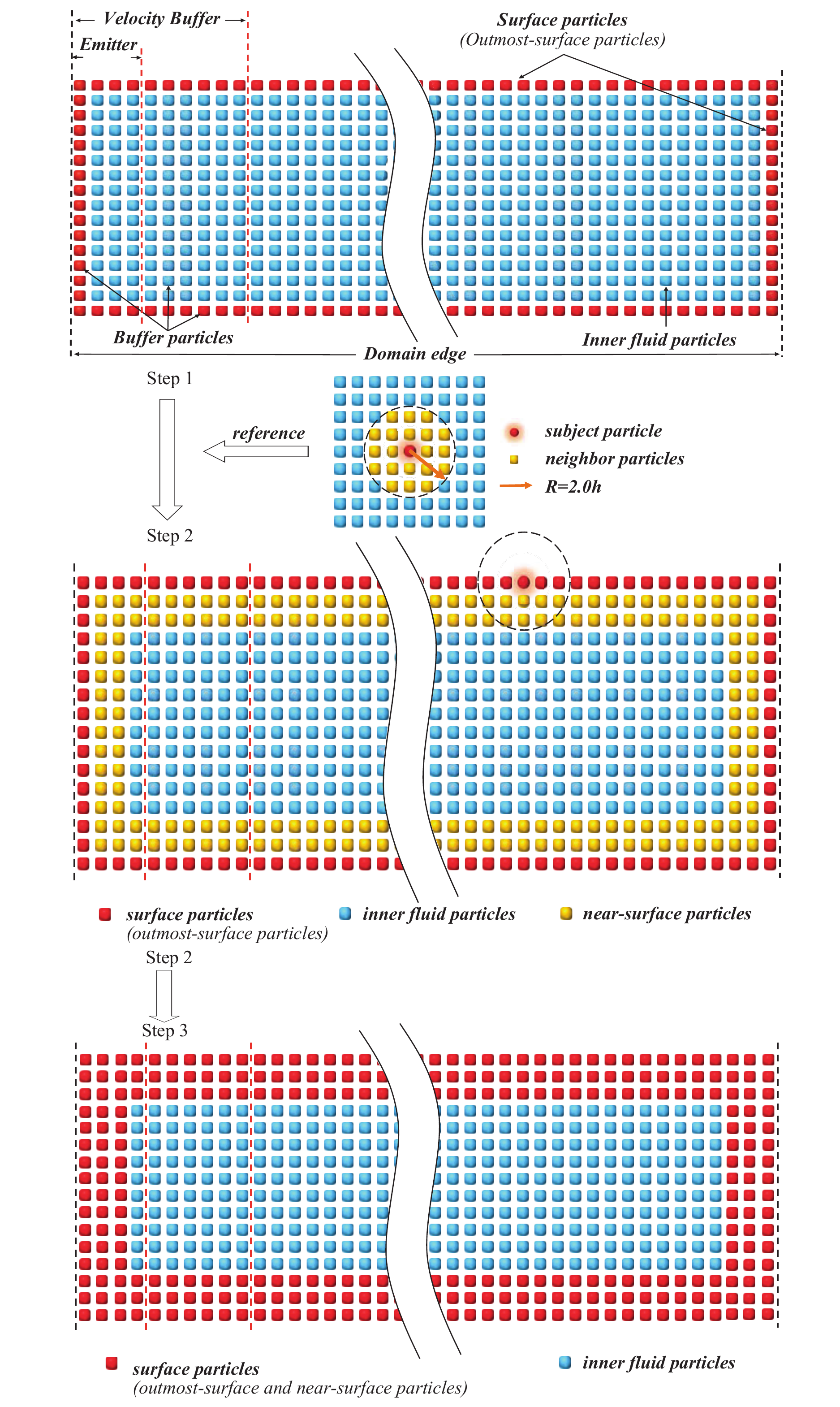}
    	\caption{Flow diagram of surface particles classification.}
    	\label{fig2}
    \end{figure}

    In order to solve the issue of missed non-regular surface particles, as shown in Fig.\ref{fig1a}, an ensuring technique is introduced. After applying the spatio-temporal identification, we have detected all surface particles with $ \nabla \cdot \mathbf{r}_{i} < 1.5 $, the ensuring technique includes all the neighbours of such particles as surface particles, as shown in Fig.\ref{fig2}.  
    Note that, while such a technique ensures all surface particles are detected and hence numerical stability, it slightly increases the surface layer's thickness. Because the fluid states near the free-stream boundary are smooth as defined by the boundary condition, a slight increase of free surface thickness does not smear the solution considerably.
	\subsection{Far-field corrections}
	While the density of inner fluid particles is still re-initialized by Eq.(\ref{eq6}), for all surface particles and their neighbours, to ensure numerical stability, their density is corrected according to Refs. \cite{Zhang2020,rezavand2021generalised} 
	
	\begin{equation} \label{eq10}
		\rho_{i}=\rho_{0}\dfrac{\sum W_{ij}}{\sum W^{0}_{ij}}+ max(0.0, (\rho_{i} -\rho_{0}\dfrac{\sum W_{ij}}{\sum W^{0}_{ij}})) * \rho_{0}/\rho_{i}, 
	\end{equation} 
	which smooths the summation approximation of density with the values updated with the continuity equation. Note that the assumption of smooth pressure distribution on surface particles here due to the weakly compressible formulation.
	
	Then, the velocities of all surface particles are also smoothed as follows
	
	\begin{equation} \label{eq11}
		\mathbf{v}=( v_{x}, v_{y})=
		\begin{cases} 
			v_{x} = U_{\infty}+ min(\rho_{i},\rho_{0})*(v_{x} - U_{\infty})/\rho_{0}   \\[3mm]
			v_{y} = v_{y}
		\end{cases},
	\end{equation} 
	where $ v_{x} $ and $ v_{y} $ are the velocity components in the stream and transverse directions, respectively. $ \rho_{i} $ is the re-initialized density obtained from Eq.(\ref{eq10}), and $ U_{\infty} $ the free-stream velocity. 
	Note that the transverse component $ v_{y} $ is actually free from the correction. 
	\subsection{In- and outflow boundary conditions}
	\label{section3-4}
	Due to the far-field corrections (density and velocity corrections), the fluid particles can freely flow out the domain bound and then be deleted immediately without introducing large fluctuation (as they are also identified as surface particles before being deleted). Therefore, as in previous work, a special outlet region is unnecessary, and only the inflow region is discussed. 
	
\begin{figure}[htbp]
		\centering  
		\subfigure[Initial configuration]{
			\begin{minipage}{0.8\linewidth}
				\includegraphics[width=1\textwidth]{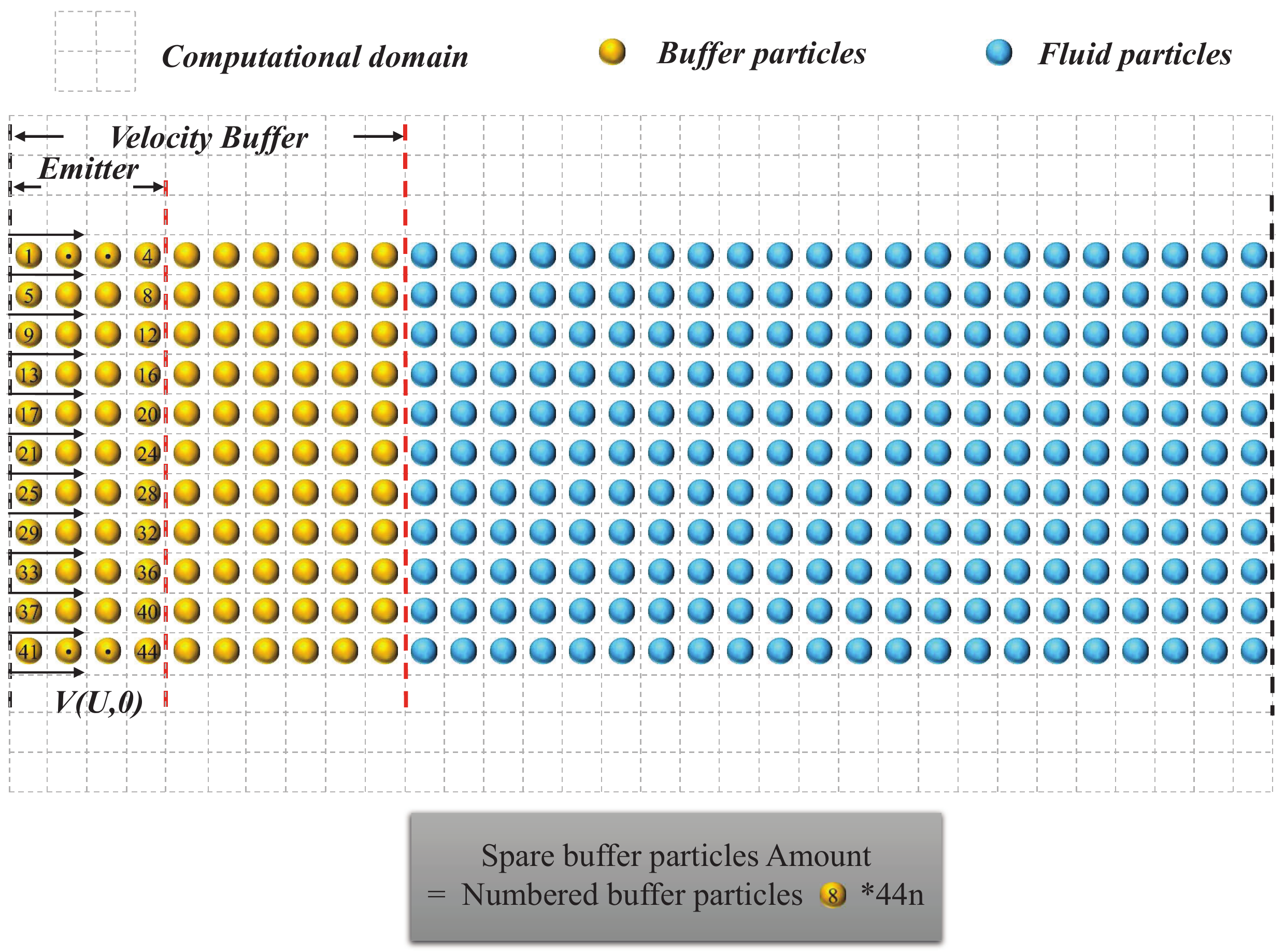}
			\end{minipage}
			\label{fig3a}
		}
		
		\subfigure[Working process]{
			\begin{minipage}{0.8\linewidth}
				\includegraphics[width=1\textwidth]{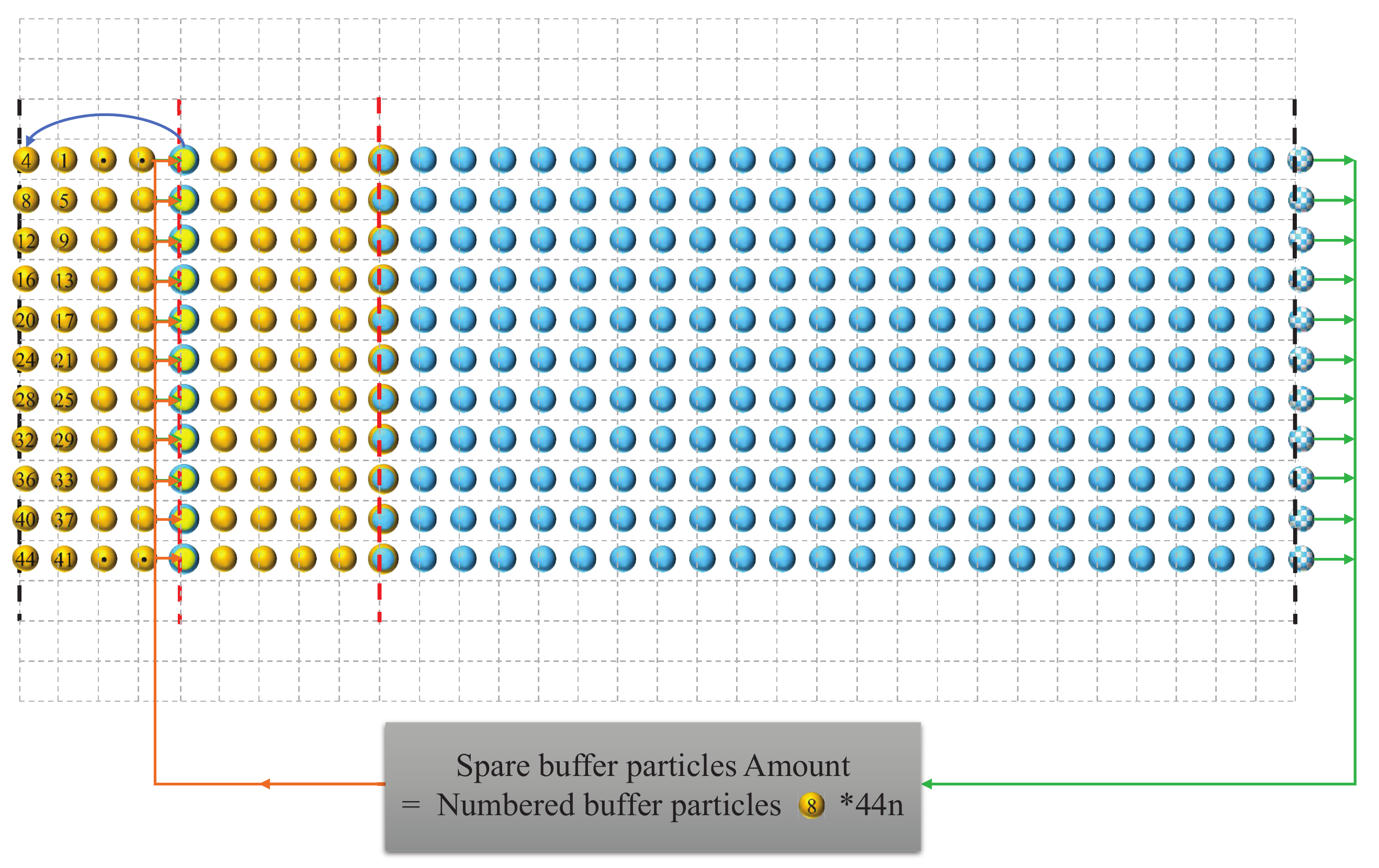}
			\end{minipage}
			\label{fig3b}
		}
		\caption{Schematic of the particle injection and deletion procedures.}
		\label{fig3}
	\end{figure}

	As shown in Fig.\ref{fig3a}, the inlet region is composed of two overlapped parts, namely the velocity buffer and emitter, respectively, for imposing the inflow velocity condition and particle generation. In order to ensure full support of particles near the boundary, the emitter should consist of at least four-layer particles in case of the smoothing length $ h=1.3dx $. Furthermore, to relax the possible velocity and eliminate the pressure fluctuations near the buffer-fluid interface, the velocity buffer width needs to be at least four times that of the emitter, and the buffer particle velocity should also be further relaxed to the target value $ \mathbf{v}_{target} =(U_{\infty}, 0)$, i.e. the free-stream velocity, as in Ref. \cite{Zhang2020}
	\begin{equation} \label{eq12}
		\mathbf{v_{i}}	\leftarrow \lambda\mathbf{v_{i}}+\mathbf{v}_{target}(1-\lambda),
	\end{equation} 
	where $ \lambda =0.7 $ is the relaxation strength. Note that there are some particles in the velocity buffer, as shown in Fig. \ref{fig2}, that are identified as surface particles and are also subject to the far-field density and velocity corrections.
	
	Fig.\ref{fig3b} demonstrates the detailed working procedure of the emitter as in Refs. \cite{Tafuni2018,Martin2009}. When a particle crosses the emitter bound, a new particle will be generated with the same state, and the original particle is recycled into the emitter with the following periodic position
	\begin{equation} \label{eq13}
		\mathbf{r^{*}}=\mathbf{r_{0}}-\mathbf{L},
	\end{equation} 
	where $ \mathbf{r_{0}} $ is the position of the particle at the instant leaving the emitter bound, and $ \mathbf{L}=(l,0) $ the distance vector with the emitter width $ l $. After fully relaxed in the velocity buffer, the new particle will be treated as an ordinary one entering the regular flow region. 
	
	\section{Results and discussion}
	\label{}
	In this section, we first consider two typical examples with free-stream boundary conditions, namely laminar flow over a flat plate and flow past a circular cylinder, to validate the accuracy and advantages of the proposed method. Then, another two test cases, including flow-induced oscillation of an elastic beam attached to a cylinder and flow past a sphere in three-dimension space, are computed to evaluate the performance and versatility of the present method for complex flow problems. 
	
	For these non-steady test cases, the designated free-stream velocity is smoothly imposed on the particle in the velocity buffer in the following manner \cite{Turek2007}
	
	\begin{equation} \label{eq14}
		U_{\infty} (t) =
		\begin{cases} 
			0.5U_{\infty}(1-cos(\frac{\pi t}{t_{0}}))     & t<t_{0}\\[3mm]
			U_{\infty}         & otherwise
		\end{cases},
	\end{equation} 
	where $ t_{0} $ is the acceleration time. 
	
	However, even if the Eq.(\ref{eq14}) is adopted, particles near the buffer-fluid interface still experience a severe change at the simulation start-up. Since there are no upper and lower solid walls, the flow domain may deform severely. In order to reduce this start-up effect, both fluid and buffer particles are accelerated as
	
	\begin{equation} \label{eq15}
		\frac{d\mathbf{v}}{dt}=(a_{x}, 0)=
		\begin{cases} 
			0.5U_{\infty}\frac{\pi}{t_{0}}sin(\frac{\pi t}{t_{0}})     & t<t_{0}\\[3mm]
			0        & otherwise
		\end{cases},
	\end{equation} 
	where $ a_{x} $ is the acceleration in $ x $- direction. 
	
	Furthermore, since the global background pressure in the transport-velocity formulation will lead to force imbalance on the surface particles, the transport-velocity formulation is only implemented on inner fluid particles. In practice, the velocity buffer and emitter placed at the left inlet zone generally consist of 20 and 8 particle layers, respectively. The 5th-order Wendland kernel \cite{Wendland1995} is adopted herein, and the smoothing length and support radius are set to be $ h=1.3dx $ and $ r=2.0h $, respectively. All the above parameters remain unchanged in all the subsequent simulations unless there are specific indications.
	\subsection{Laminar flow over a flat plate}
	\label{section4-1}
\begin{figure}[htbp]
		\centering     
		\includegraphics[width=0.8\textwidth]{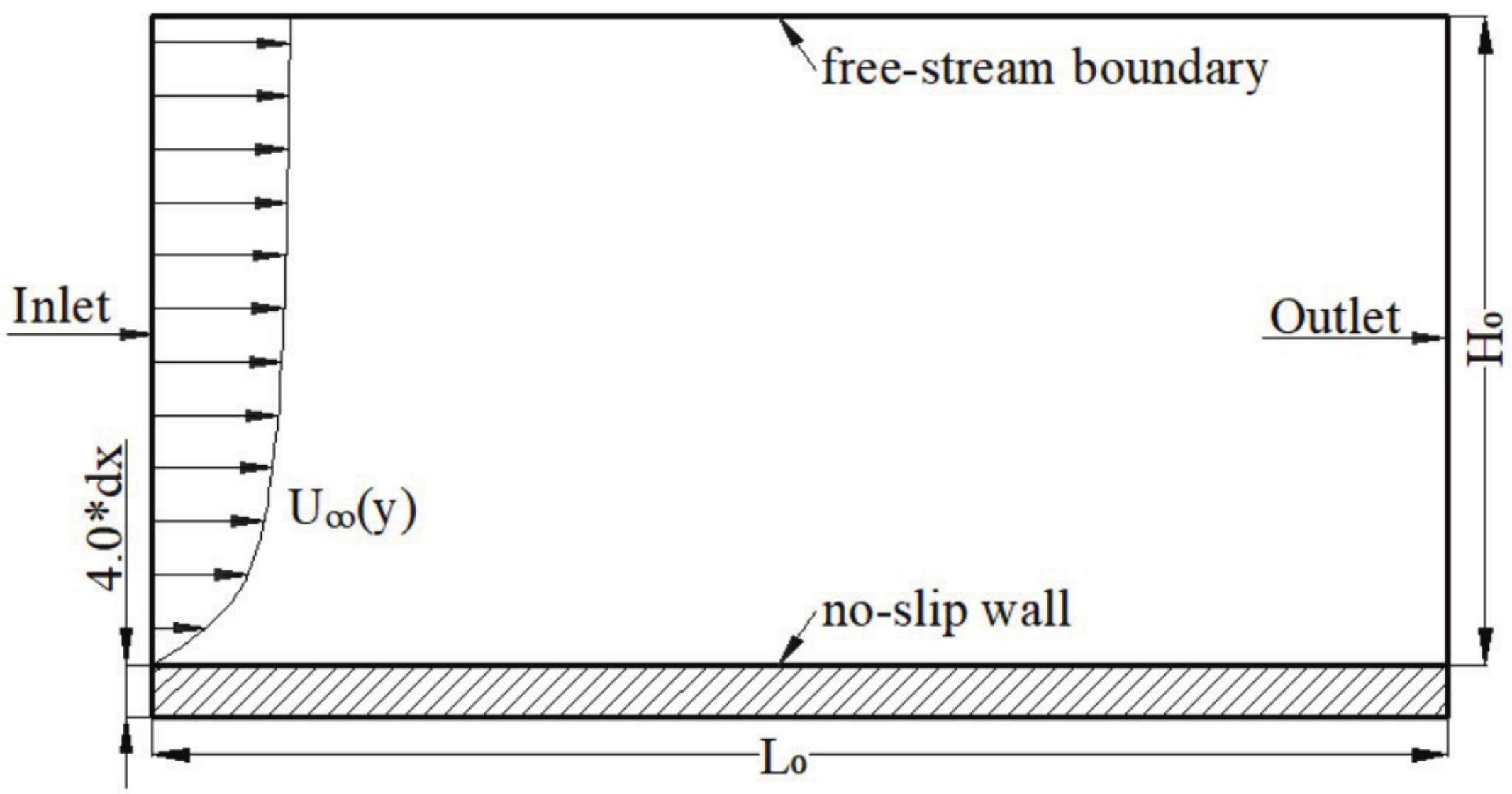}
		\caption{Schematic of the 2-D laminar flow over a flat plate with free-stream boundary.}
		\label{fig4}
	\end{figure}

    Referring to the gravity-driven flow over a sloping channel bed \cite{Federico2012}, the 2-D laminar flow over a flat plate with a free-stream boundary is computed to validate the proposed method. Implementation details, including the distribution and detection of surface particles, are also examined. The schematic of the initial setting is shown in Fig.\ref{fig4}.  
	Initially, a rectangular fluid domain is chosen with a length of $ L_{0}=2H_{0} $, where $ H_{0}=0.05m $ is the water depth. The fluid particles are placed on a Cartesian lattice with a uniform particle spacing of $ dx=0.0025 m $. The solid flat plate has a thickness of 4 particle layers. With the non-slip condition on the flat plate, the velocity profile at the inlet can be given by the following parabolic equation
	\begin{equation} \label{eq16}
		U_{\infty} (y)=\dfrac{\rho g S_{0}}{2\mu}(2H_{0}y-y^{2}),
	\end{equation} 
	where the gravity acceleration $ g=9.8 m/s^{2} $, the bottom slope $ S_{0}=0.001 $, $ \mu $ the dynamics viscosity, and $ y $ the span-wise distance to plate. 
	In this case, the Reynolds number $ Re=\rho U_{ave} H_{0} / \mu = 100$ is fixed, and the fluid density is initialized with a constant value $ \rho=1000kg/m^{3} $. Thus, the dynamics viscosity $ \mu $ can be calculated by
	\begin{equation} \label{eq17}
		\begin{cases} 
			\mu=\dfrac{\rho U_{ave} H_{0}}{Re}	\\[3mm]
			U_{ave}=\dfrac{1}{H_{0}}\int_{0}^{H_{0}} \dfrac{\rho g S_{0}}{2\mu}(2H_{0}y-y^{2})\, dy=\dfrac{\rho gS_{0}H_{0}^{2}}{3\mu}
		\end{cases}
	\end{equation}
	Namely, $ \mu = 0.0639 Pa\cdot s$, and the corresponding free-stream velocity can be obtained by $ U_{\infty} (H_{0})=\rho g S_{0}H_{0}^{2}/2\mu=0.1917m/s $. 
	
\begin{figure}[htbp]
		\centering     
		\subfigure[Time: $ t = 0.0 s $ ]{
			\begin{minipage}{0.45\linewidth}
				\includegraphics[width=1\textwidth]{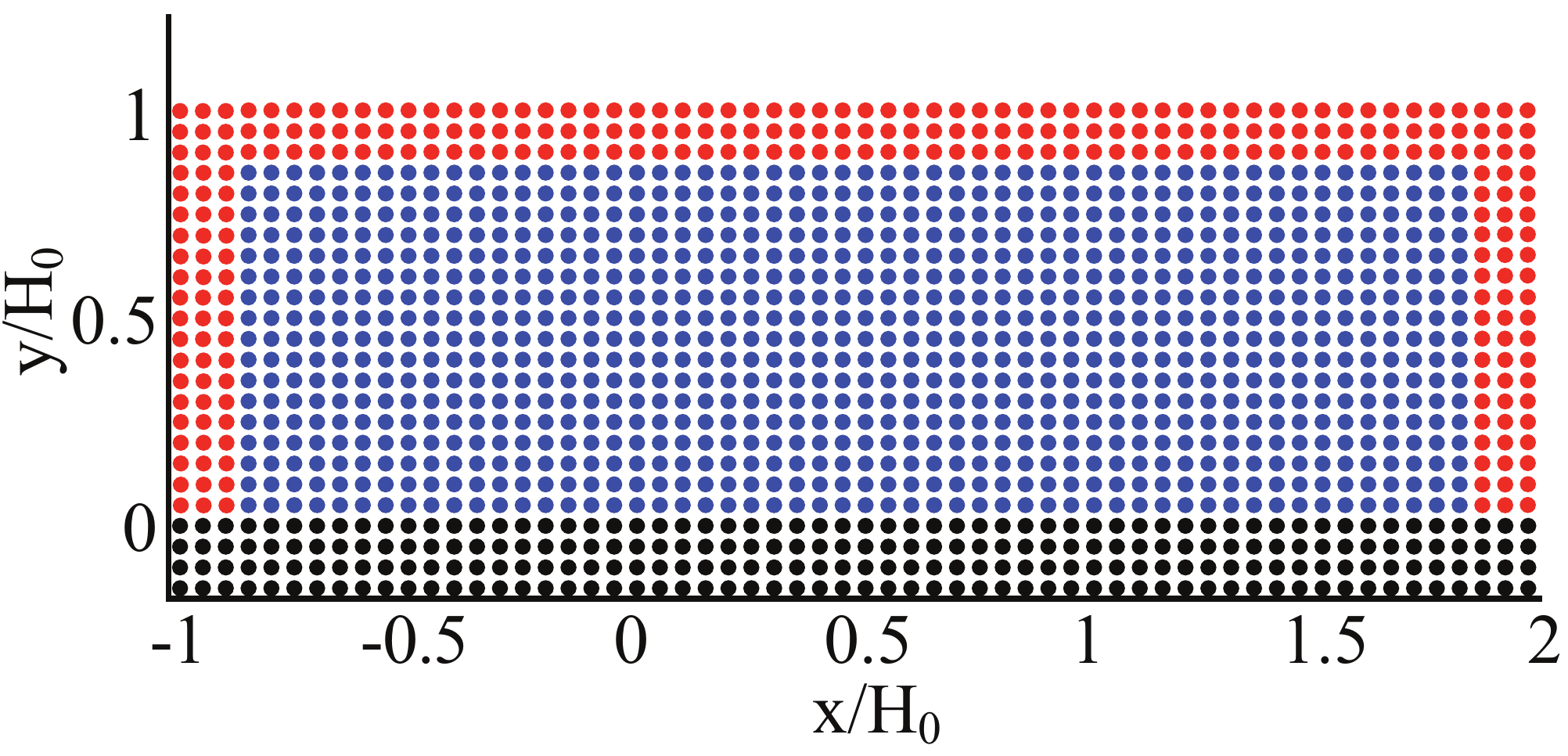}
			\end{minipage}
			\label{fig5a}
		}	
		\subfigure[Time: $ t = 20.0 s $]{
			\begin{minipage}{0.45\linewidth}
				\includegraphics[width=1\textwidth]{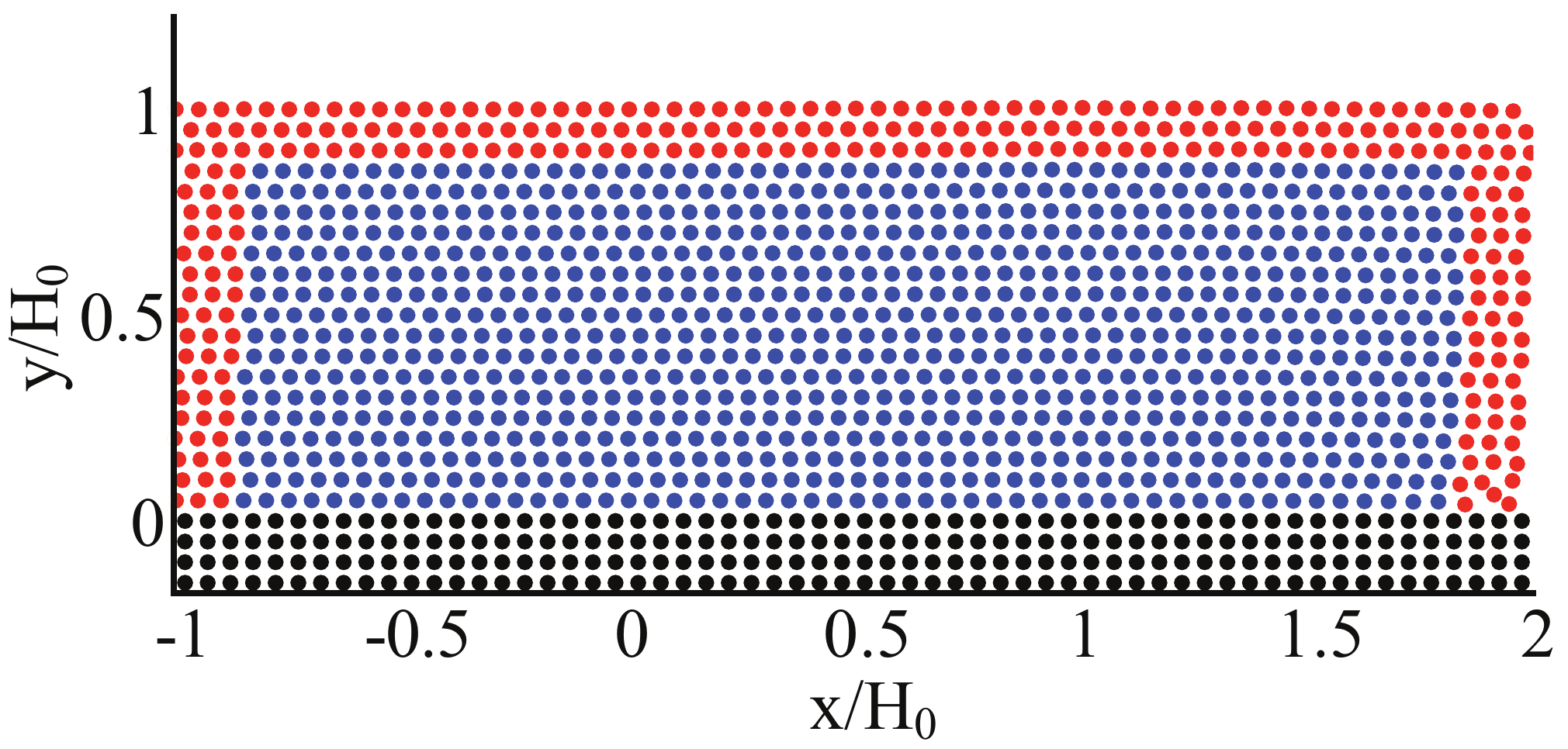}
			\end{minipage}
			\label{fig5b}
		}
		\caption{Particle identification map at two time instances. Particle categories: the blue inner fluid particles, the red surface particles, and the black solid wall particles.}
		\label{fig5}
	\end{figure}

\begin{figure}[htbp]
		\centering     
		\includegraphics[width=0.6\textwidth]{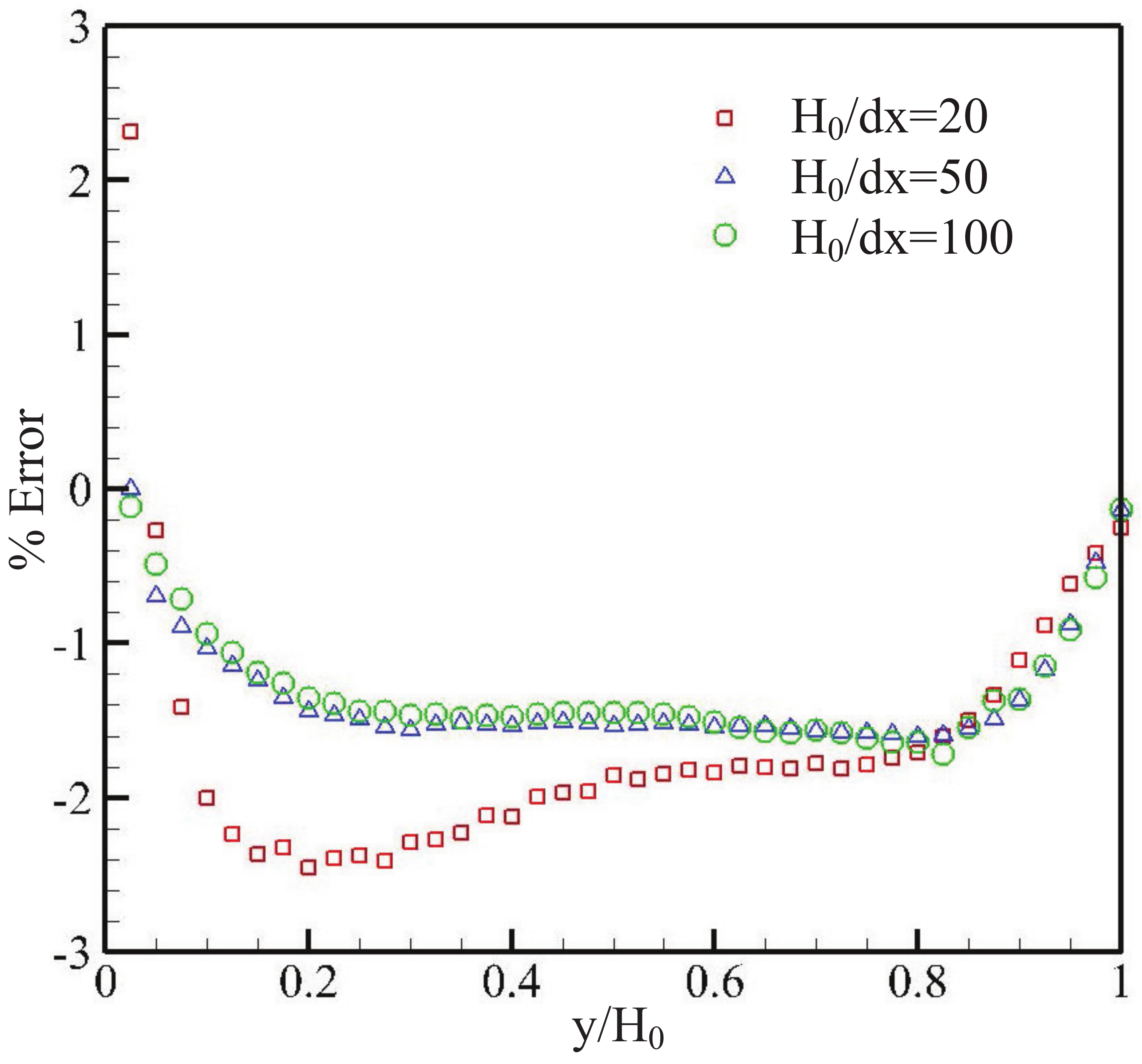}
		\caption{Errors between the numerical and analytical solutions with different particle resolutions at $ t=20.0s $.}
		\label{fig6}
	\end{figure}

\begin{figure}[htbp]
		\centering     
		\includegraphics[width=0.6\textwidth]{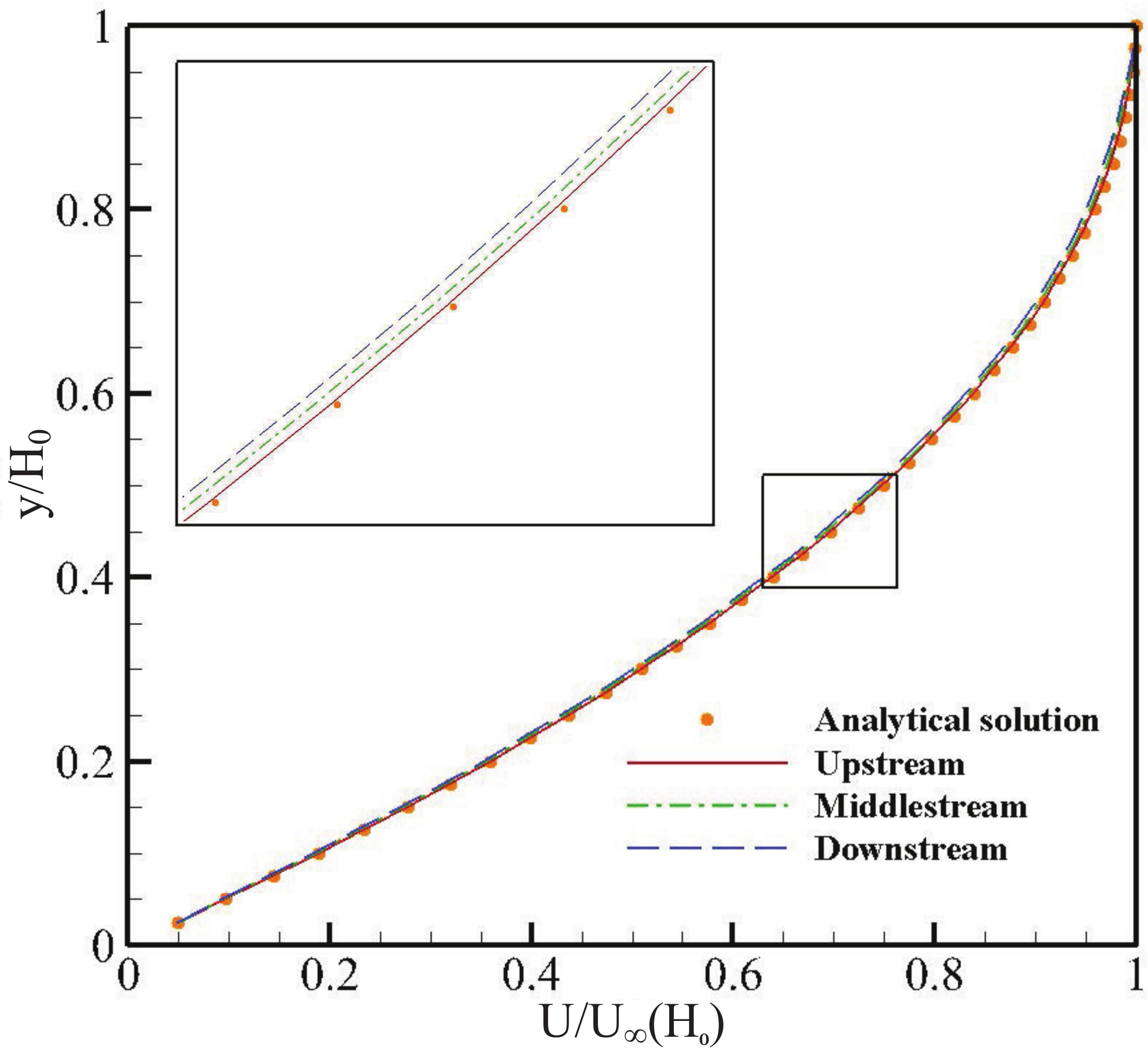}
		\caption{Comparison of the velocity profiles in numerical and analytical solutions at $ t = 30.0 s $, and the upstream, middlestream, and downstream positions, $ x=0.03m$, $ 0.06m$, and $ 0.09 m$. The initial uniform particle spacing here is $ dx=0.001m$, 
			and the velocity and height are respectively normalized by the free-stream velocity $ U_{\infty} (H_{0}) =0.1917 m/s $ and the initial depth $ H_{0} = 0.1 m $.}
		\label{fig7}
	\end{figure}

	In this flow, as indicated in Fig.\ref{fig5}, even if the fluid particles are no longer regularly arranged, they can still be identified and classified accurately. 
	When comparing with the analytical solution, the span-wise velocity profile $ U (x,y) $ at the position $ x=0.09m$ is sampled since the largest error may occur near the outlet. The related error between the numerical and analytical solutions in Fig.\ref{fig6} is calculated through
	\begin{equation} \label{eq18}
		\%\,Error = \dfrac{U (x,y)-U_{\infty} (y)}{U_{ave}} \times 100.0.
	\end{equation}
	It is known that the numerical errors of SPH method are not only introduced by discretization error proportional to $ dx/h $, but also by the smoothing error proportional to $ h/L $ ($ L $ being the characteristic length of the problem geometry, $ L=H_{0} $ here) \cite{Quinlan2006, Springel_2010, litvinov2015towards}. With the maximum resolution $ H_{0}/dx=100 $ adopted here, 
	the solution reaches the saturation regime, and is sufficiently close to the theoretical solution \cite{Federico2010, Tafuni2016} with an overall error 
	within 1.5$ \% $ observed in Fig.\ref{fig6}. Therefore, it is not necessary to achieve the full convergence with $ dx/h $ going to zero \cite{Springel_2010,Ellero2011SPHSO, MARRONE2013456}. As shown in Fig.\ref{fig7}, the span-wise velocity profiles at another two positions in stream-wise direction are also extracted and validated against the theoretical solution. 
	At these three different cross-sections, all velocity distributions from the numerical simulation are in accordance with the theoretical solution. Compared with the results in previous work \cite{Federico2010, Tafuni2016}, the present free-stream condition offers an overall smoother velocity profile and smaller deviation from the theoretical solution and guarantees the velocity consistency of the surface particles. 
	\subsection{2-D flow past a circular cylinder}
	\label{section4-2}
\begin{figure}[htbp]
		\centering     
		\includegraphics[width=0.7\textwidth]{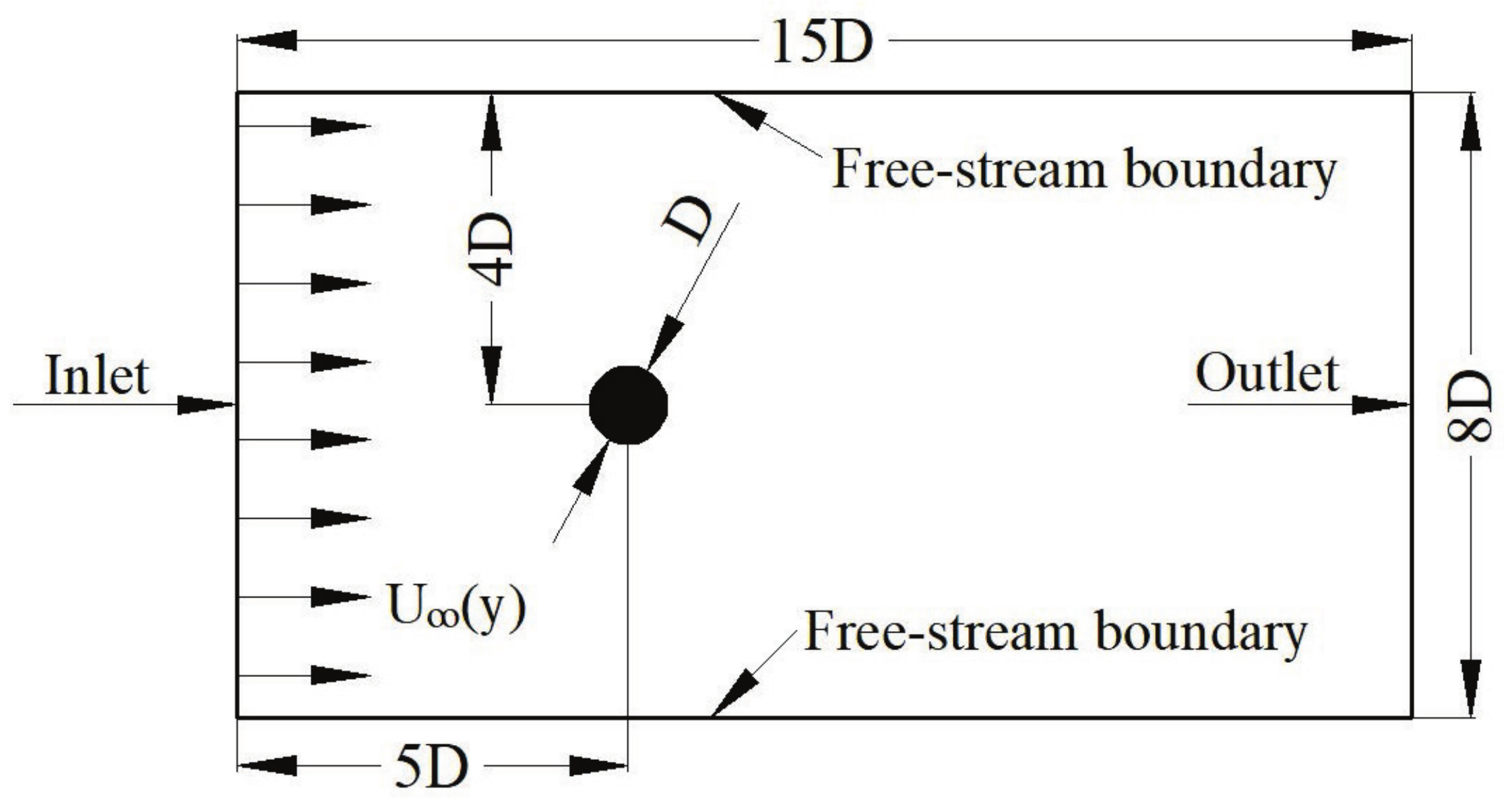}
		\caption{Schematic of the 2-D flow past a circular cylinder.}
		\label{fig8}
	\end{figure}

	The flow past a circular cylinder with free-stream boundary conditions is computed to verify the present method further. This problem has been extensively studied through numerical simulation and experiments. As shown in Fig.\ref{fig8}, a cylinder with the diameter of $ D=0.02 m $ is located on the horizontal axis of symmetry and $ 5D $ away from the inlet, and the initial length and height of the computational domain are $ L_{0} = 15 D $ and $ H_{0} = 8 D $, respectively.  
	The uniform particle spacing is initially set to $ dx = 2.5 \times 10^{-4}m $, 
	giving the total number of SPH particles approximately $ 8 \times 10^{5} $. 
	The fluid density $ \rho=1000kg/m^{3} $, and the uniform inlet velocity profile is targeted as a constant value in y-direction $ U_{\infty}(y)=1m/s $. The dynamic viscosity is calculated by $ \mu=\rho U_{\infty} D/Re $, and we test several cases with different Reynolds numbers of $ Re = 20 $, 50, 100, and 200. 

	\begin{figure}[htbp]
	\centering     
	\includegraphics[width=0.45\textwidth]{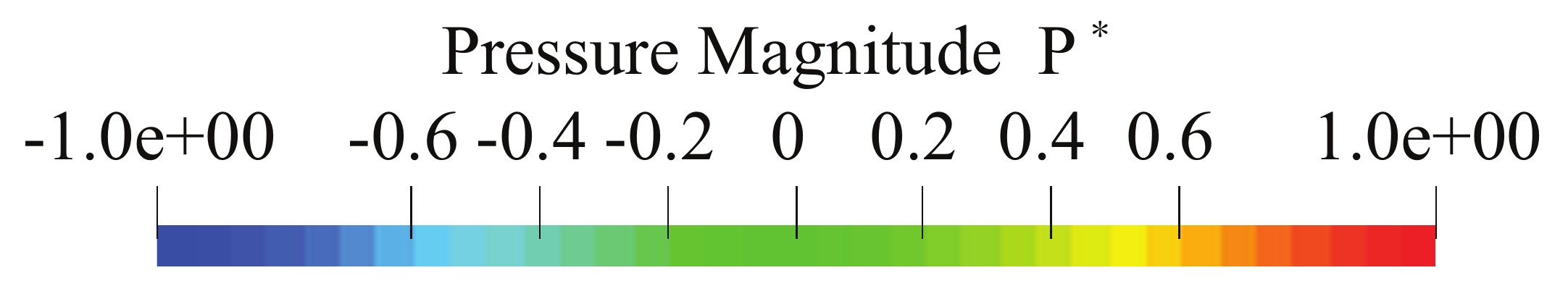}
	
	\centering     
	\subfigure[$ Re=20 $]{
		\begin{minipage}[b]{0.45\linewidth}
			\includegraphics[width=1\textwidth]{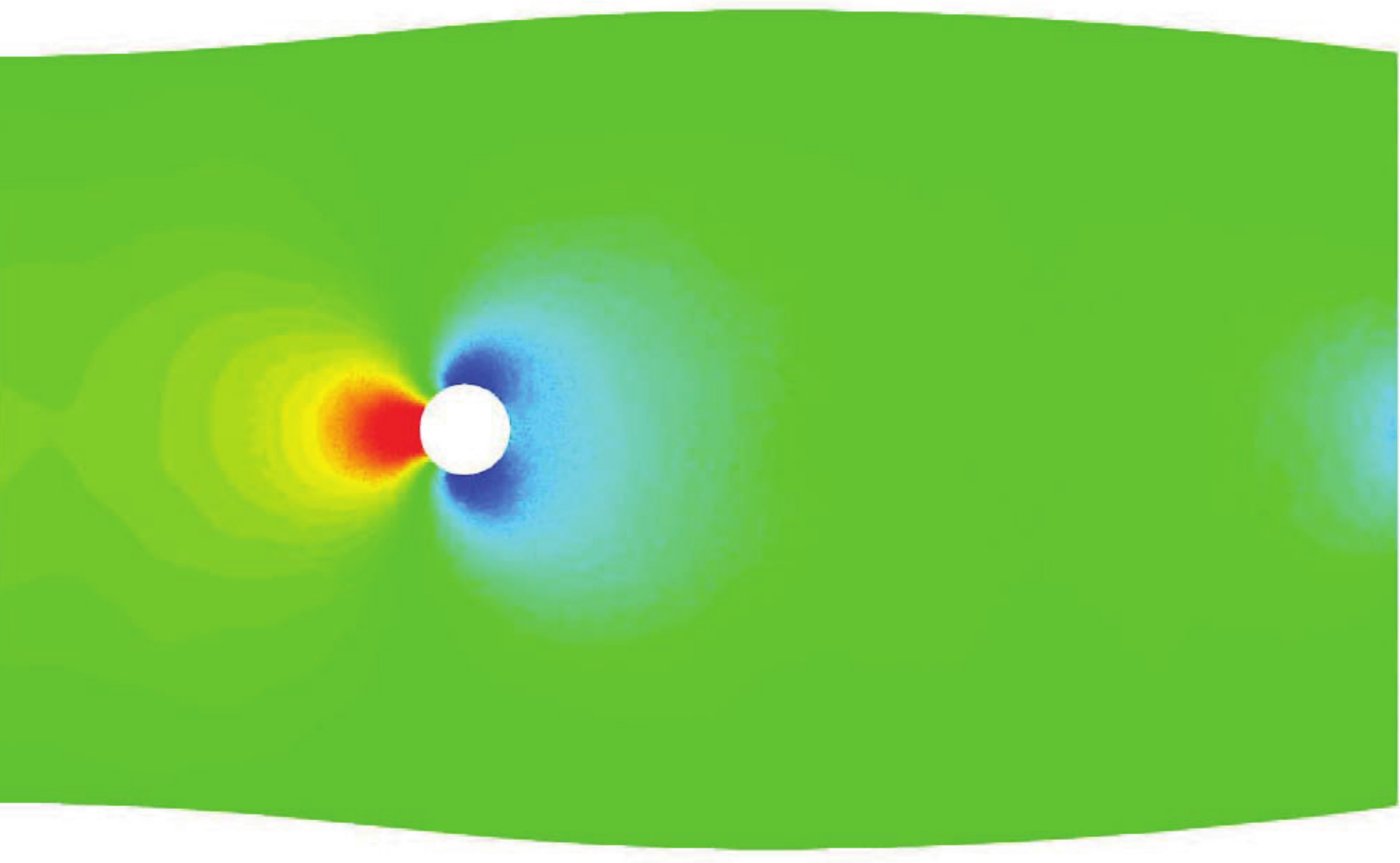}
		\end{minipage}
	}	
	\subfigure[$ Re=50 $]{
		\begin{minipage}[b]{0.45\linewidth}
			\includegraphics[width=1\textwidth]{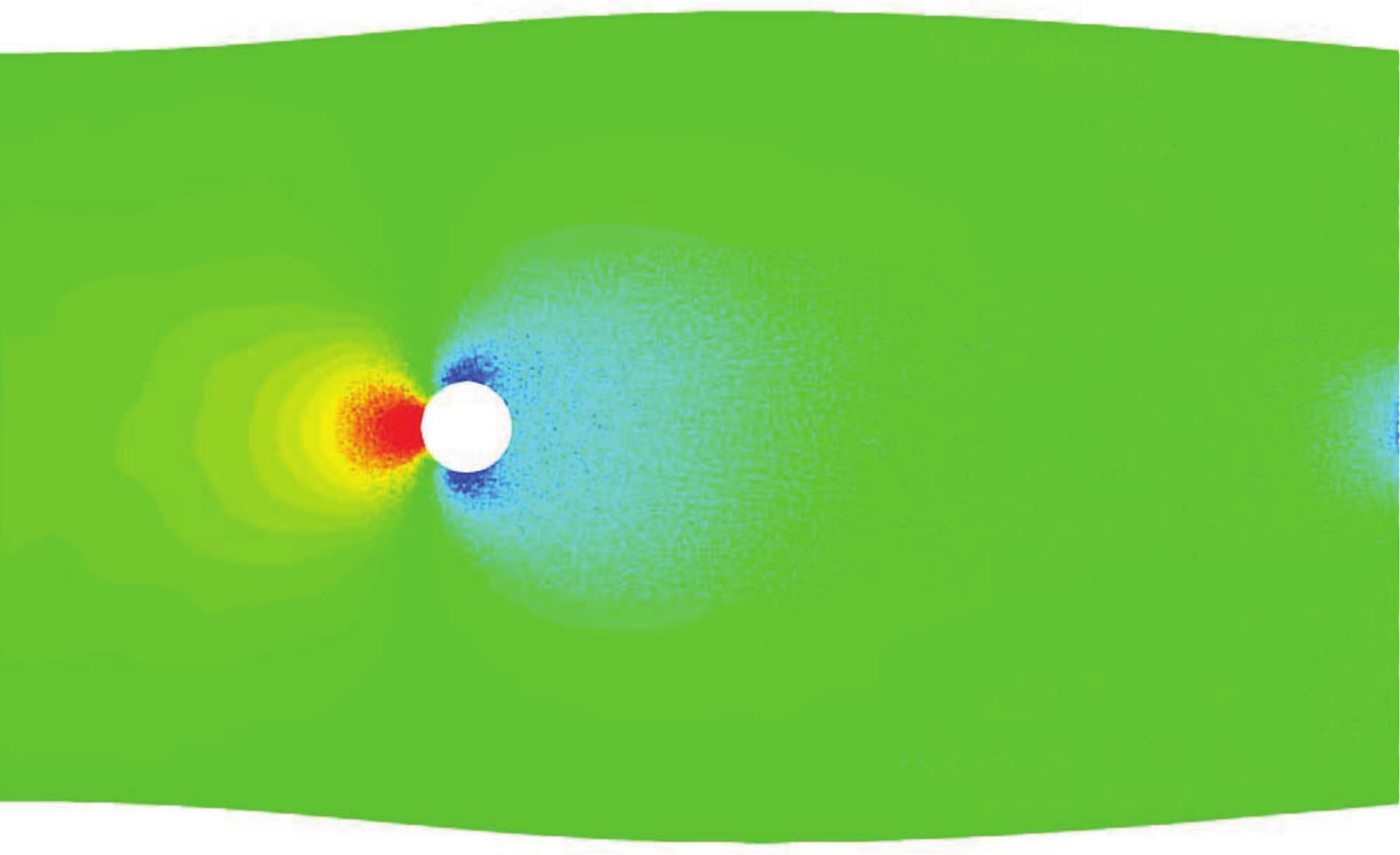}
		\end{minipage}
	}
	\subfigure[$ Re=100 $]{
		\begin{minipage}[b]{0.45\linewidth}
			\includegraphics[width=1\textwidth]{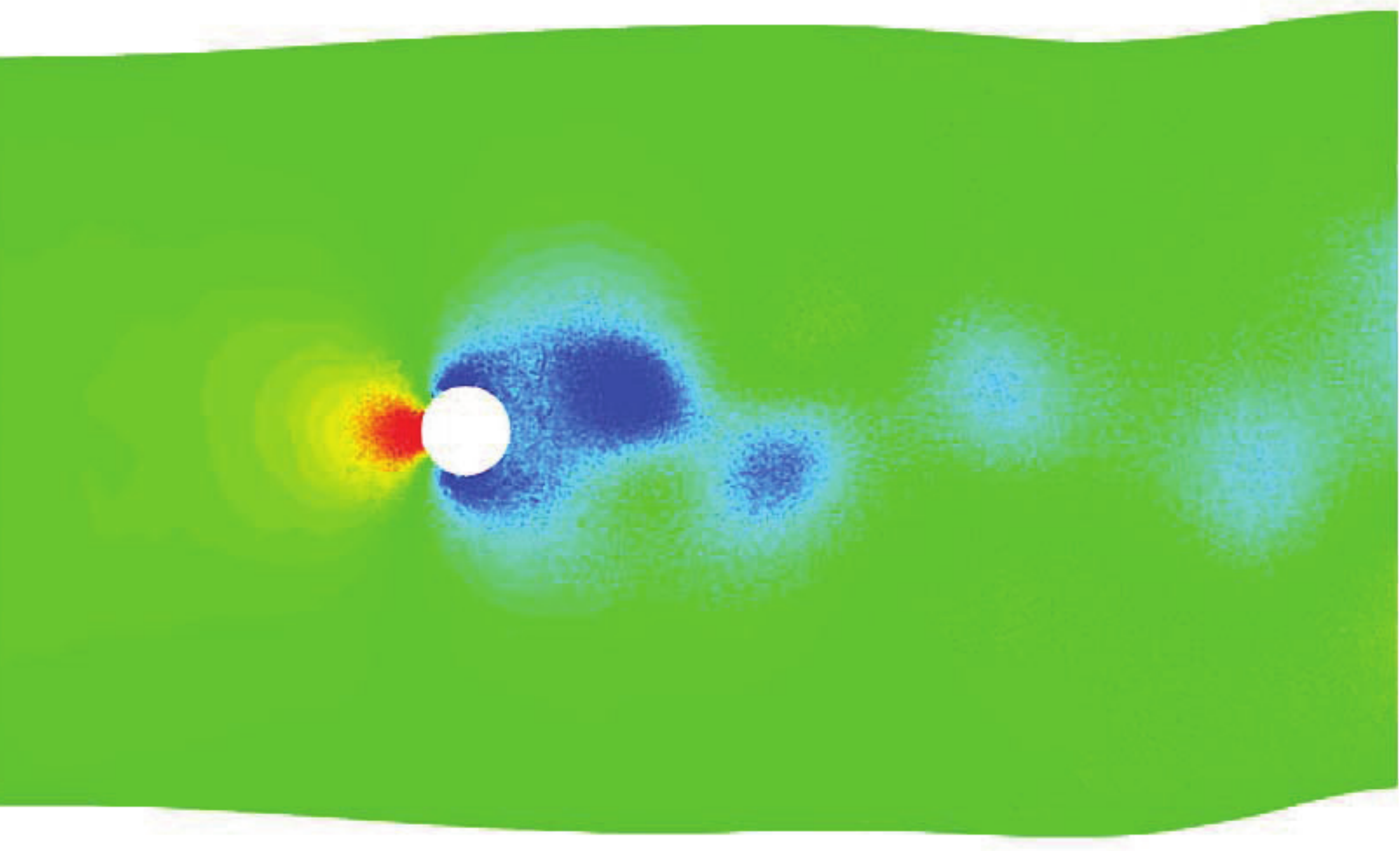}
		\end{minipage}
	}	
	\subfigure[$ Re=200 $]{
		\begin{minipage}[b]{0.45\linewidth}
			\includegraphics[width=1\textwidth,height=0.59\linewidth]{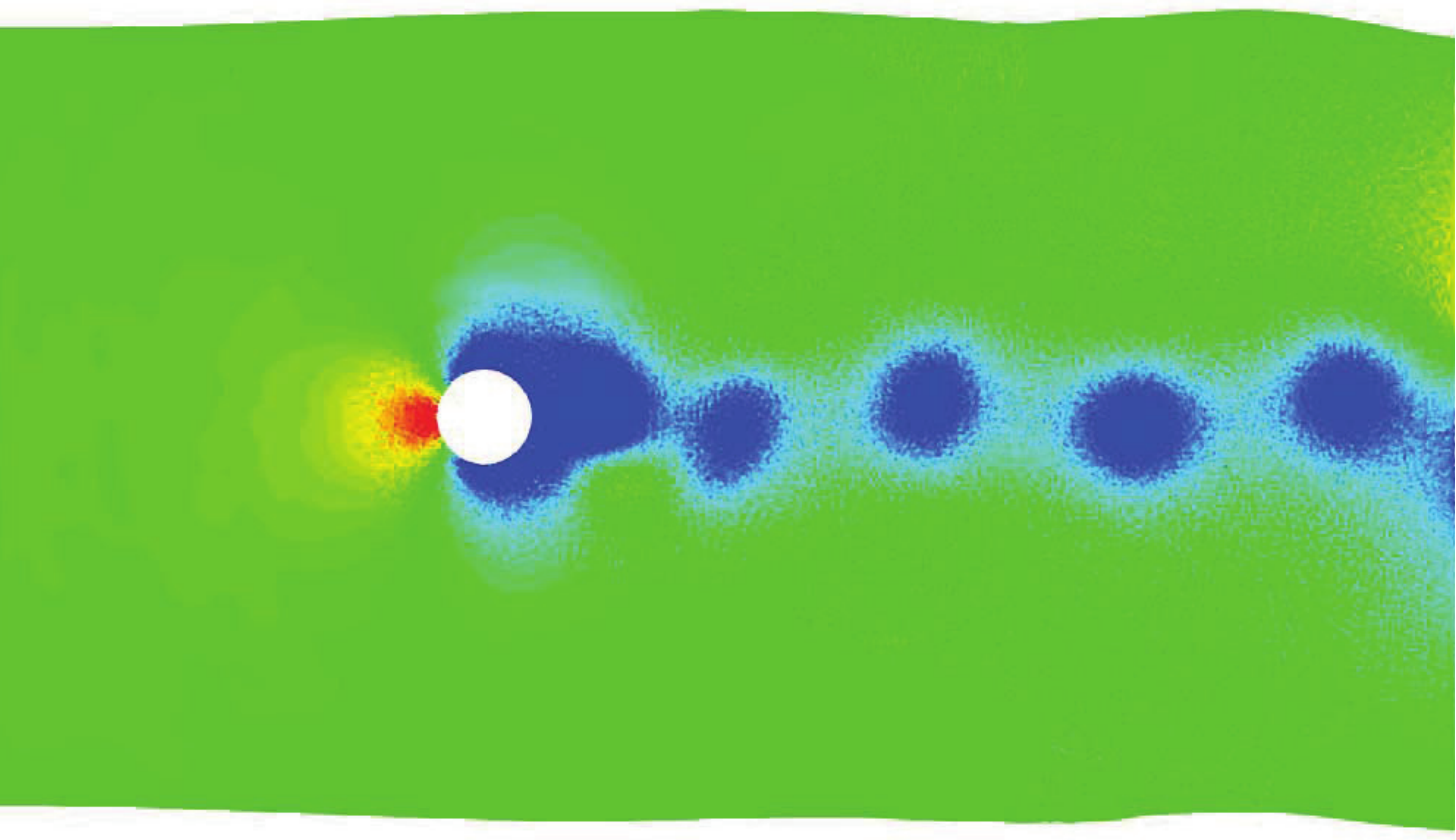}
		\end{minipage}
	}
	\caption{Pressure contours at $ t = 100.0 s $ for different Reynolds numbers ranging from $ Re = 20 $ to 200.}
	\label{fig9}
    \end{figure} 
	\begin{figure}[htbp]
		\centering     
		\includegraphics[width=0.45\textwidth]{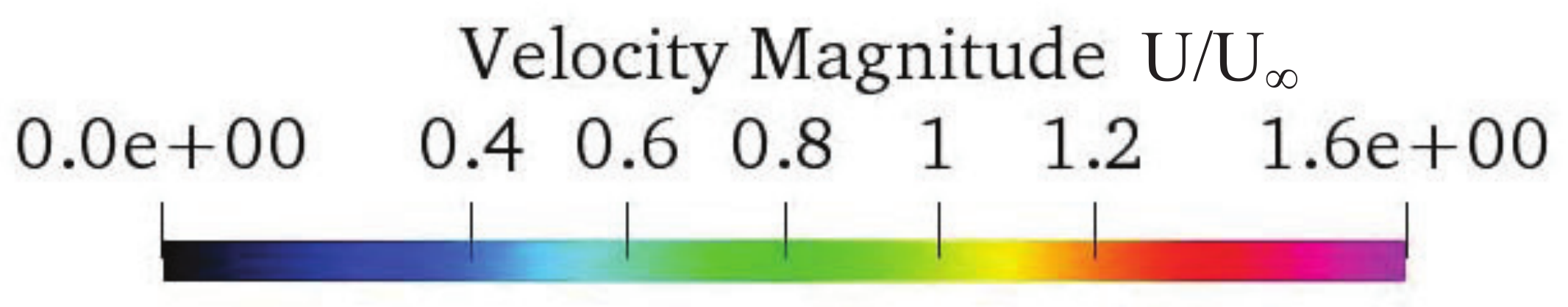}
		
		\centering     
		\subfigure[$ Re=20 $]{
			\begin{minipage}[b]{0.45\linewidth}
				\includegraphics[width=1\textwidth]{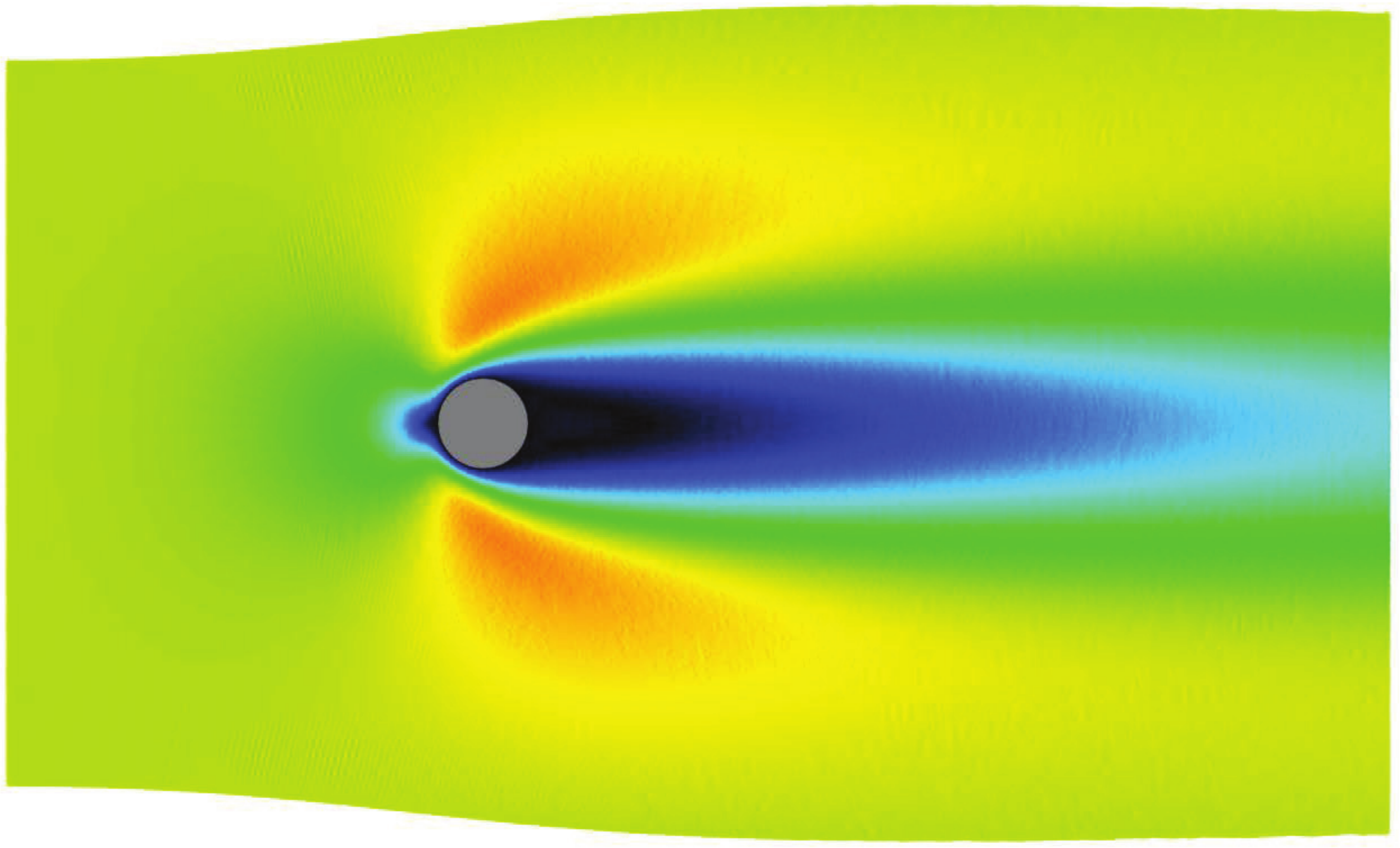}
			\end{minipage}
		}	
		\subfigure[$ Re=50 $]{
			\begin{minipage}[b]{0.45\linewidth}
				\includegraphics[width=1\textwidth]{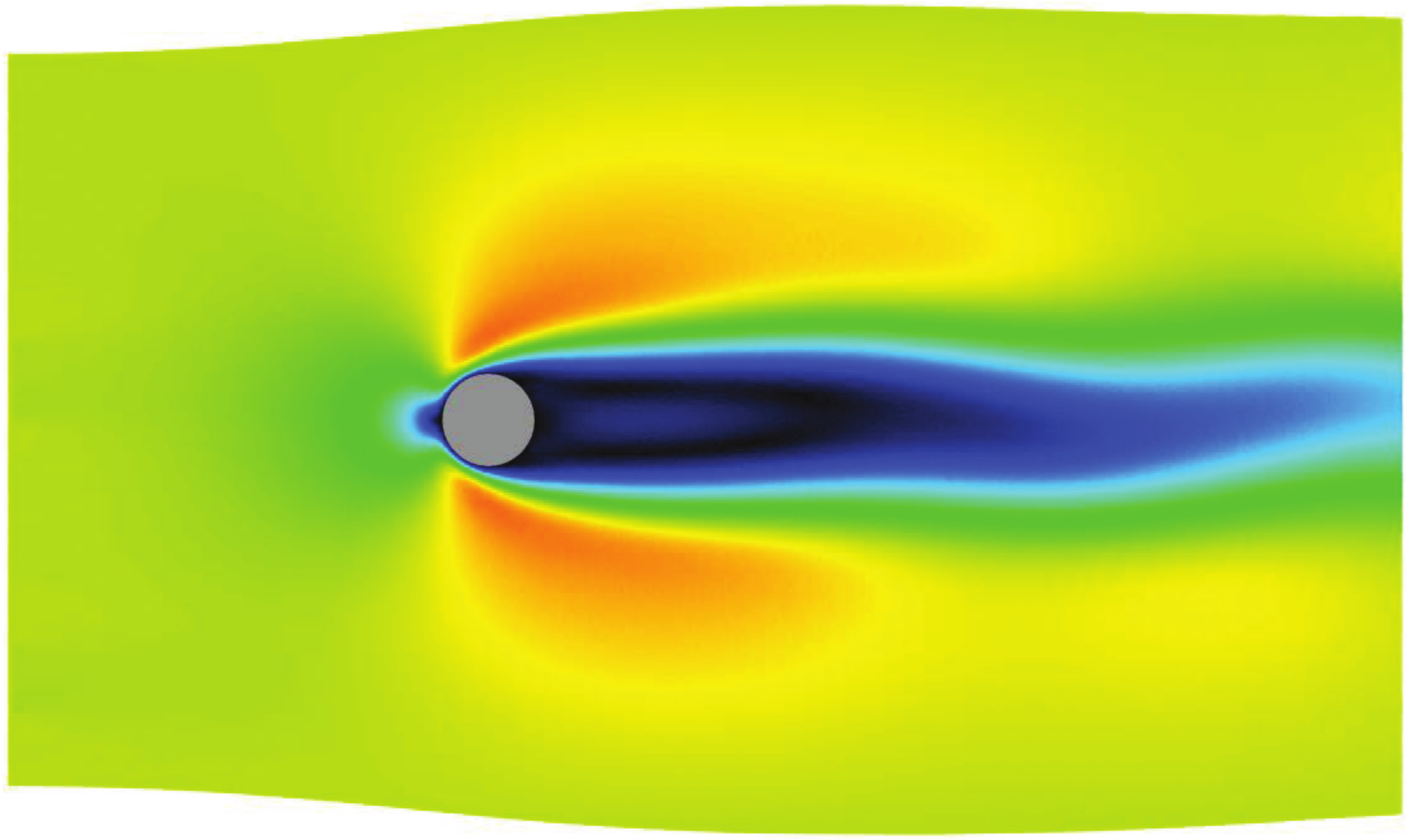}
			\end{minipage}
		}
		\subfigure[$ Re=100 $]{
			\begin{minipage}[b]{0.45\linewidth}
				\includegraphics[width=1\textwidth]{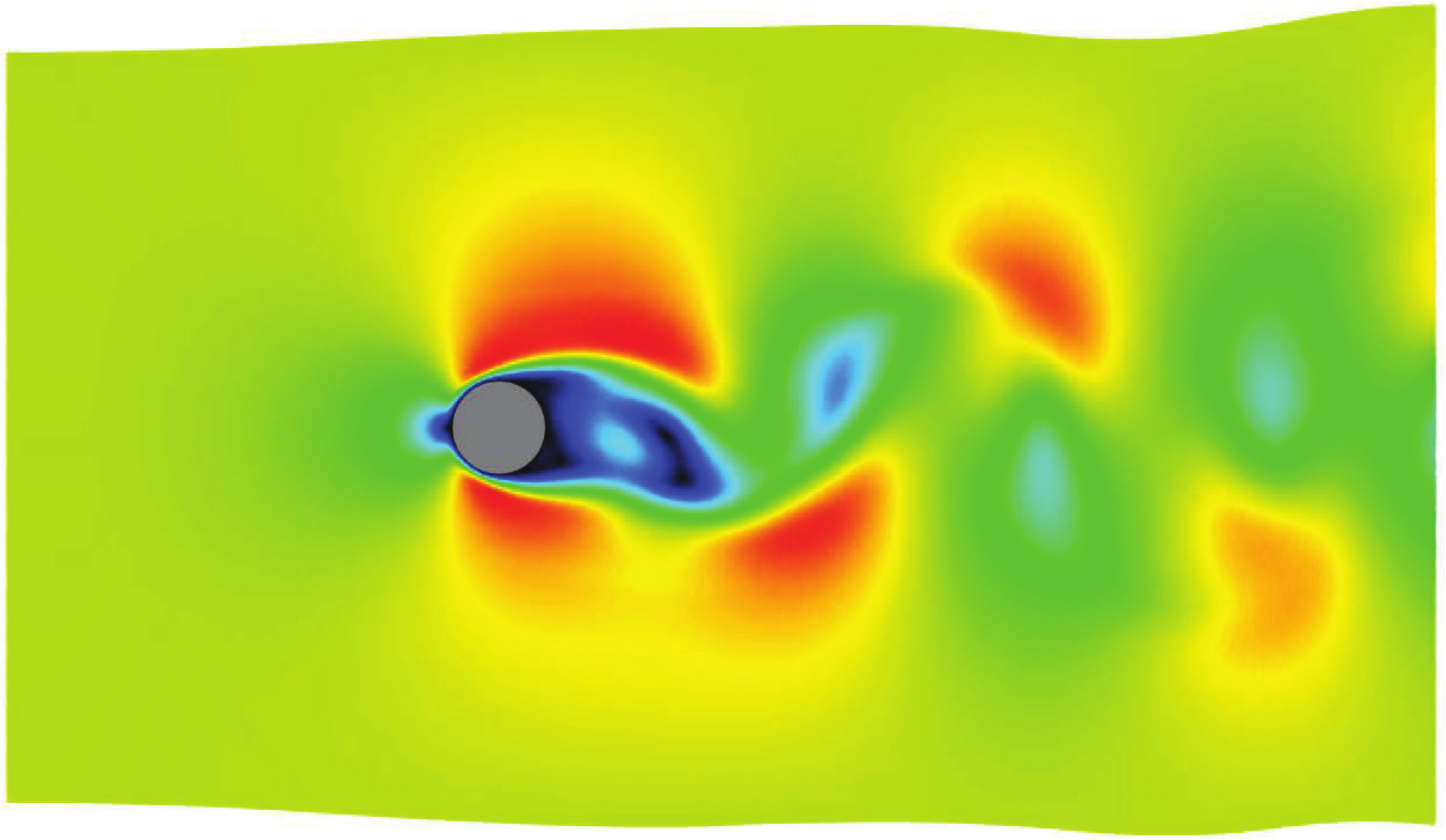}
			\end{minipage}
		}	
		\subfigure[$ Re=200 $]{
			\begin{minipage}[b]{0.45\linewidth}
				\includegraphics[width=1\textwidth,height=0.59\linewidth]{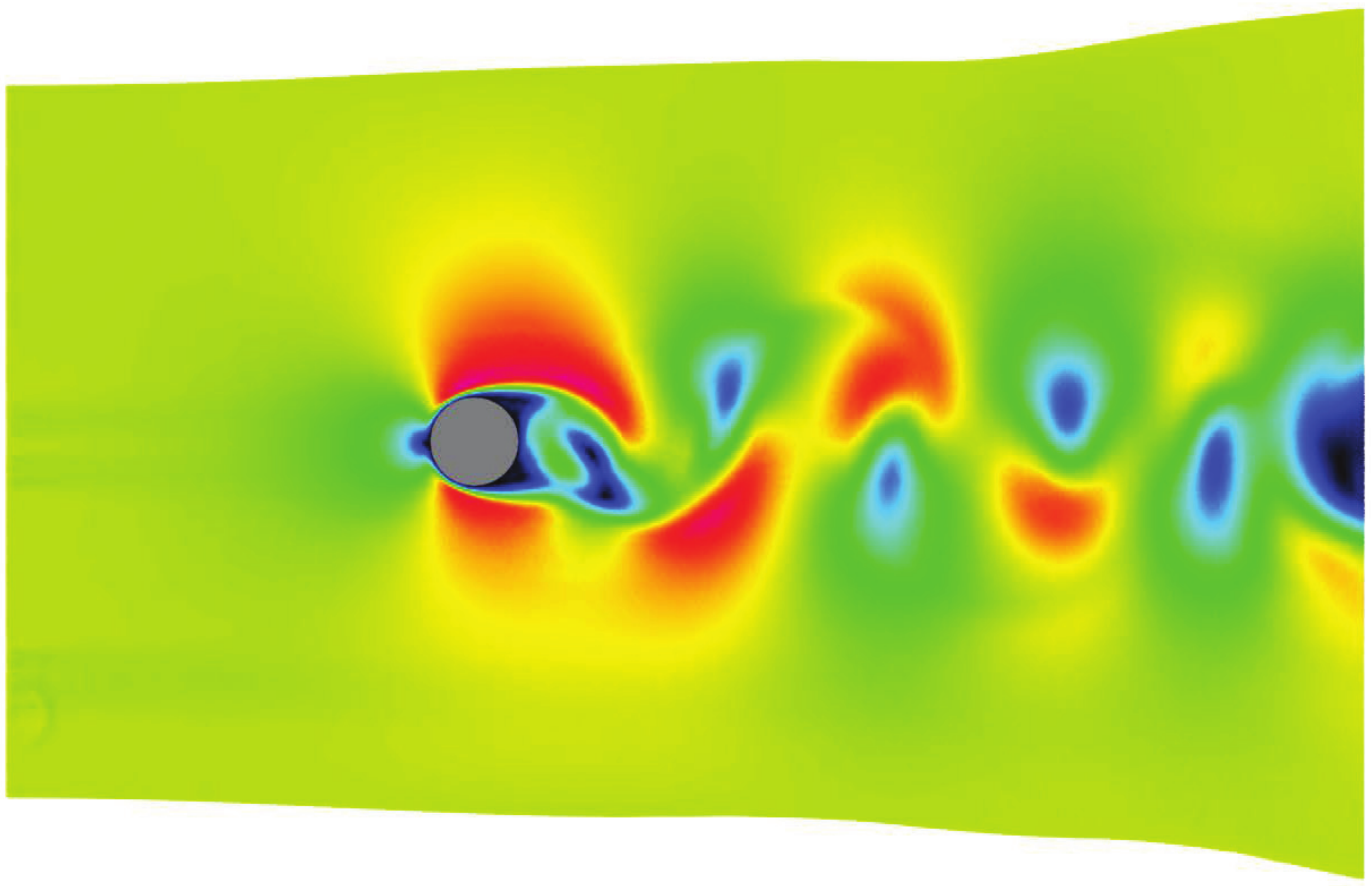}
			\end{minipage}
		}
		\caption{Velocity contours at $ t = 200.0 s $ for different Reynolds numbers ranging from $ Re = 20 $ to 200.}
		\label{fig10}
	\end{figure} 
    Fig.\ref{fig9} shows the smooth pressure contours for all cases, $P^{*}=2p\rho_{0}^{-1}U_{\infty}^{-2}$, and the previous common spurious pressure fluctuation at the boundary around \cite{Ferrand2017,Colagrossi2013, Wang2019,Khorasanizade2015, Alvarado-Rodriguez2017, Negi2020} is suppressed well by the far-field corrections on surface particles. Figure \ref{fig10} exhibits the flow speed contours at $ t = 200.0 s $ for all the cases. For the steady case of $ Re = 20 $, both the deformations of the inner flow field and the free-stream boundary are maintained symmetrically, and stagnation areas can be clearly found at up and downstream of the cylinder. As for the transition case of $ Re = 50 $, 
	the flow becomes unstable and asymmetric, and the free-stream boundary deforms with the evolution of the flow. In the unsteady cases $ Re = 100 $ and $ Re = 200 $, the vortex sheds off the cylinder periodically, and the Karman vortex street behind the cylinder can be clearly observed. Simultaneously, the boundary deformation is more violent in these two cases, whereas the boundary velocity still keeps smooth. In general, all the present results are qualitatively in agreement with the corresponding results in Ref. 
	\cite{White2006, Brehm2015, Almarouf2017, Liu1998, Le2006, Russell2003, Taira2007, Jin1993}.
	
\begin{figure}[htbp]
		\centering    
		\subfigure[$ Re=20 $]{
			\begin{minipage}[b]{0.45\linewidth}
				\includegraphics[width=1\textwidth]{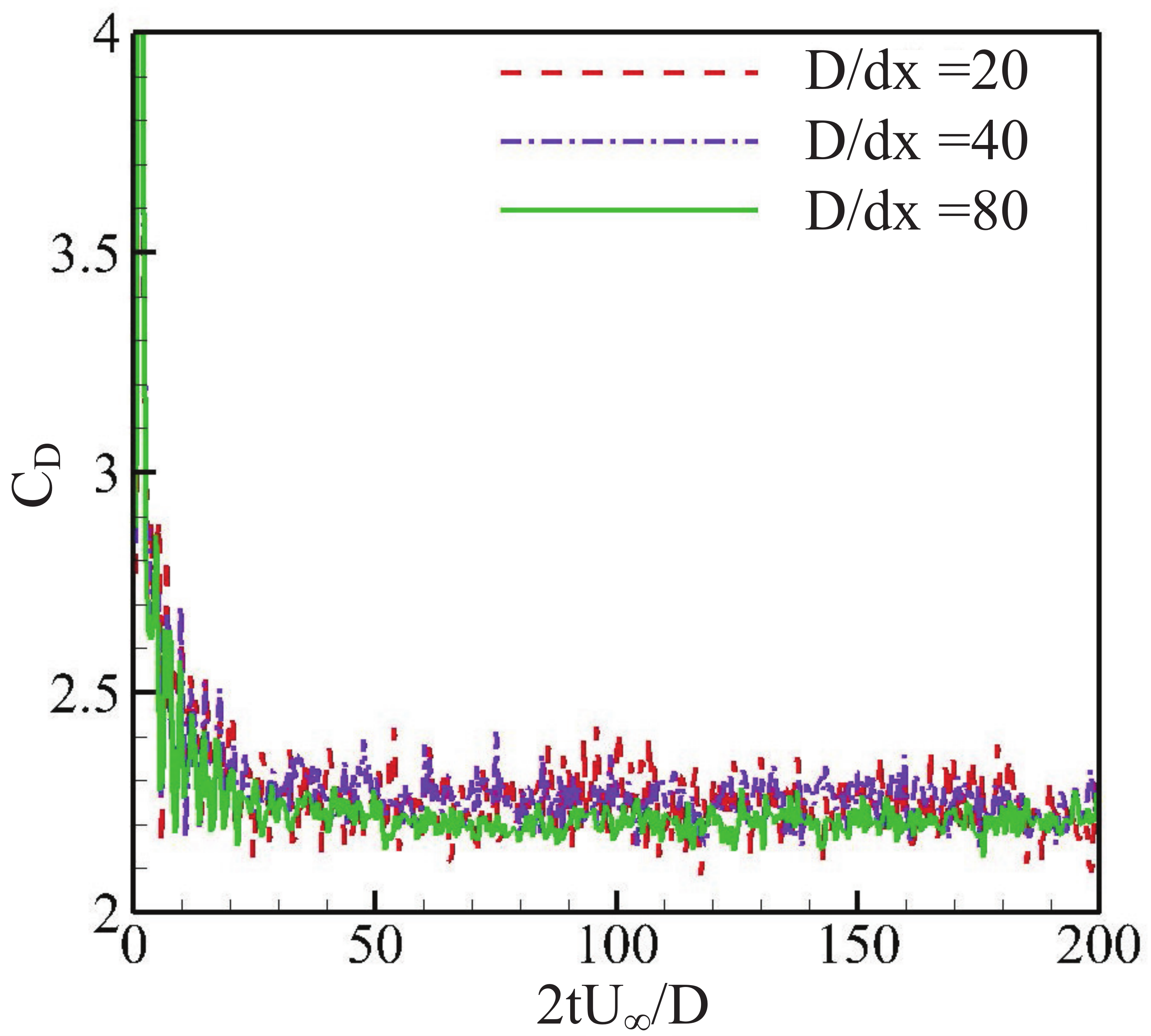}
			\end{minipage}
			\begin{minipage}[b]{0.45\linewidth}
				\includegraphics[width=1\textwidth]{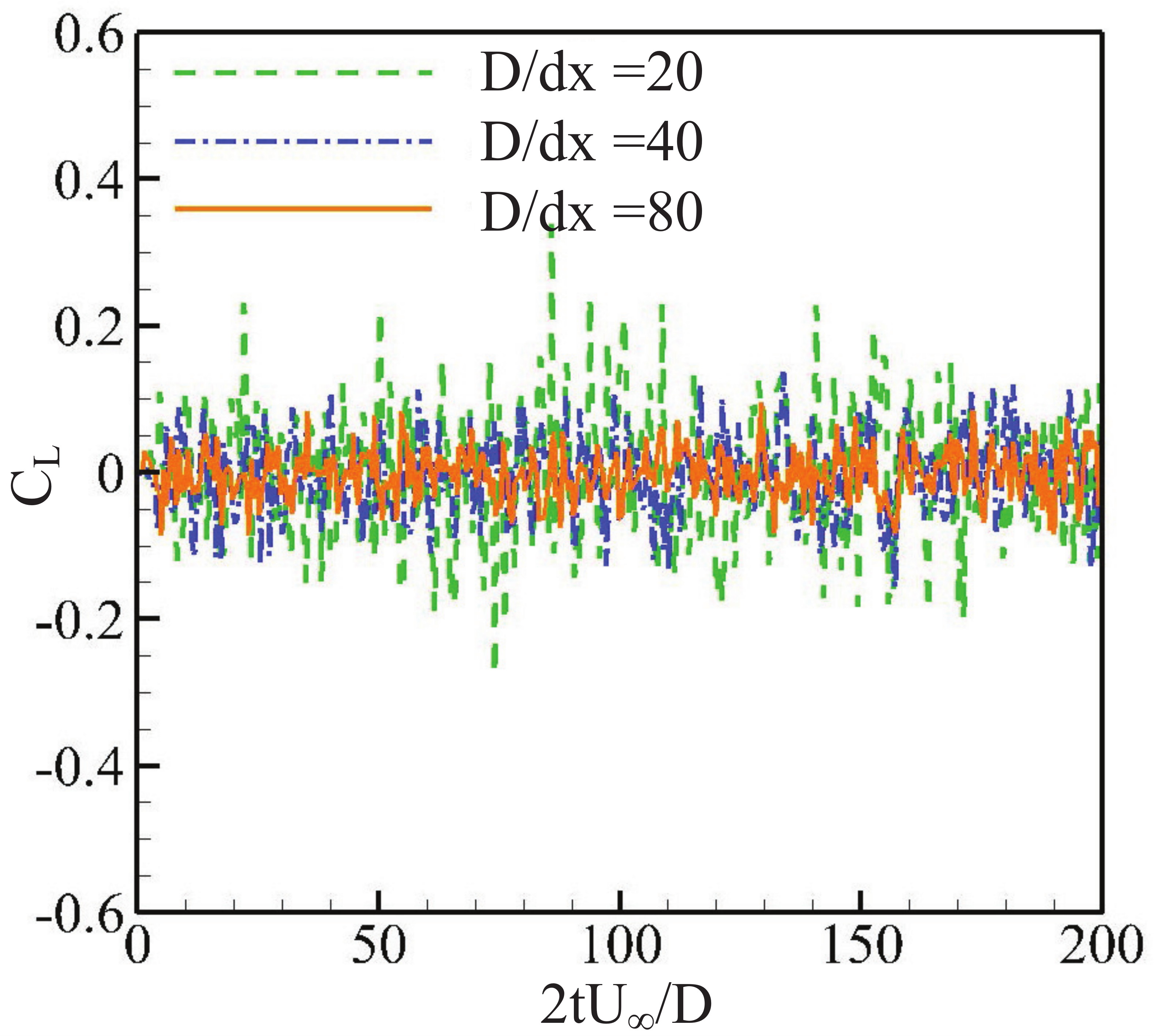}
			\end{minipage}
			\label{}
		}	 	
		\subfigure[$ Re=100 $]{
			\begin{minipage}[b]{0.45\linewidth}
				\includegraphics[width=1\textwidth]{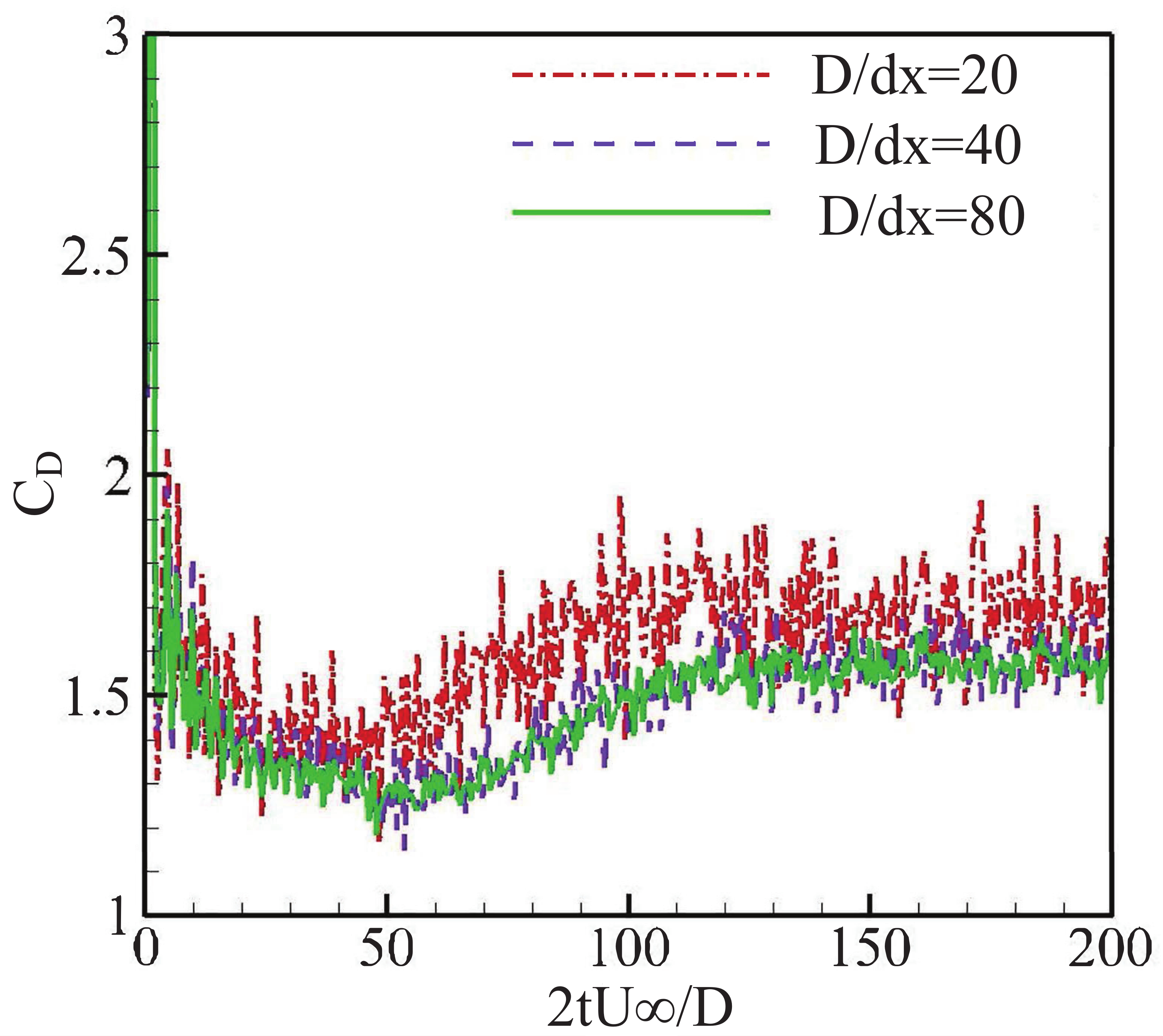}
			\end{minipage}
			\begin{minipage}[b]{0.45\linewidth}
				\includegraphics[width=1\textwidth]{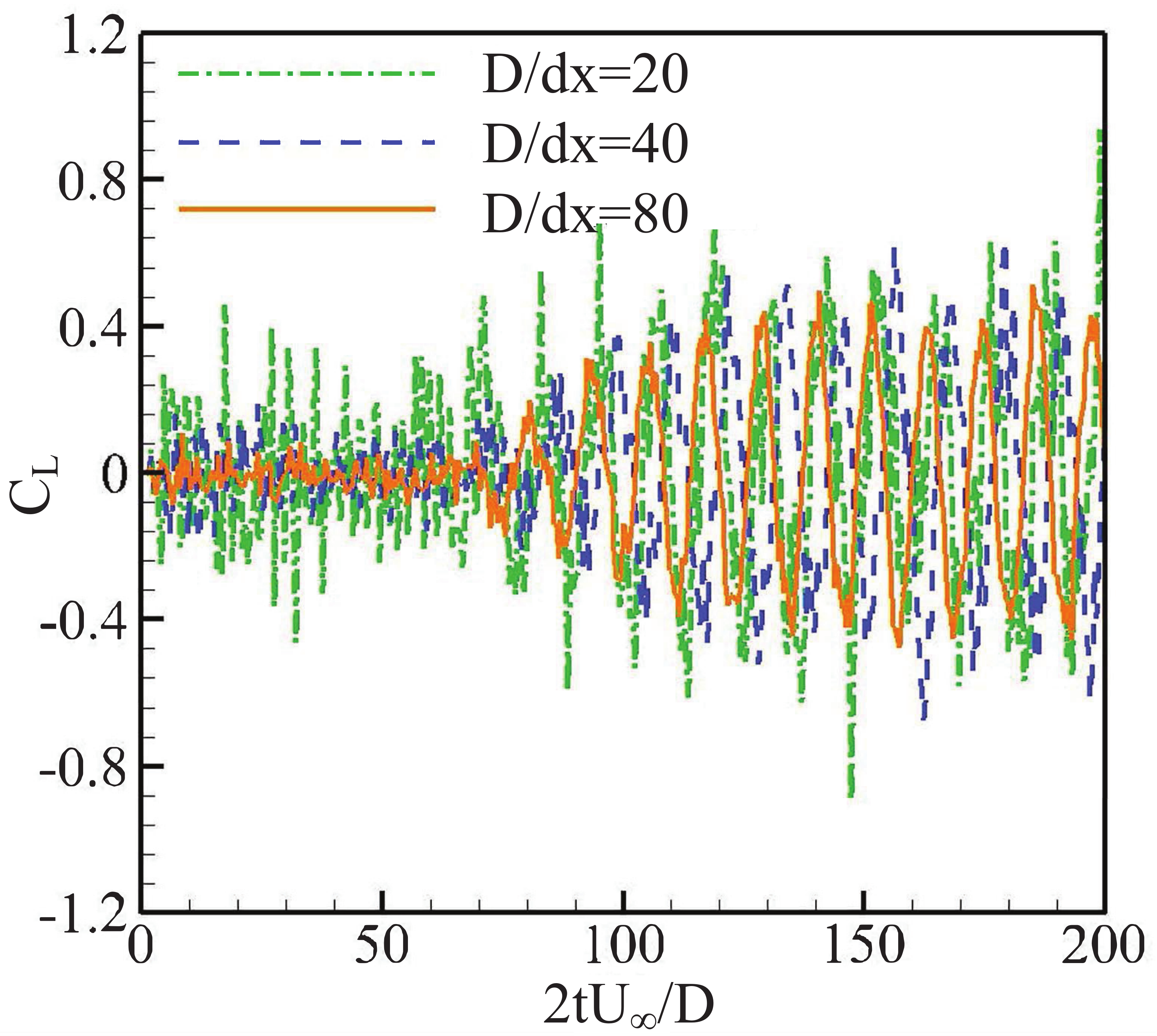}
			\end{minipage}
			\label{}
		}	
		\subfigure[$ Re=200 $]{
			\begin{minipage}[b]{0.45\linewidth}
				\includegraphics[width=1\textwidth]{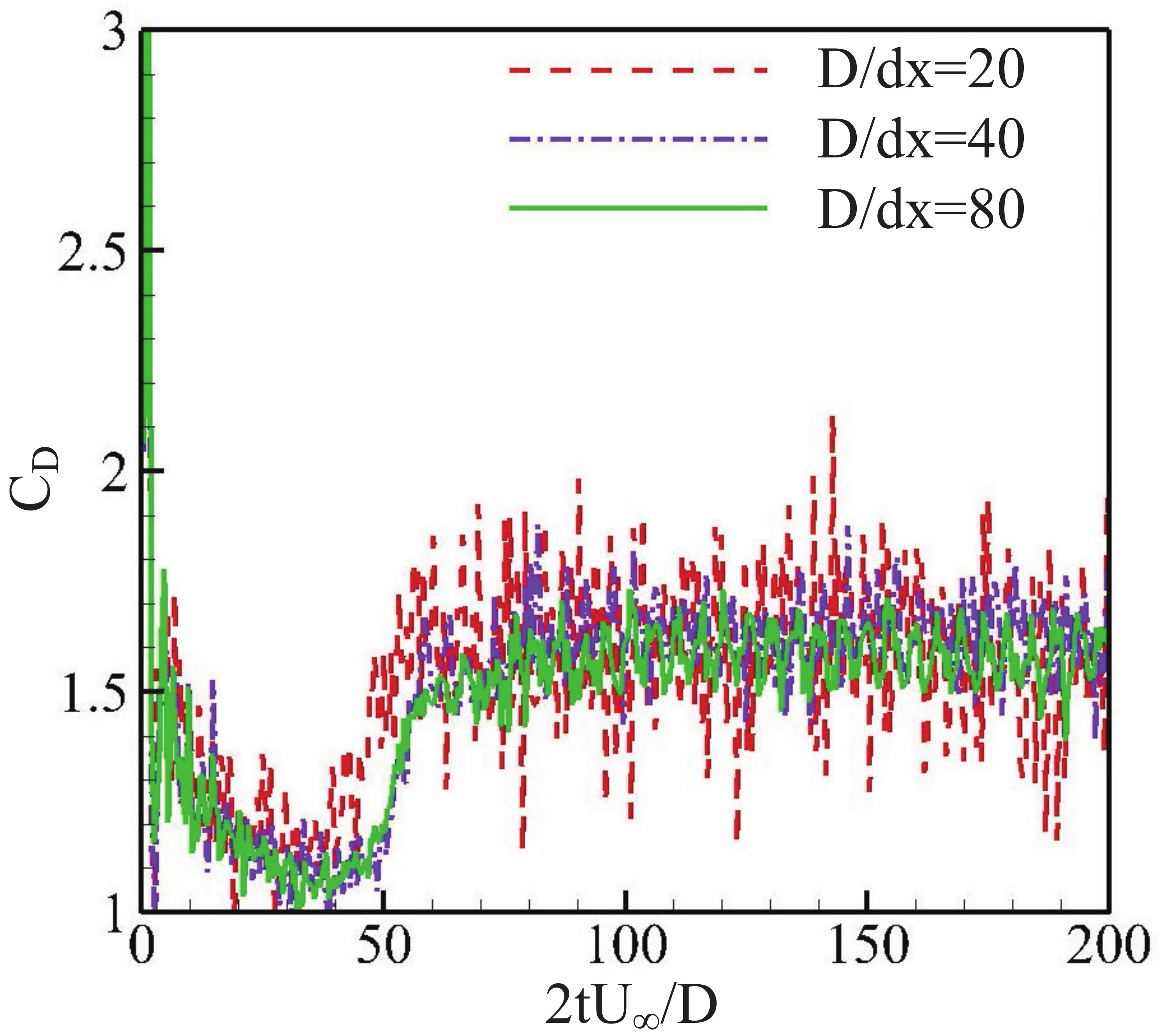}
			\end{minipage}
			\begin{minipage}[b]{0.45\linewidth}
				\includegraphics[width=1\textwidth]{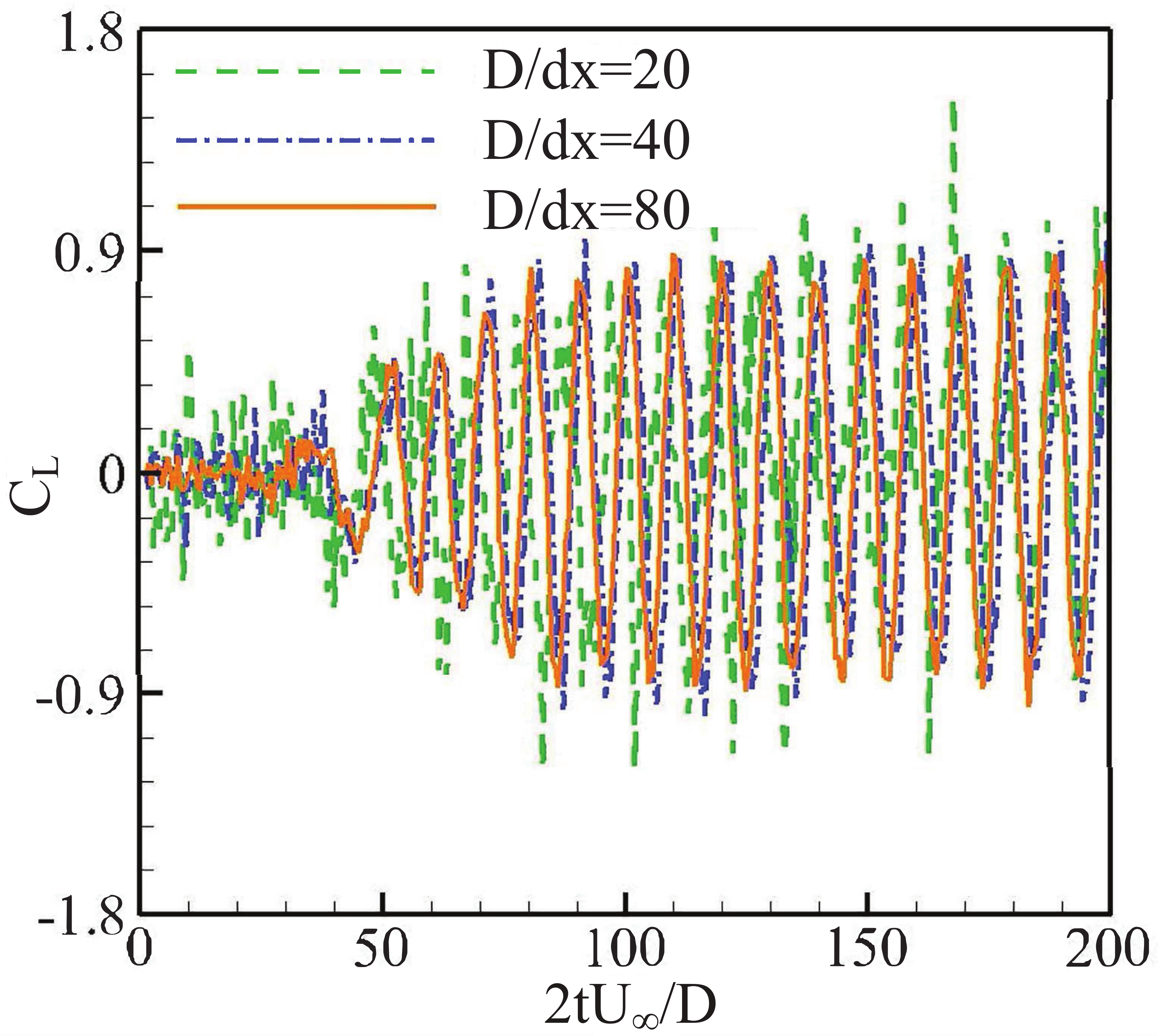}
			\end{minipage}
			\label{}
		}
		\caption{Convergence study of drag and lift coefficients $ C_{D} $ and $ C_{L} $ for different particle resolutions. }
		\label{fig11}
	\end{figure}

	To quantitatively assess the proposed algorithm, the drag and lift coefficients $ C_{D}=2F_{D}\rho^{-1}_{\infty}U^{-2}_{\infty}D^{-1} $ and $ C_{L}=2F_{L}\rho^{-1}_{\infty}U^{-2}_{\infty}D^{-1} $ are computed by using different spatial resolutions and plotted in Fig.\ref{fig11}. 
	Here, $ F_{D} $ and $ F_{L} $ are the drag and lift forces on the cylinder, respectively. With $ dx/h $ remains as a constant, the convergent tendency in all chosen cases of $ Re = 20  $, $ 100 $, and $ 200 $ is observed when the particle spacing decreases to $ dx=2.5 \times 10^{-4}m $ 
	\cite{MARRONE2013456, Quinlan2006}, and all the obtained results (drag and lift coefficients, $ C_{D} $ and $ C_{L} $) approach the reference data well, 
	as can be seen in Tabs. \ref{tab1} to \ref{tab3}.
	\begin{table}[htbp]
		\centering
		\caption{Comparison of $ C_{D} $, $ C_{L} $ and $ St $ for the case of $ Re=100 $.}
		\resizebox{0.8\textwidth}{!}{
			\begin{tabular}{ccccc}
				\toprule 
				\textbf{ } & \multicolumn{1}{c}{\textbf{$ C_{D} $}} & \multicolumn{1}{c}{\textbf{$ C_{L} $}} & \multicolumn{1}{c}{\textbf{$ St $}} \\
				\midrule 
				\textbf{White \cite{White2006}} & 1.46 &  $ - $  & $ - $ \\
				\textbf{Brehm et al. \cite{Brehm2015}} &  1.32$ \pm $0.0100 & $ \pm $0.320 & 0.165 \\
				\textbf{Al-Marouf and Samtaney \cite{Almarouf2017}} & 1.34$ \pm $0.0089 & $ \pm $0.325 & 0.166 \\	
				\textbf{Liu et al. \cite{Liu1998}} & 1.35$ \pm $0.0120 & $ \pm $0.339 & 0.165 \\
				\textbf{Le et al. \cite{Le2006}} & 1.37$ \pm $0.0090 & $ \pm $0.323 & 0.160 \\
				\textbf{Russell and Wang \cite{Russell2003}} & 1.38$ \pm $0.0070 & $ \pm $0.300 & 0.172 \\		
				\textbf{Present} & 1.57$ \pm $0.0050 & $ \pm $0.419 & 0.172 \\
				\bottomrule
		\end{tabular}}  
		\label{tab1}                                    
	\end{table}
	\begin{table}[htbp]
		\centering
		\caption{Comparison of $ C_{D} $, $ C_{L} $ and $ St $ for the case of $ Re=200 $.}
		\resizebox{0.8\textwidth}{!}{
			\begin{tabular}{ccccc}
				\toprule 
				\textbf{ } & \multicolumn{1}{c}{\textbf{$ C_{D} $}} & \multicolumn{1}{c}{\textbf{$ C_{L} $}} & \multicolumn{1}{c}{\textbf{$ St $}} \\
				\midrule 
				\textbf{Liu et al. \cite{Liu1998}} & 1.31$ \pm $0.049 & $ \pm $0.69& 0.192 \\						
				\textbf{Taira and Colonius \cite{Taira2007}} & 1.35$ \pm $0.048 & $ \pm $0.68 & 0.196 \\
				\textbf{Marrone et al. \cite{Colagrossi2013}} & 1.38$ \pm $0.05 & $ \pm $0.680 & 0.200 \\			
				\textbf{Tafuni et al. \cite{Tafuni2018}} & 1.46 & $ \pm $0.693 & 0.206 \\		
				\textbf{Negi et al. \cite{Negi2020}}  & 1.524$ \pm $0.05 & $ \pm $0.722 & $ \pm $0.210 \\
				\textbf{Jin and Braza \cite{Jin1993}}  & 1.532$ \pm $0.05 & $ \pm $0.744 & $ \pm $0.210 \\			
				\textbf{Present} & 1.571$ \pm $0.05 & $ \pm $0.81 & 0.203\\
				\bottomrule
		\end{tabular}}   
		\label{tab2}                                     
	\end{table}
	\begin{table}[htbp]
		\centering
		\caption{Comparison of $ C_{D} $ for the case of $ Re=20 $.}
		\resizebox{0.45\textwidth}{!}{
			\begin{tabular}{ccccc}
				\toprule 
				\textbf{ } & \multicolumn{1}{c}{\textbf{$ C_{D} $}} \\
				\midrule 
				\textbf{Tritton \cite{Tritton1959}} & 2.09 \\		
				\textbf{Taira and Colonius \cite{Taira2007}} & 2.06 \\
				\textbf{Tafuni et al. \cite{Tafuni2018}} & 2.29 \\
				\textbf{Negi et al. \cite{Negi2020}} & 2.317 \\
				\textbf{Present} & 2.20 \\				
				\bottomrule
		\end{tabular}}   
		\label{tab3}                                     
	\end{table}
	For the unsteady cases, the Strouhal numbers $ St = f D U^{-1}_{\infty} $ is also considered, where $ f $ is the vortex shedding frequency. For $ Re = 100 $ and 200, the calculated drag and lift coefficients are slightly larger than the numerical results in the literature, yet the difference is narrowed compared with the corresponding experimental data. Meanwhile, the Strouhal numbers in the above two cases are very close to the other results. For the steady case of $ Re = 20 $, the drag coefficient $ C_{D} $ is well in accordance with the experimental and numerical data in the literature. Overall, except for the slight overestimation of drag force in the unsteady cases, the present free-stream boundary algorithm demonstrates good accuracy. 
	
\begin{figure}[htbp]	
		\centering     
		\subfigure[Time evolution of $ C_{D} $]{
			\begin{minipage}[b]{0.45\linewidth}
				\includegraphics[width=1\textwidth]{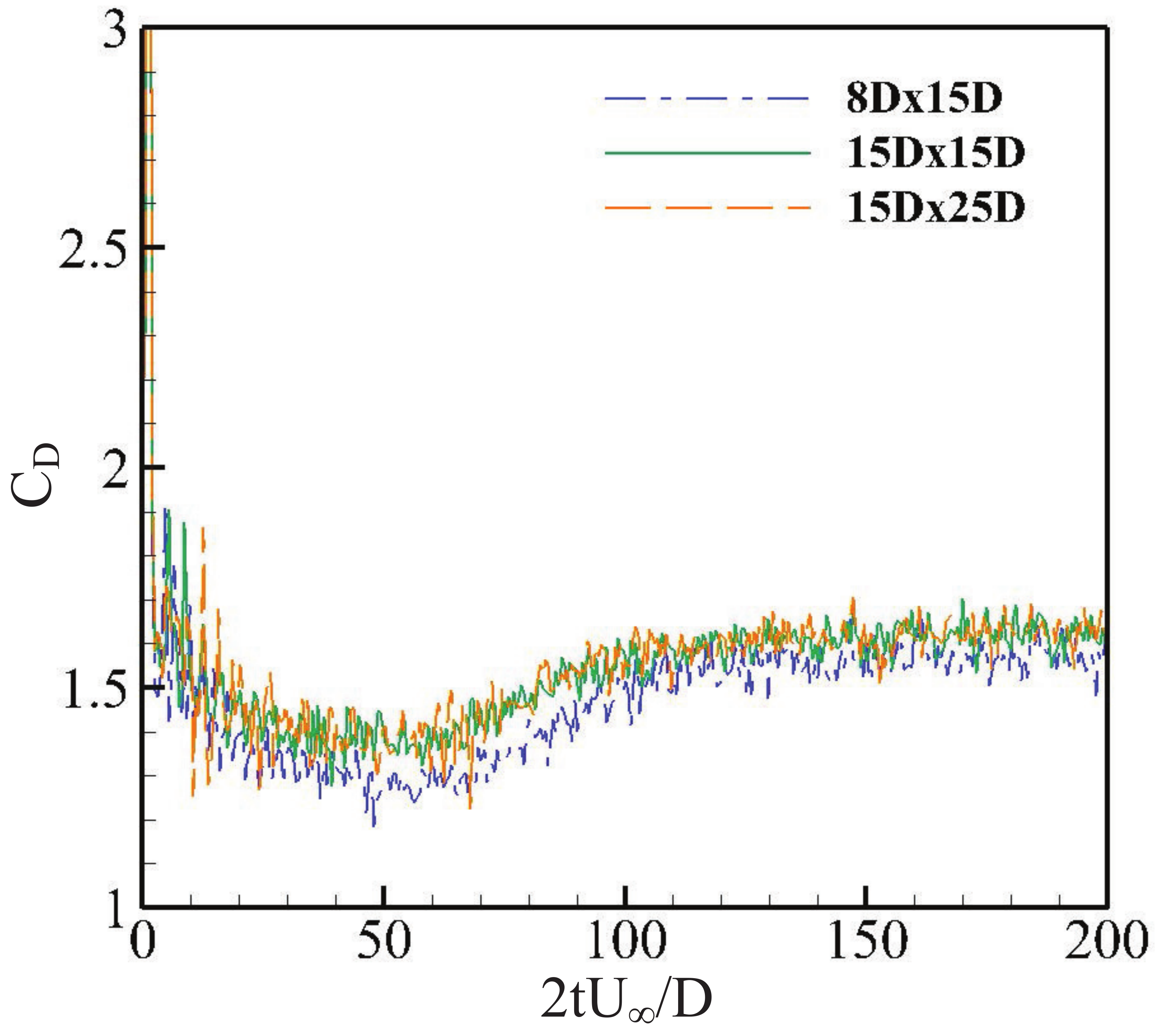}
			\end{minipage}
		}	
		\subfigure[Time evolution of $ C_{L} $]{
			\begin{minipage}[b]{0.45\linewidth}
				\includegraphics[width=1\textwidth]{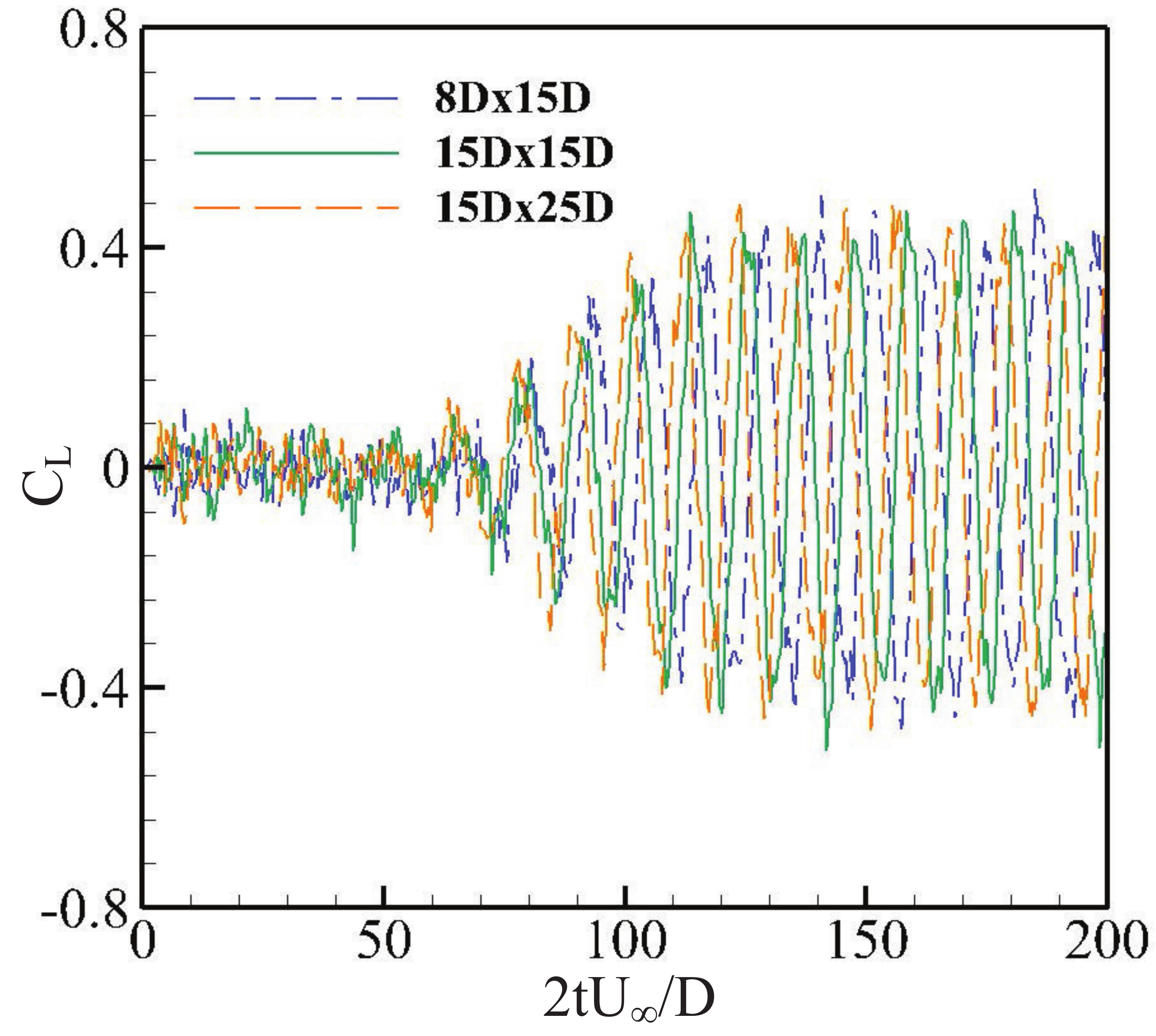}
			\end{minipage}
		}
		\caption{Comprison of drag and lift coefficients $ C_{D} $ and $ C_{L} $ with different domain sizes at $ Re=100 $.}
		\label{fig12}
	\end{figure}

	In addition, we further test the case $ Re=100 $ with different initial domain sizes $(L_{0} = 15D, H_{0}= 15D)$ and $(L_{0} = 25D, H_{0}= 15D)$, to investigate the effects of domain size on the results. Fig.\ref{fig12} shows the time evolution of the drag and lift coefficients for each case, where the time is normalized by $ 2tU_{\infty}D^{-1} $. 
	As shown herein, after the start-up process, the drag coefficient $ C_{D}$ gradually approaches a constant value, while the lift coefficient $ C_{L}$ oscillates around zero over time. For the three chosen domain sizes, the corresponding drag coefficients are the same after the flow stabilizing, and the oscillation frequencies and amplitudes of the lift coefficients are also consistent. This implies that the flow features in the simulation are not related to the chosen domain sizes. Thus, the proposed algorithm costs less computational resources with the free-stream boundary condition to obtain acceptable results compared to the larger domain sizes chosen in other WCSPH simulations \cite{Tafuni2018} and mesh methods \cite{Almarouf2017}.
	\subsection{Flow-induced oscillation of an elastic beam attached to a cylinder}
\begin{figure}[htbp]
		\centering     
		\includegraphics[width=0.7\textwidth]{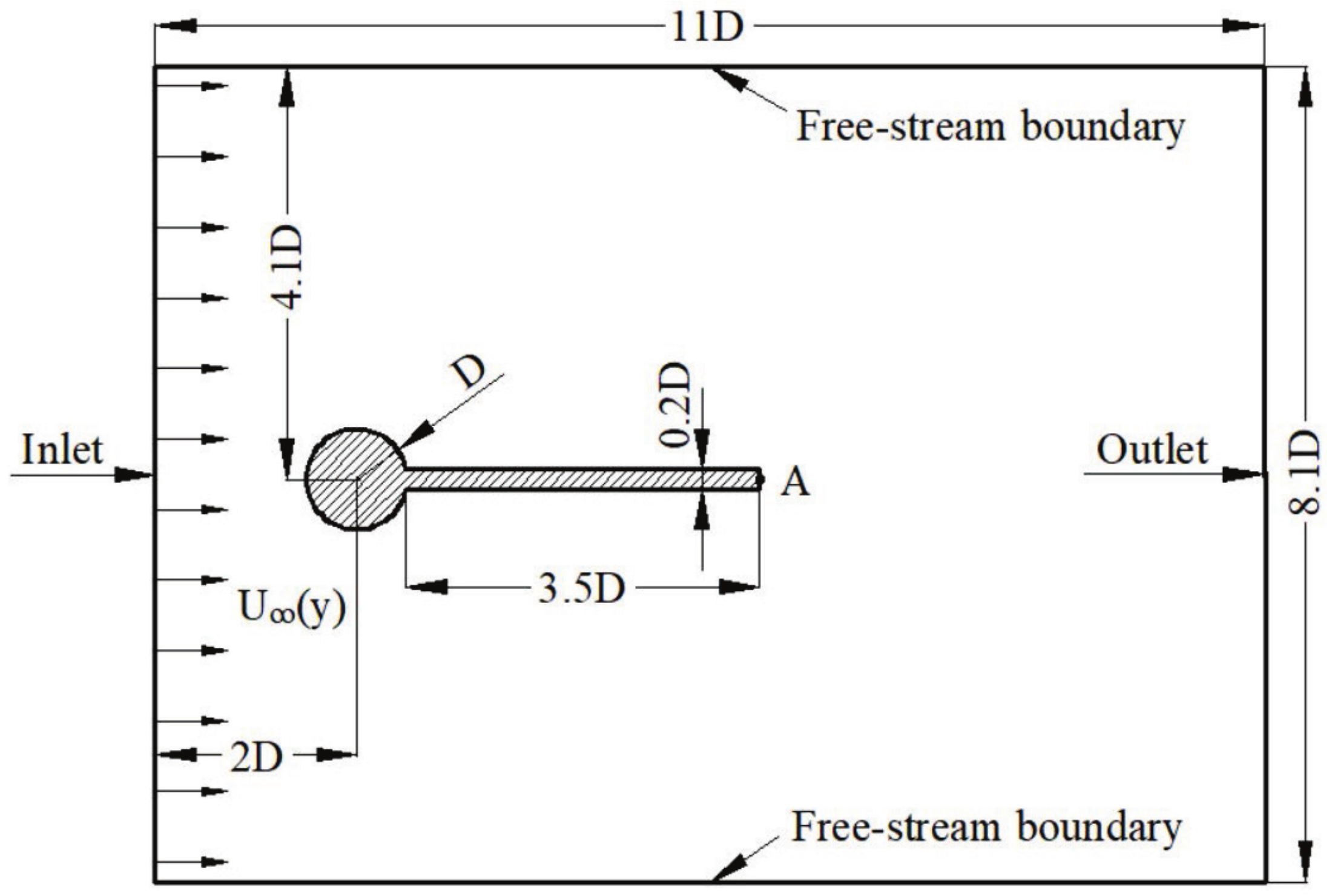}
		\caption{Schematic of 2-D flow-induced oscillation of an elastic beam attached to a cylinder.}
		\label{fig13}
	\end{figure}

	This part considers the 2D flow-induced oscillation of an elastic beam attached to a cylinder in the free-stream environment. This classic fluid-structure interaction (FSI) problem \cite{Turek2007} is usually simulated by setting the span-wise boundaries as solid walls. To the authors' best knowledge, this is the first time the WCSPH free-stream boundary conditions have been adopted to simulate this FSI problem. Fig.\ref{fig13} depicts the initial conditions with given geometric parameters. 
	The length and height of domain are respectively $ L_{0} = 1.1m $ and $ H_{0} = 0.81m $, and the circular cylinder with the diamter of $ D=0.1m $ is located at the point ($ 2D $, $ 4D $). It should be mentioned that the domain is placed in a horizontally symmetrical position, while the center of the cylinder is not on the symmetric axis. Unlike the initial velocity distribution in Ref. \cite{Turek2007}, the velocity profile in the buffer is targeted to an uniform value in y-direction $ U_{\infty}(y) = 1m/s $, since we assume here is a free-stream flow. With preliminary numerical tests, the uniform particle spacing $ dx =0.005m $ adopted here is fine enough to reach the saturation regime \cite{Quinlan2006}, and the artificial sound speed $ c_{0} $ is modified to $ 20U_{0} $ due to the anticipated large velocity at the tip of the elastic beam. The fluid density is $ \rho_{f}=1000kg/m^{3} $, and the fluid-structure density ratio $ \rho_{f}/\rho_{s} $ is $ 1:10 $. The dynamic viscosity is calculated by $ \mu=\rho_{f} U_{\infty} D/Re $, where the Reynolds number $ Re=100 $. Dimensionless Young's modulus $ E^{*}=E/(\rho_{f}U_{0}^{2})= 1.4\times10^{3} $, 
	and Poisson ratio $ \nu^{s}=0.4 $. 
	
\begin{figure}[htbp]
		\centering     
		\subfigure{
			\begin{minipage}[b]{0.88\linewidth}
				\includegraphics[width=1\textwidth]{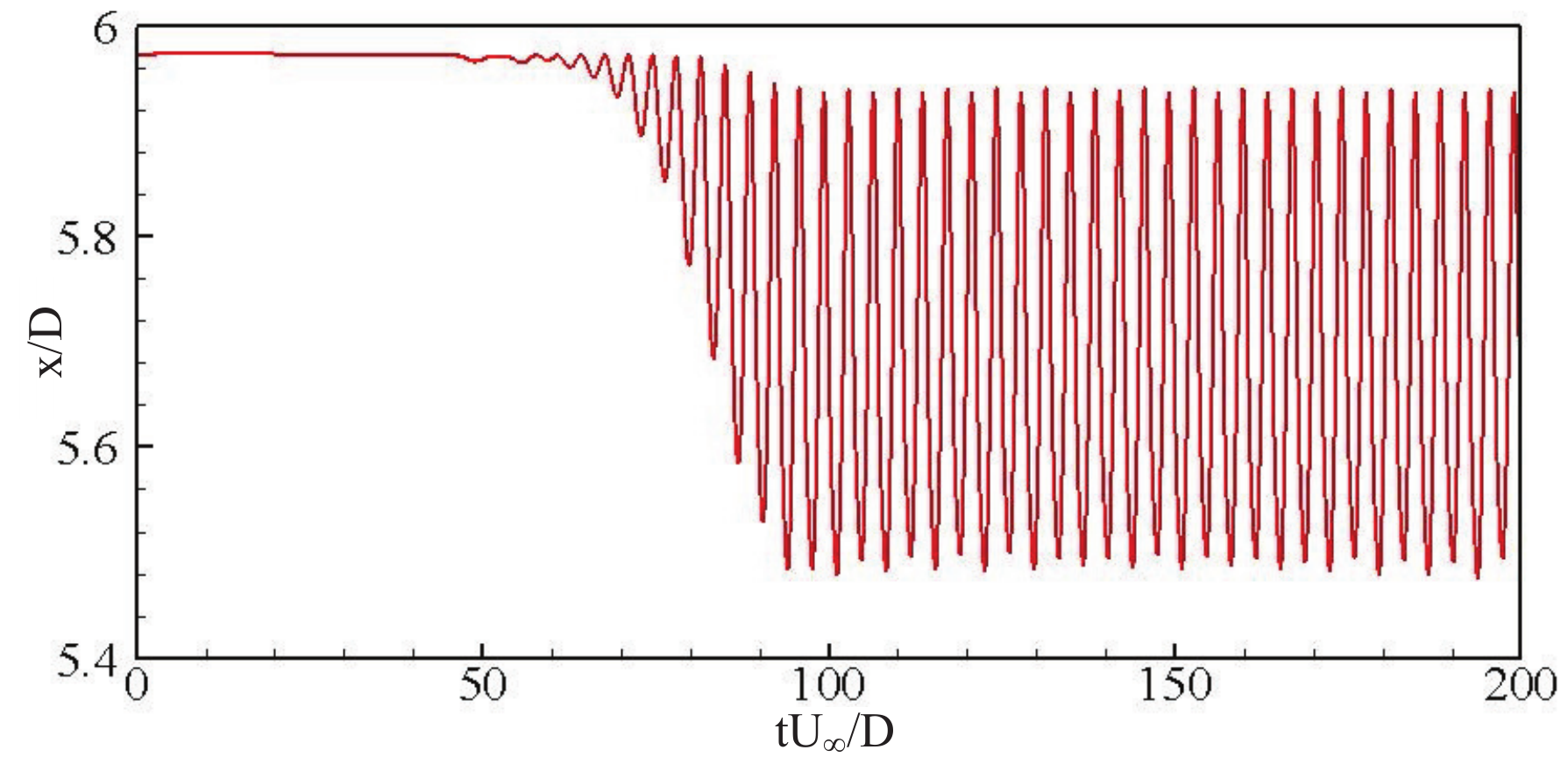}
			\end{minipage}
		}	
		\subfigure{
			\begin{minipage}[b]{0.88\linewidth}
				\includegraphics[width=1\textwidth]{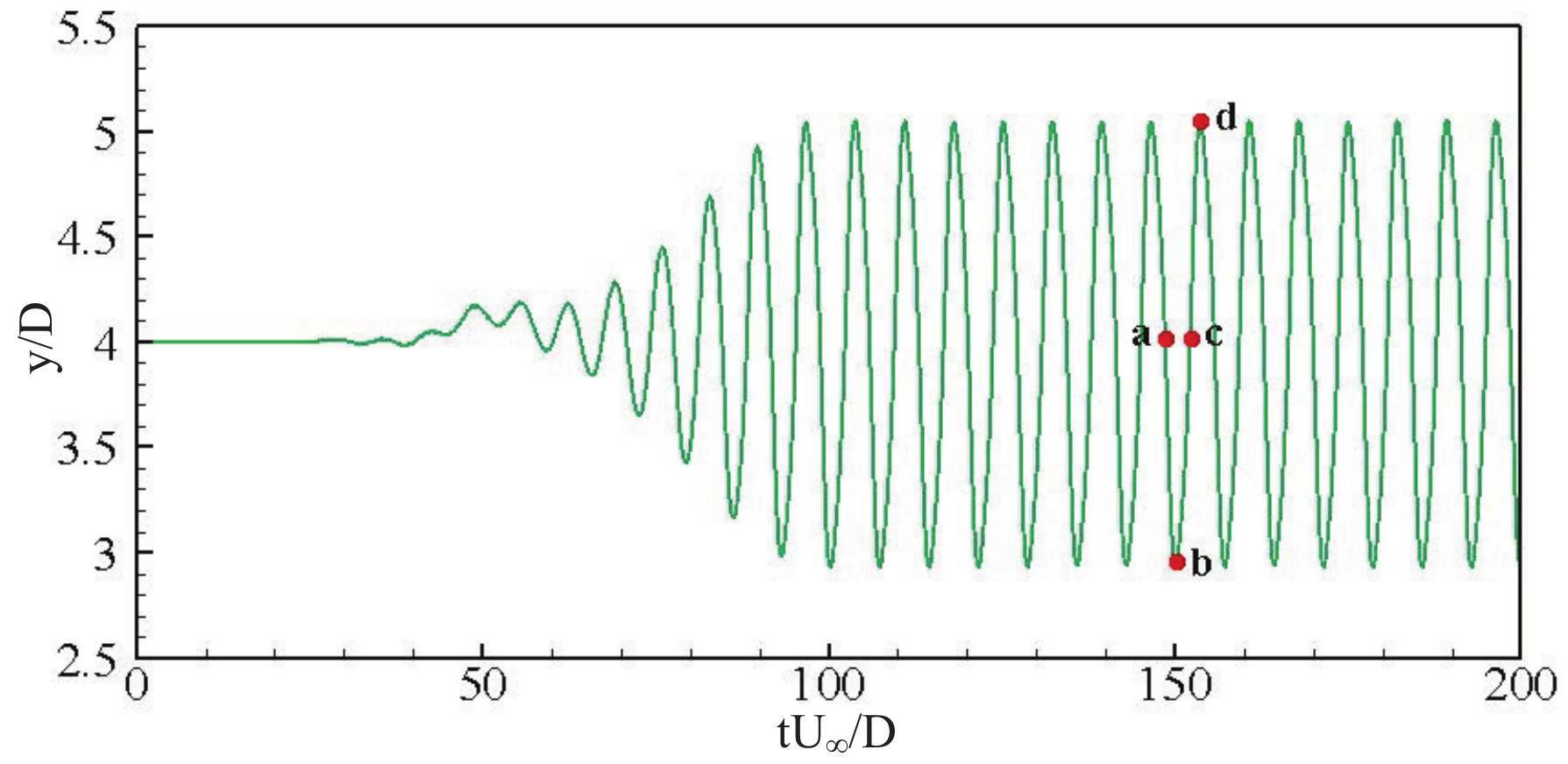}
			\end{minipage}
		}
		\subfigure{
			\begin{minipage}[b]{0.88\linewidth}
				\includegraphics[width=1\textwidth]{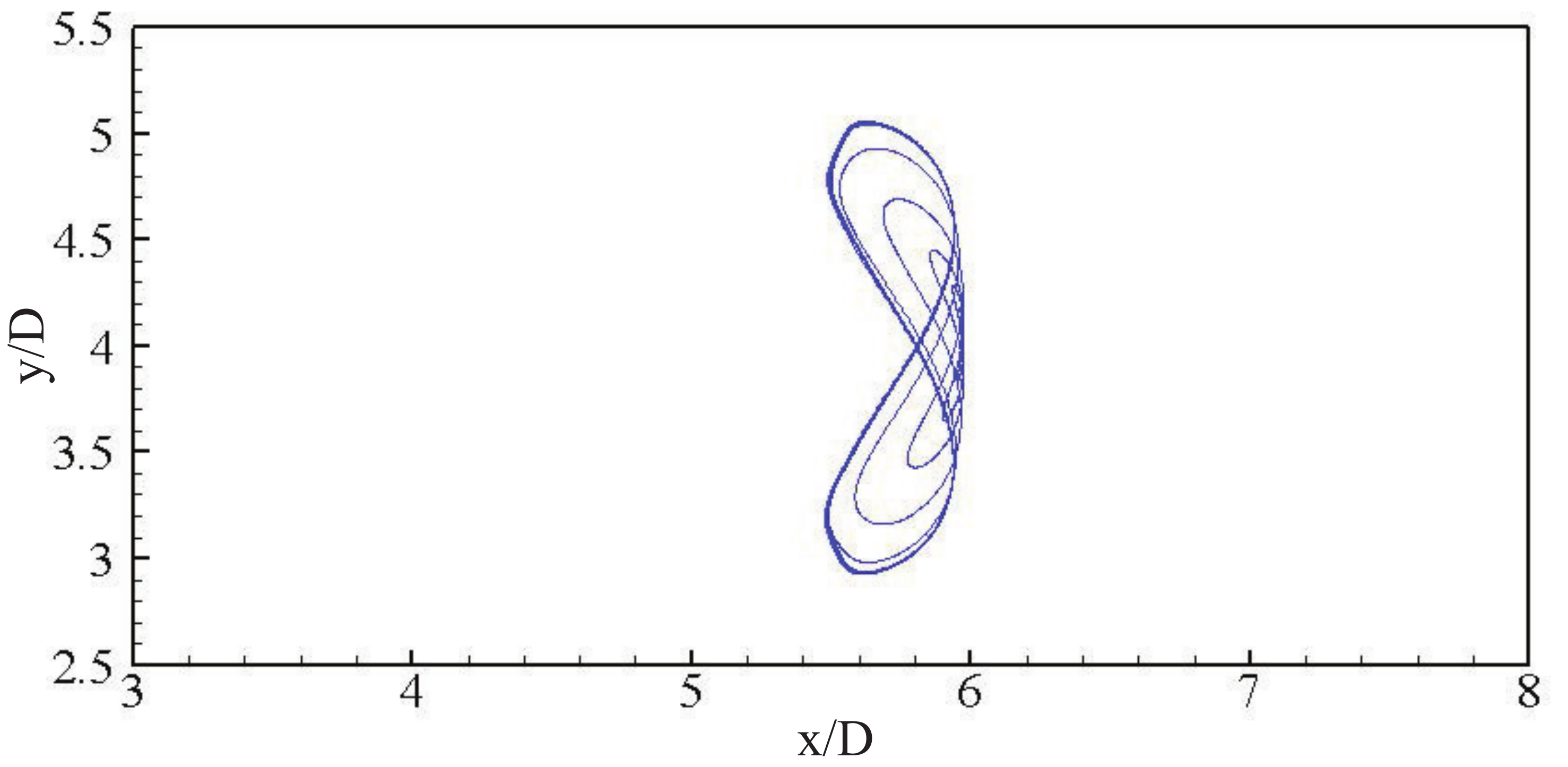}
			\end{minipage}
		}
		\caption{Flow-induced vibration of an elastic beam attached to a cylinder: $ x $-direction displacement (top panel), $ y $-direction displacement (middle panel) and the trajectory (bottom panel) of point A.}
		\label{fig14}
	\end{figure}

\begin{figure}[htbp]
		\centering   
		\begin{minipage}[]{0.45\linewidth}
			\includegraphics[width=1\textwidth]{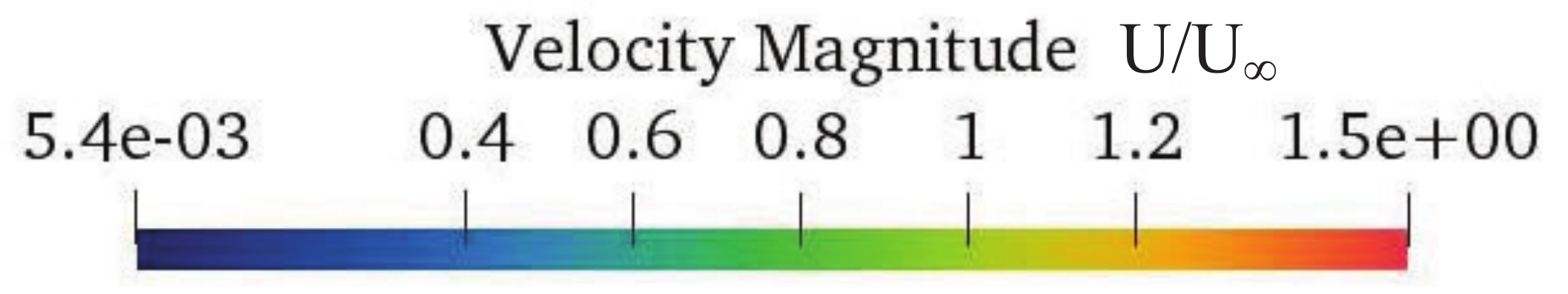}
		\end{minipage}
		
		\subfigure[$ Time\; instance\ a $ ]{
			\begin{minipage}[b]{0.45\linewidth}
				\includegraphics[width=1\textwidth]{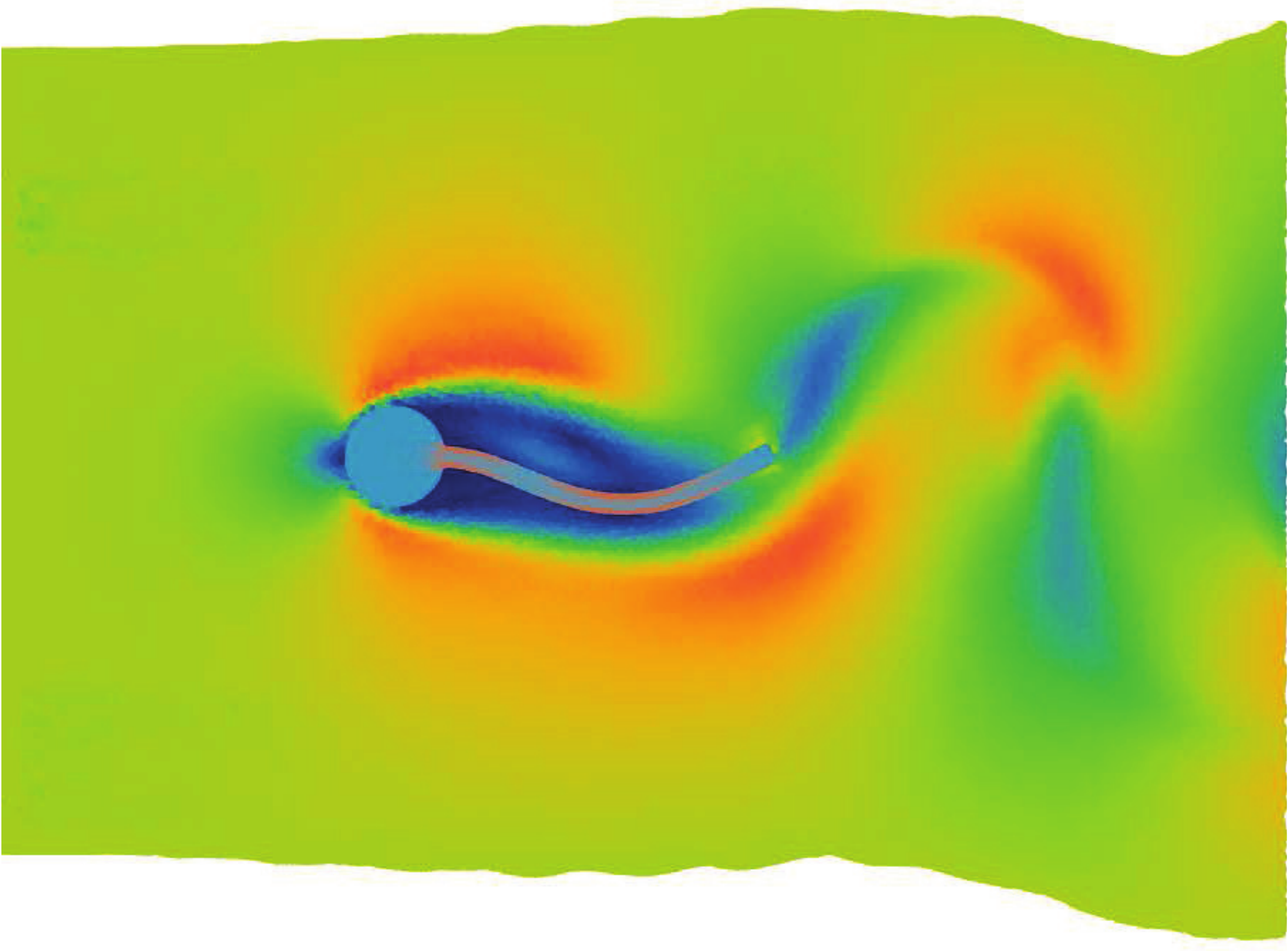}
			\end{minipage}
		}	
		\subfigure[$ Time\; instance\ b $]{
			\begin{minipage}[b]{0.45\linewidth}
				\includegraphics[width=1\textwidth]{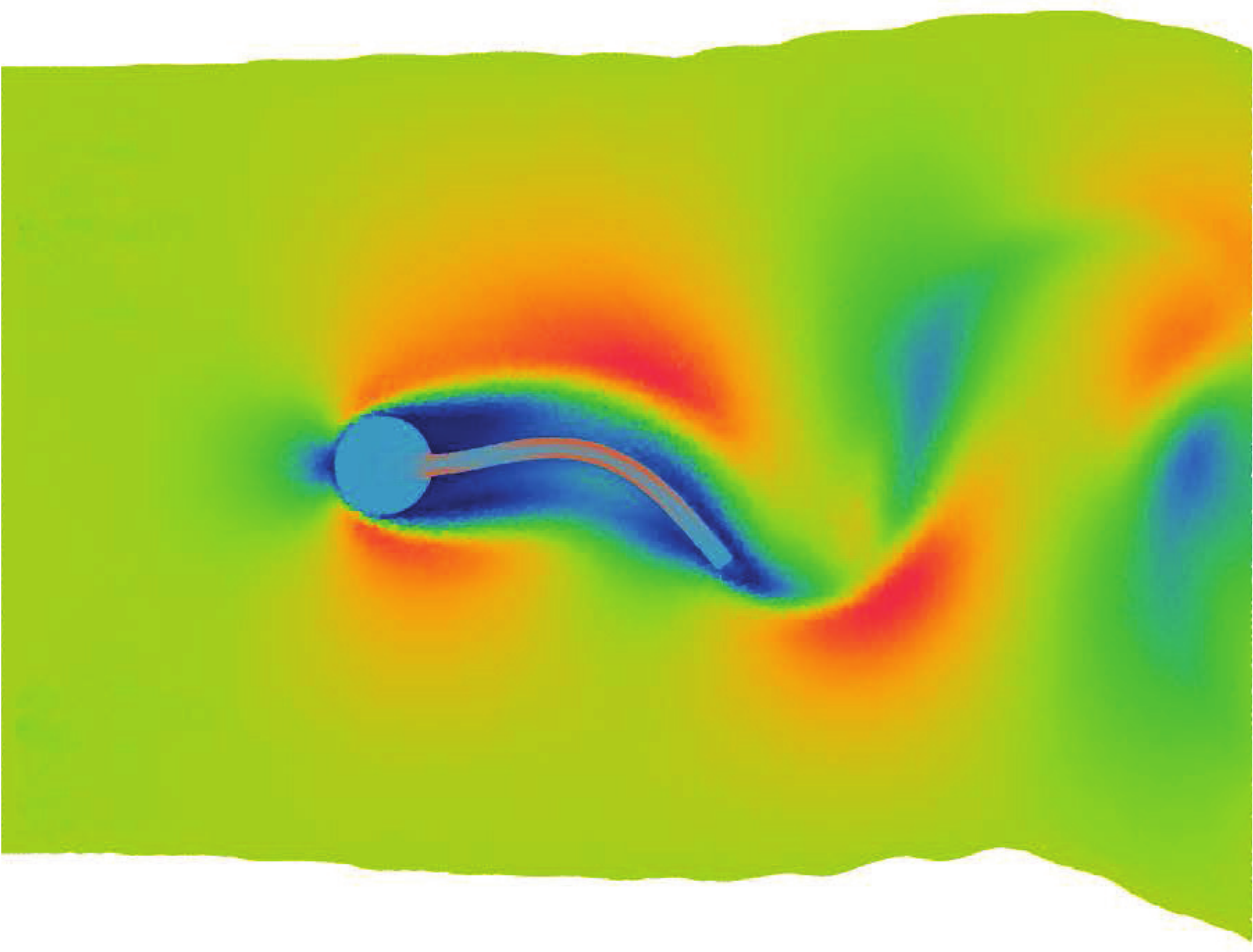}
			\end{minipage}
		}
		
		\subfigure[$ Time\; instance\ c $]{
			\begin{minipage}[b]{0.45\linewidth}
				\includegraphics[width=1\textwidth]{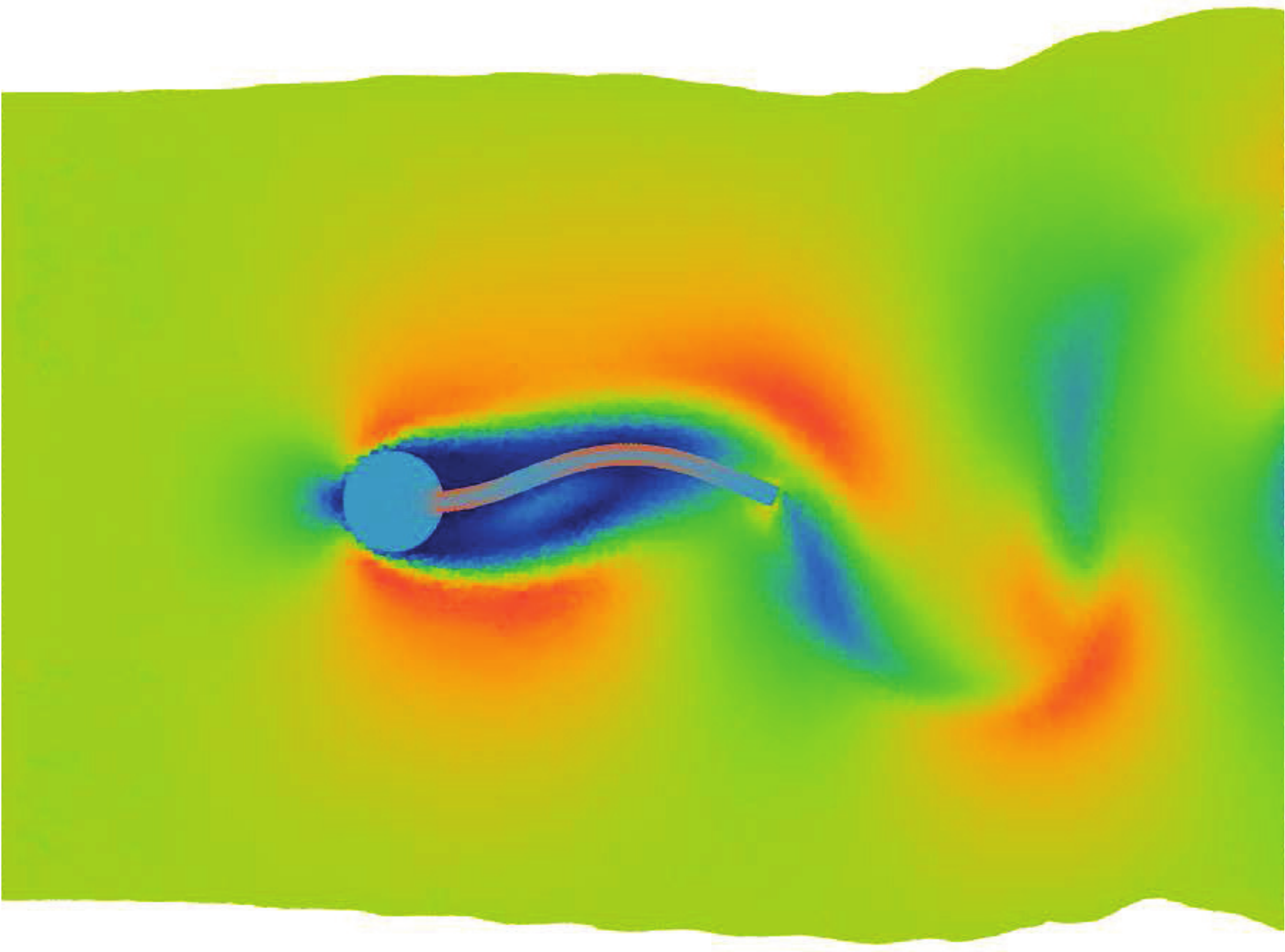}
			\end{minipage}
		}	
		\subfigure[$ Time\; instance\ d $]{
			\begin{minipage}[b]{0.45\linewidth}
				\includegraphics[width=1\textwidth]{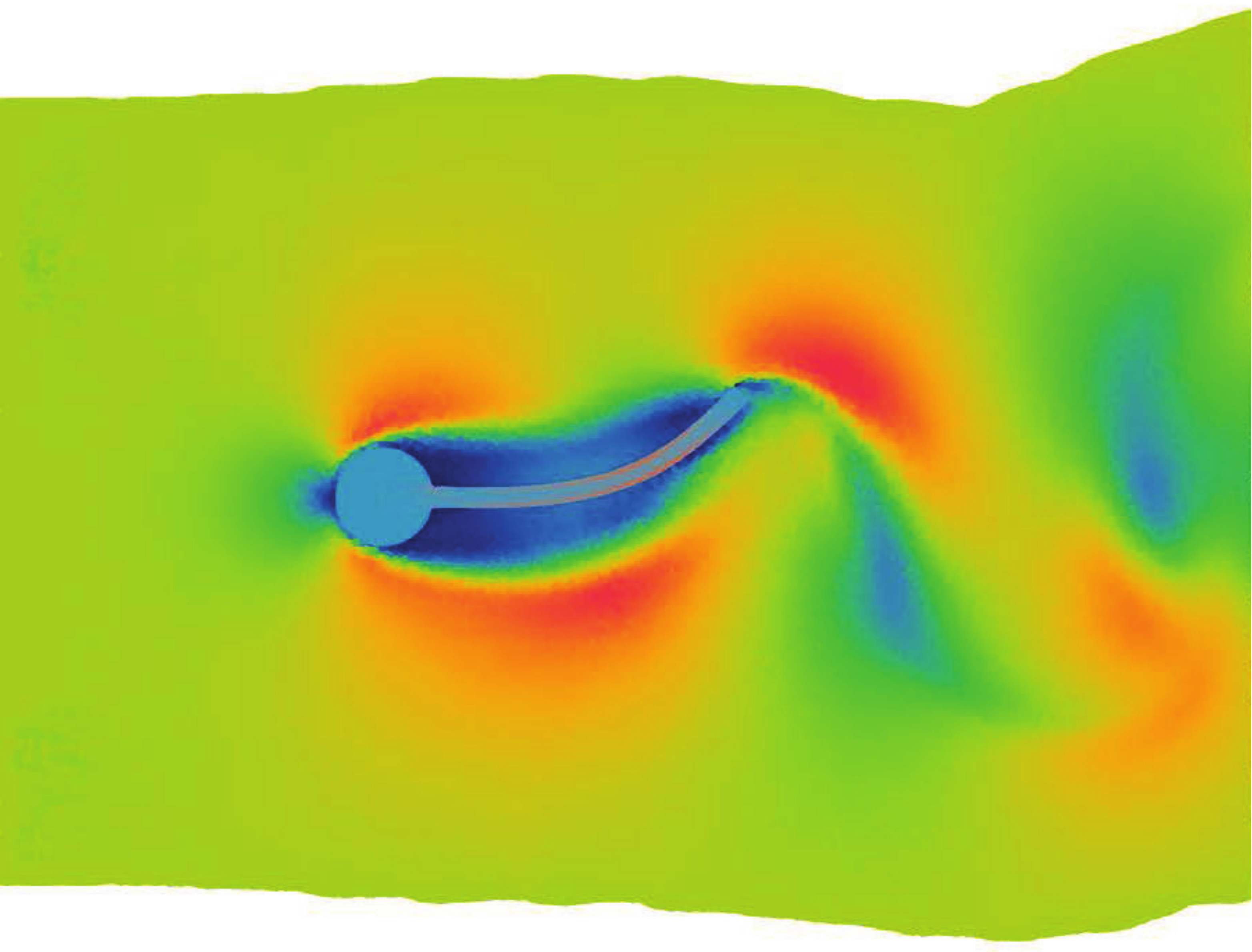}
			\end{minipage}
		}
		\begin{minipage}[]{0.45\linewidth}
			\includegraphics[width=1\textwidth]{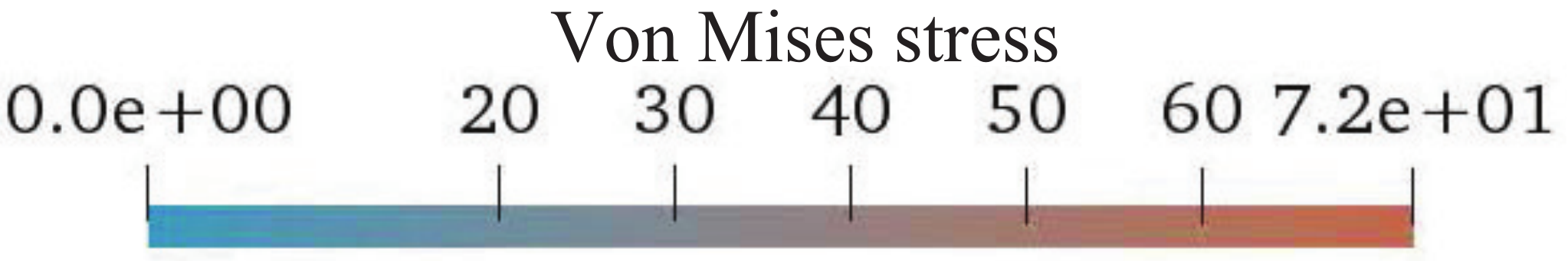}
		\end{minipage}
		\caption{The fluid velocity magnitude contour at different instances marked in Fig.\ref{fig14}, with the corresponding beam deformation colored by Von Mises stress.}
		\label{fig15}
	\end{figure}

	Fig.\ref{fig14} illustrates horizontal and vertical displacements of point A, 
	which is fixed at the beam end as marked in Fig.\ref{fig13}. 
	After the initial time $ tU_{\infty}/D \geq 100 $, the beam reaches a periodic oscillation with stable amplitude and frequency, and its trajectory is a Lissajous curve which has the frequency ratio of 2:1 between horizontal and vertical motions. Fig.\ref{fig15} displays snapshots of the velocity contour in one motion period, together with the profile of the deformed beam. 
	As shown here, with the flow evolving, the generated vortexes behind the cylinder induce the deformation and sway of the elastic beam. Meanwhile, the motion of the beam disturbs the flow field as well. Compared to the results with wall restriction in Refs. \cite{Turek2007, Han2018, Bhardwaj2012, Tian2014}, the beam motion in the free-stream flow is more violent. To be more specific, the amplitude in the $ y $- direction is more prominent in the present case while the oscillation frequency is lower, as shown in Table \ref{tab4}. 
	\begin{table}[htbp]
		\centering
		\caption{Comparison of oscillation amplitude in $y$- direction and frequency for the FSI test case, Present-1 ($ L_{0} = 1.1m $, $ H_{0}=0.81m$), Present-2 ($ L_{0} = 1.1m $, $ H_{0}=1.21m$), Present-3 ($ L_{0} = 1.1m $, $ H_{0}=1.61m$)}
		\resizebox{\textwidth}{!}{
			\begin{tabular}{ccccc}
				\toprule 
				\textbf{Reference} & \multicolumn{1}{c}{\textbf{Amplitude in y direction(/D)}} & \multicolumn{1}{c}{\textbf{Frequency}} \\
				\midrule 
				\textbf{Turek and Hron \cite{Turek2007}} & 0.830 & 0.190 \\
				\textbf{Zhang et al. \cite{Han2018}} & 0.855 & 0.189 \\
				\textbf{Bhardwaj and Mittal \cite{Bhardwaj2012}} & 0.920 & 0.190 \\
				\textbf{Tian et al. \cite{Tian2014}}  & 0.784 & 0.190 \\
				\textbf{Present-1} & 1.056 & 0.141 \\
				\textbf{Present-2} & 1.076 & 0.141 \\
				\textbf{Present-3} & 1.060 & 0.142 \\
				\bottomrule
		\end{tabular}} 
		\label{tab4}                                          
	\end{table}
	This may be due to the fact that the free-stream boundaries on the upper and lower sides could render more free motion space for the elastic beam.
	
	Similar to the example in the previous section \ref{section4-2}, we also investigate the influence of domain size on the simulated results. Two more different initial fluid heights $ H_{0}=1.21m$ and 1.61m are chosen, and the length $ L_{0} = 1.1m $ remains constant. The corresponding results are also listed in Table \ref{tab4}, and it is found that both the $ y $- direction amplitude and frequency of the oscillation remain almost unchanged as the domain height varies. 
	\subsection{3D flow past a sphere}
\begin{figure}[htbp]
		\centering     
		\includegraphics[width=0.9\textwidth]{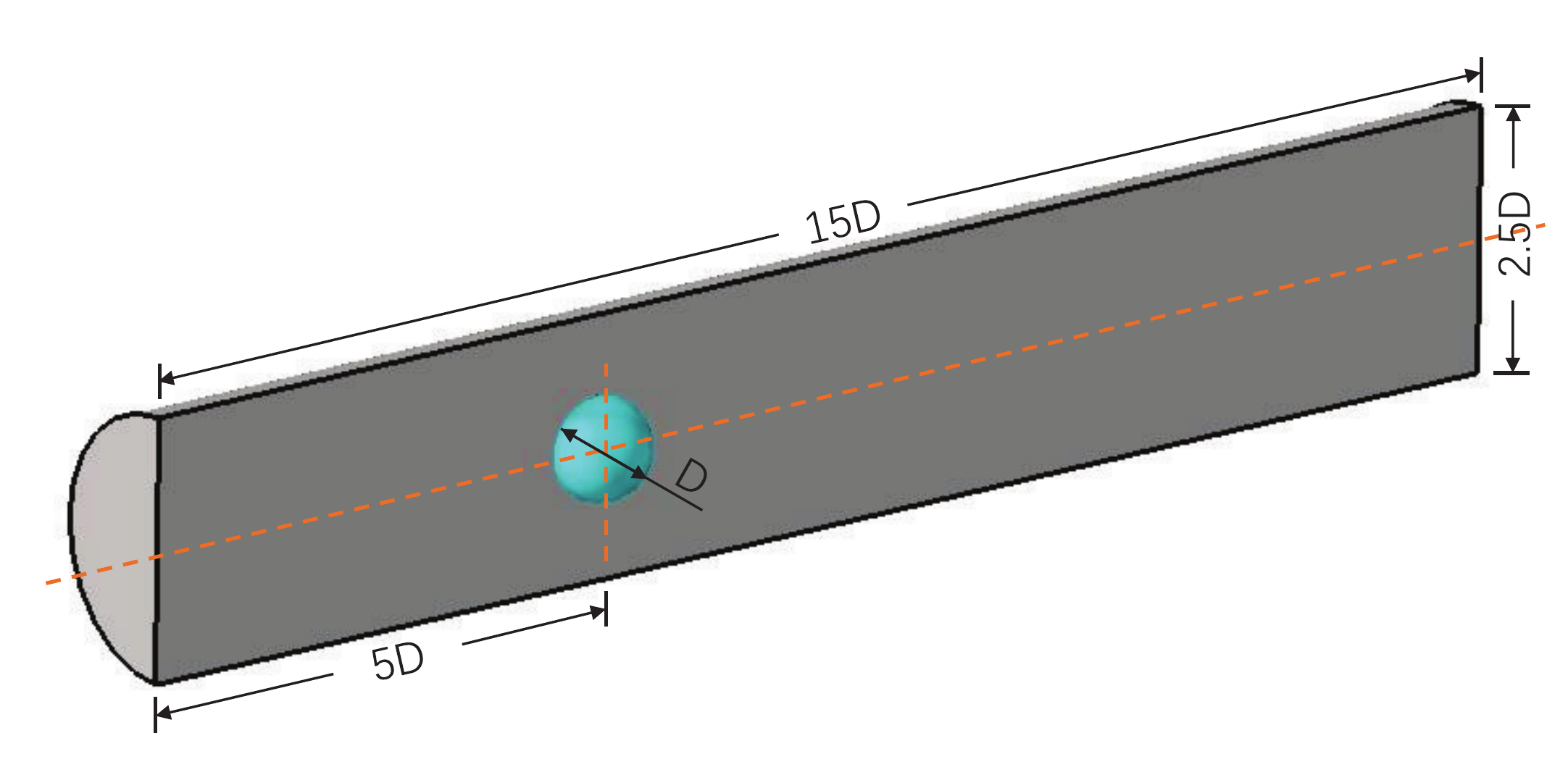}
		\caption{Schematic of the 3-D flow past a sphere.}
		\label{fig16}
	\end{figure}

	At last, we consider the flow past a sphere with the free-stream boundary condition to demonstrate the proposed algorithm's versatility for practical 3D problems. As depicted in Fig.\ref{fig16}, initially, a rigid solid sphere is placed on the symmetric center axis of a cylindrical domain. 
	The initial cylinder diameter and length are respectively $ d = 2.5 D $ and $ L_{0} = 15 D $, where $ D=0.02 m $ is the sphere diameter. The uniform particle spacing $ dx = 5.0 \times 10^{-4}m $, and the other computational parameters are same as those in Section \ref{section4-2}. We simulate four cases with different Reynolds numbers ranging from $ Re = 20 $ to 1500, where $ Re=\rho U_{\infty} D/\mu $, and the dynamic viscosity is varied to acquire different Reynolds numbers.
	
\begin{figure}[htbp]
		\centering   
		\begin{minipage}[]{0.5\linewidth}
			\includegraphics[width=1\textwidth]{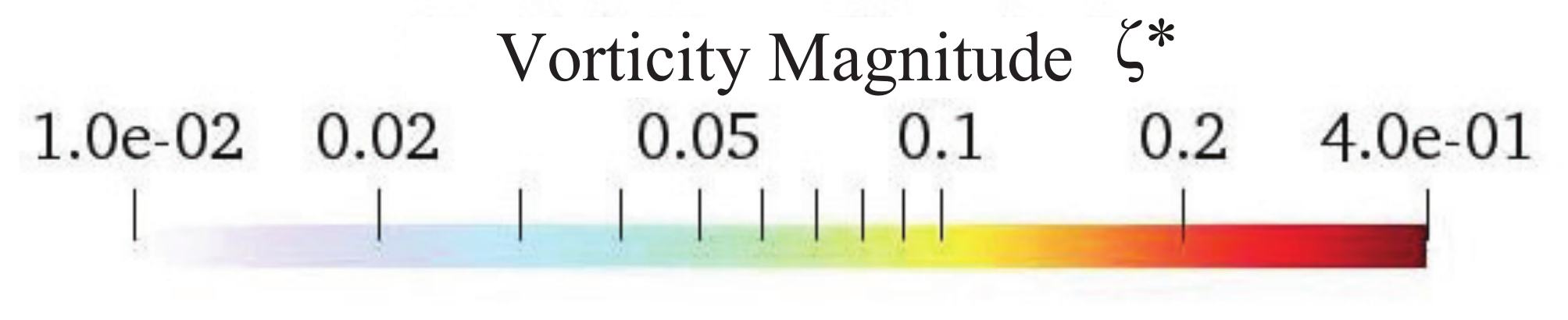}
		\end{minipage}
		
		\subfigure[$ Re=20 $ ]{
			\begin{minipage}[b]{0.45\linewidth}
				\includegraphics[width=1\textwidth]{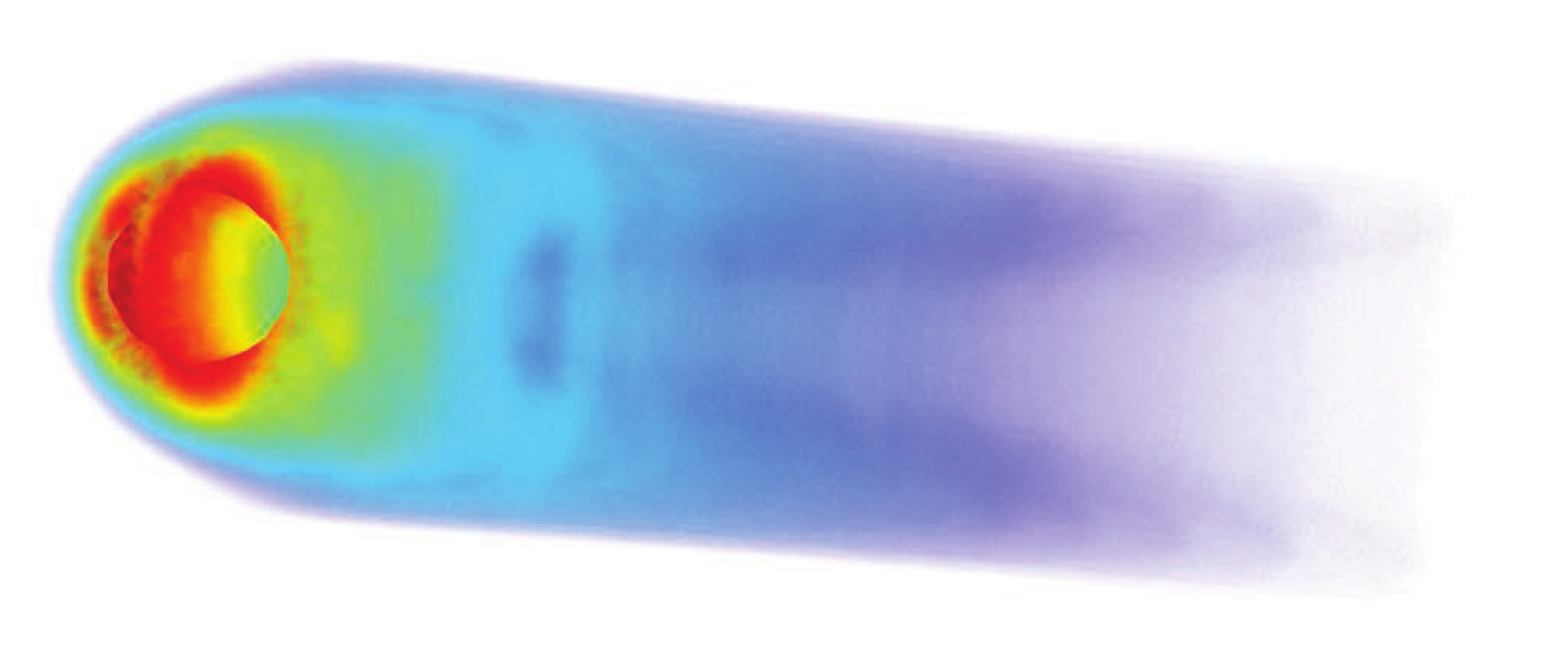}
			\end{minipage}
		}	
		\subfigure[$ Re=100 $]{
			\begin{minipage}[b]{0.8\linewidth}
				\includegraphics[width=1\textwidth]{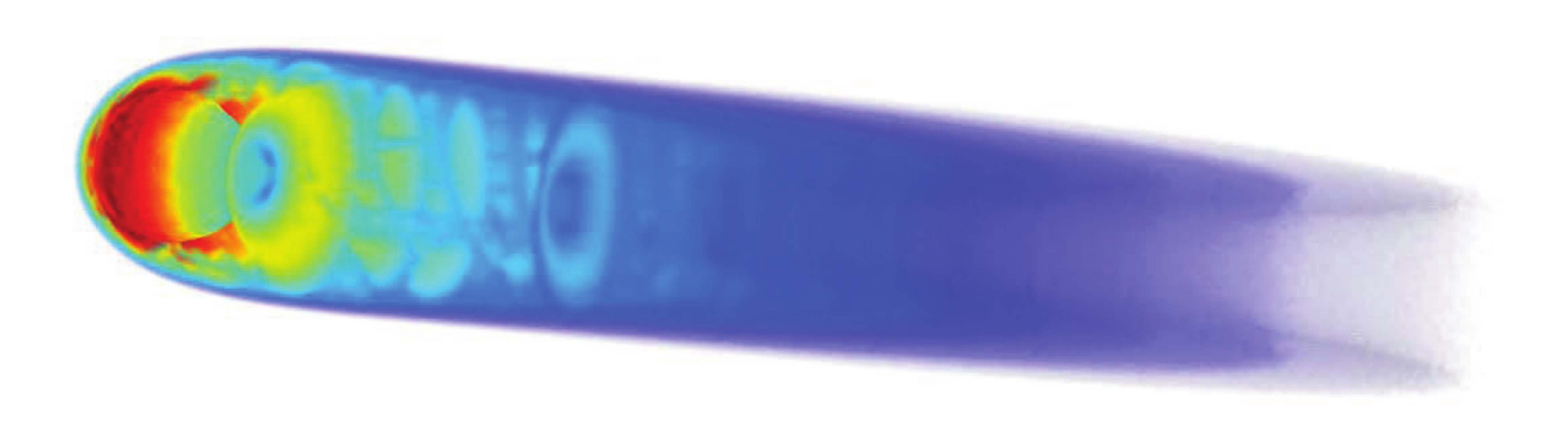}
			\end{minipage}
		}
		
		\subfigure[$ Re=200 $]{
			\begin{minipage}[b]{0.9\linewidth}
				\includegraphics[width=1\textwidth]{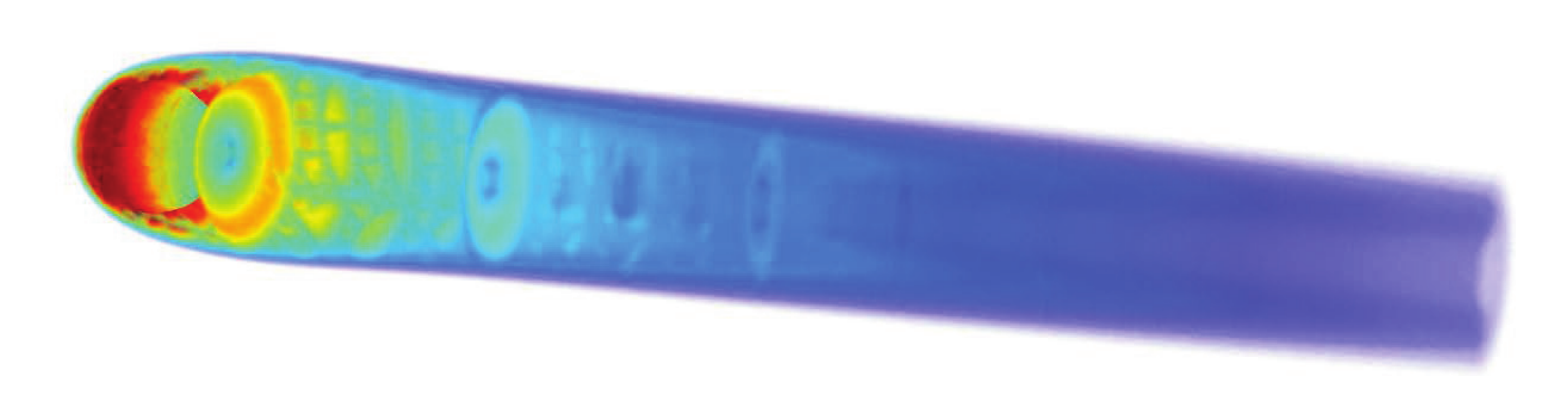}
			\end{minipage}
		}	
		\caption{The visualization of the vorticity magnitude for different Reynolds numbers.}
		\label{fig17}
	\end{figure}

\begin{figure}[htbp]
		\centering  
		\subfigure[Vorticity magnitude contour]{
			\begin{minipage}[b]{0.8\linewidth}
				\includegraphics[width=1\textwidth]{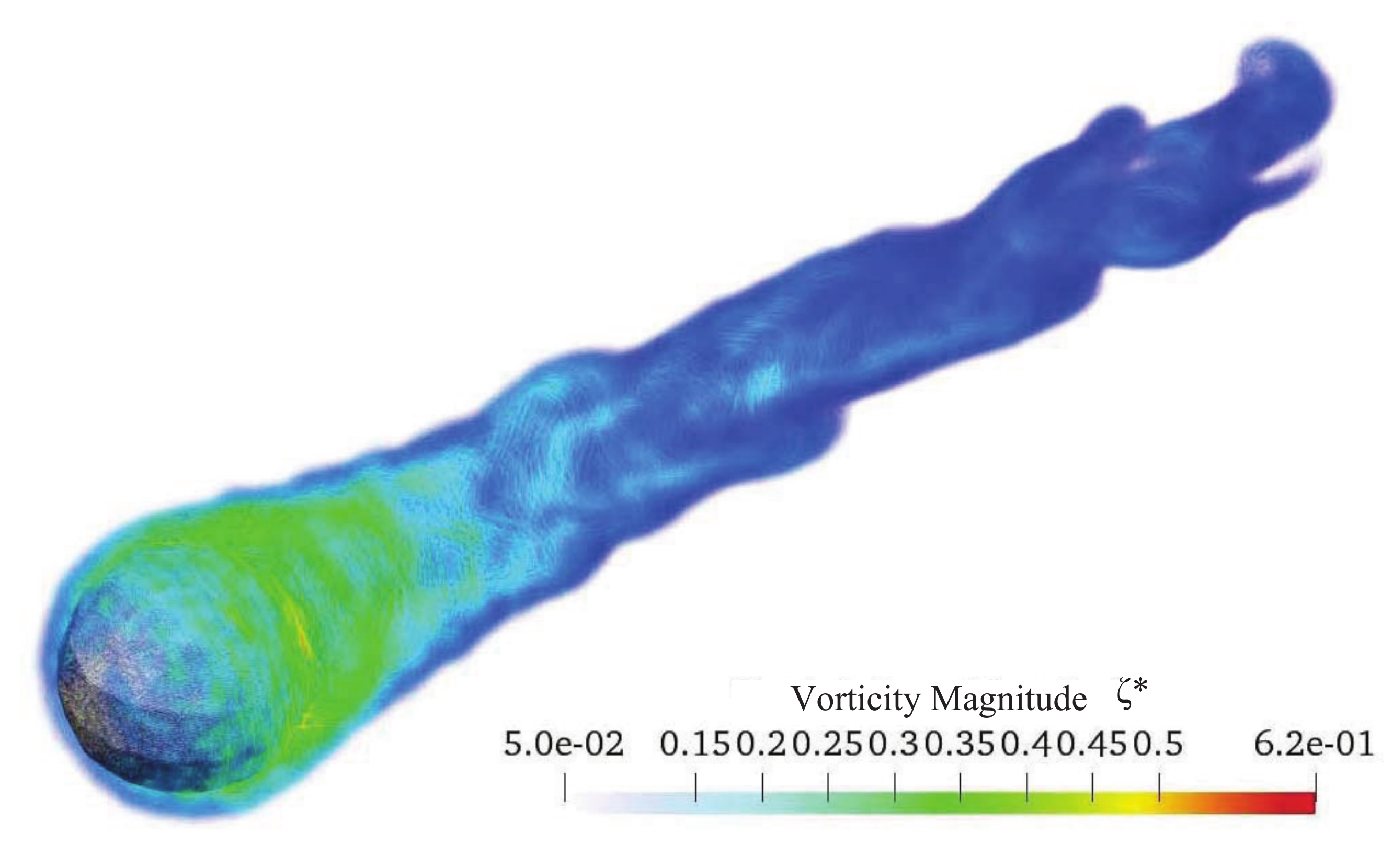}
			\end{minipage}
		}	
		\subfigure[Iso-surface of vorticity colored by velocity magnitude]{
			\begin{minipage}[b]{0.8\linewidth}
				\includegraphics[width=1\textwidth]{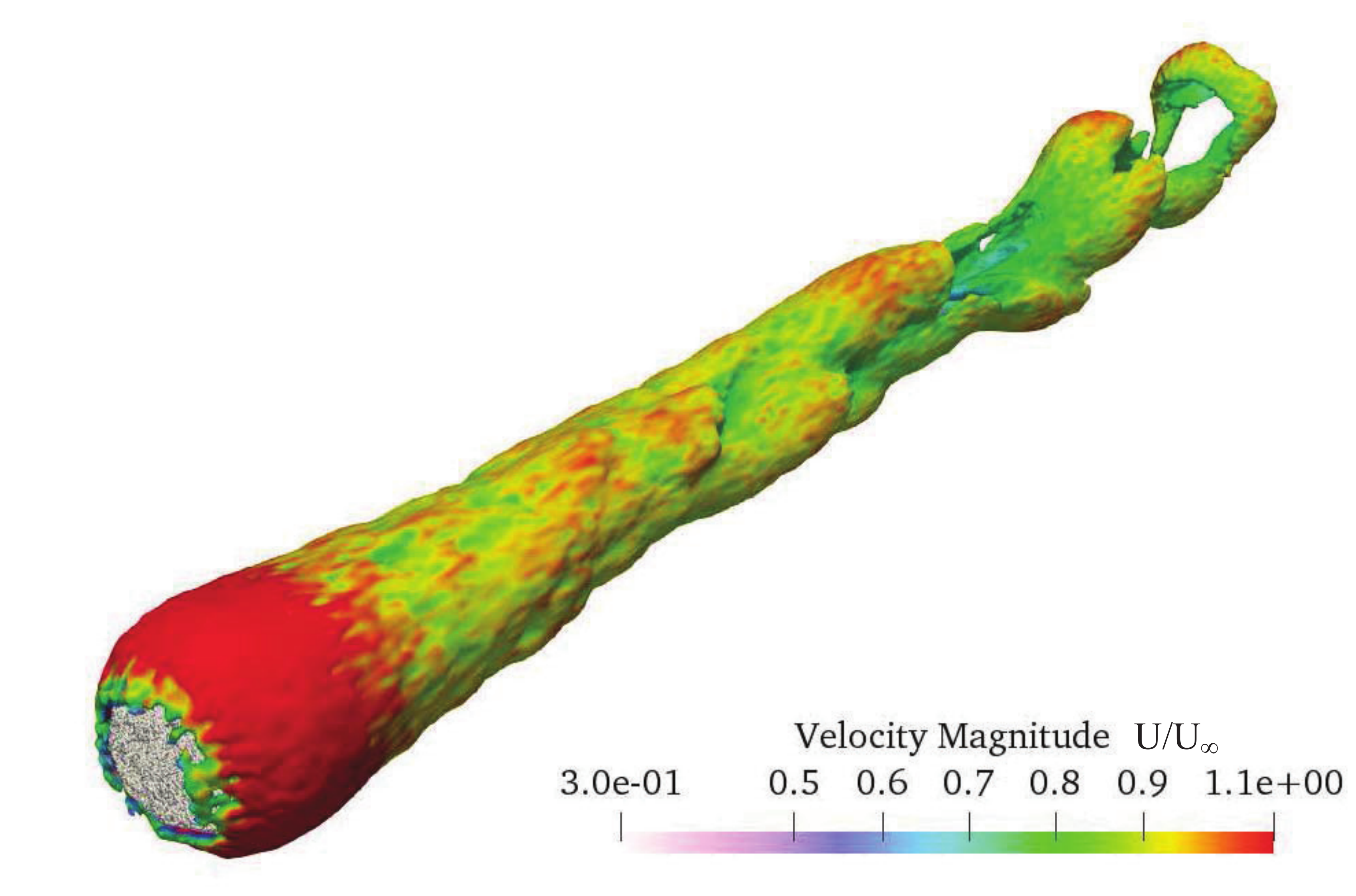}
			\end{minipage}
		}   
		\caption{The visualization of vortex structure at $ Re=1500 $.}
		\label{fig18}
	\end{figure}

	Fig.\ref{fig17} shows the visualization of the vorticity magnitude normalized 
	by $ \zeta^{*} = \zeta D / U_{\infty}(y)$ for different Reynolds numbers. 
	In the cases with $ Re=20 $ and $ Re=100 $, the toroidal vortices formed around the sphere gradually deform into pairs of counter-rotating vortexes stretched in the stream-wise direction, and a closed recirculating wake is eventually formed behind the sphere. As the Reynolds number increases ($ Re=200 $), the above flow behaviours can also be observed, while the formed vortex pairs oscillate back and forth in the stream-wise direction with time evolving. In addition, for the case with the highest Reynolds number of $ Re=1500 $, the vorticity magnitude contour and the vorticity iso-surface coloured by velocity magnitude are shown in Fig.\ref{fig18}.  
	As observed, the velocity perturbation along the sphere centerline results in the vortex rings alternately formed around the sphere. Later, the ring vortices periodically shed off the sphere, drift downstream in the wake flow, and finally degenerate. These observed phenomena are generally in accordance with the results in Ref. \cite{Campregher2009}. 
	
\begin{figure}[htbp]
		\centering     
		\includegraphics[width=0.7\textwidth]{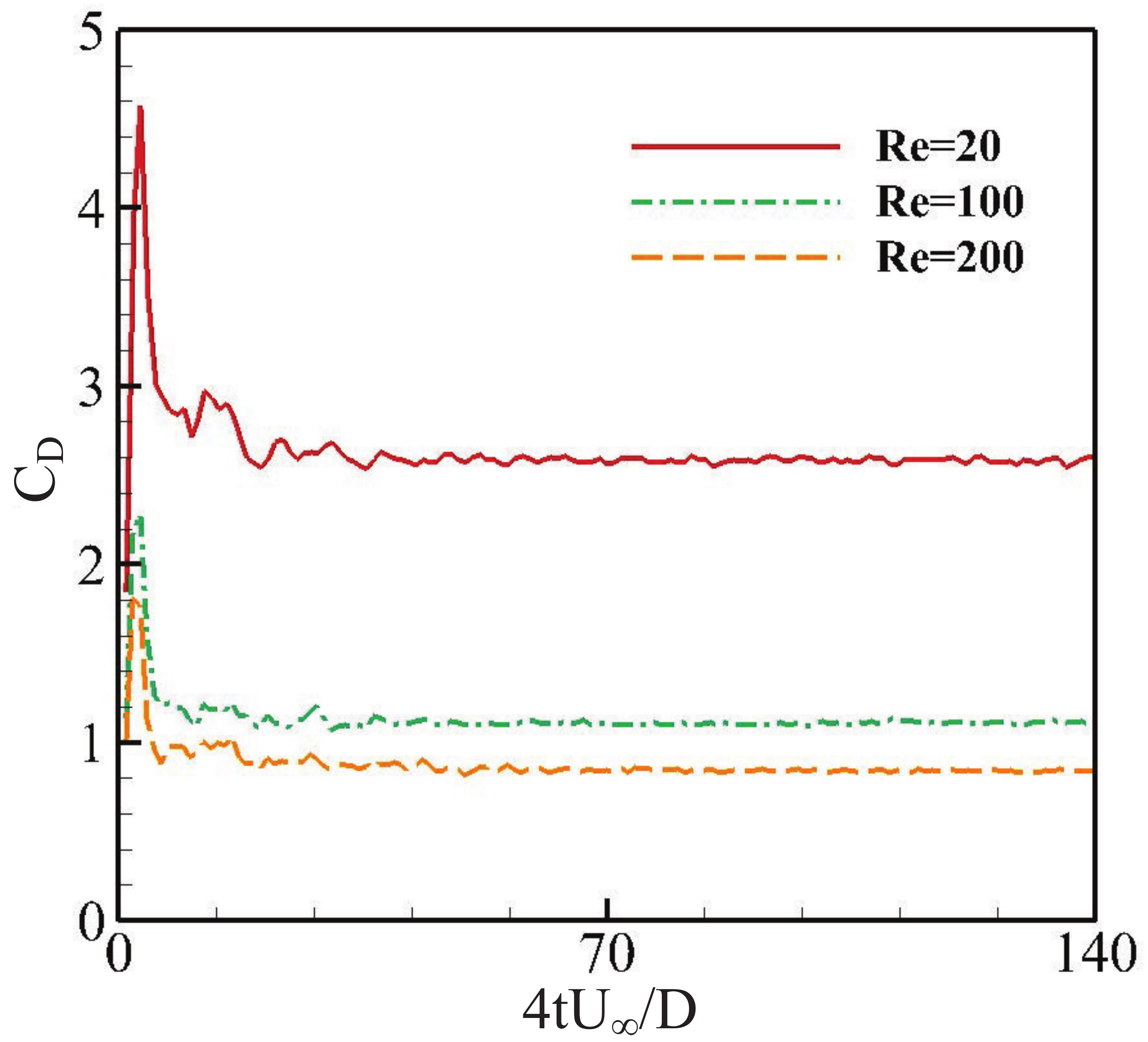}
		\caption{Time evolution of the drag coefficient for each case.}
		\label{fig19}
	\end{figure}

	For the flow past a sphere, the drag coefficient is defined as
	\begin{equation} \label{eq19}
		C_{D}=\dfrac{F_{D}}{\frac{1}{2}\rho \mathbf{U}_{\infty}(\frac{\pi D^{2}}{4})},
	\end{equation}
	where $ F_{D} $ is the total drag force on the sphere. Fig.\ref{fig19} shows the time evolution of drag coefficient $ C_{D} $ at different Reynolds numbers. 
	It is indicated that, after some start-up time, the drag coefficients of these low Reynolds number cases asymptotically approach each different constant value, 
	and they decrease with increasing the Reynolds number. Furthermore, an empirical formula for calculating the drag coefficient, in this case, is proposed as follows \cite{Campregher2009}
	\begin{equation} \label{eq20}
		C_{D}=\frac{24}{Re}[1+0.1935(Re^{0.6305})]\qquad 20\leq Re \leq 260.
	\end{equation}
	In Table \ref{tab5}, the calculated values together with numerical and experimental results are summarized. 
	\begin{table}[htb]
		\footnotesize
		\centering
		\caption{Comparison of $ C_{D} $ for each case.}
		\resizebox{\textwidth}{!}{ }{
			\begin{tabular}{p{1em}<{\centering} p{5.9em}<{\centering} p{5.9em}<{\centering} p{5.9em}<{\centering} p{5.2em}<{\centering} p{5.0em}<{\centering}}
				\toprule 
				\textbf{$ Re $}&
				\textbf{Magnaudet et al. \cite{magnaudet1995}} & \textbf{Tabata and Itakura \cite{Tabata1998}} & 
				\textbf{Roos and Willmarth\cite{ROOS1971}} & 
				\textbf{Theory \cite{Campregher2009}} & 
				\textbf{Present}\\
				\midrule 
				\textbf{20} & 2.707 & 2.7240  & 2.82 & 2.735 & 2.58  \\
				\midrule 
				\textbf{100} & 1.092 & 1.0895  & 1.08 & 1.087 & 1.108  \\
				\midrule 
				\textbf{200} & 0.765 & 0.77176  & 0.732 & 0.776 & 0.840 \\
				\bottomrule
		\end{tabular}}  
		\label{tab5}                                        
	\end{table}
	As indicated herein, the numerical results from the present simulation are close well to other reference results for cases $ Re=20 $ and 100, while only a small discrepancy exists for the case $ Re=200 $. Note that due to the limit of computational capacity, the maximum resolution of this numerical simulation and the diameter of the cylindrical domain is only set to $ D/dx=40 $ and $ d=2.5D $ here, respectively. This is much smaller than the parameters used in the previous section \ref{section4-2}, which may influence the quantitative results of the convergence study, such as the drag coefficient. 
	\section{Conclusion}
	In this paper, we present a Lagrange free-stream boundary condition for the WCSPH method. This algorithm is based on several numerical techniques, including the spatio-temporal identification and ensuring methods precisely avoid misjudgment on surface particle detection, the density and velocity corrections satisfying the far-field conditions at surface particles and the in- and outflow boundaries eliminating the velocity and pressure jump at the inflow region effectively. The numerical results from the cases of flow over a flat plate and a cylinder demonstrate the correctness and effectiveness of the present method. It is also found that the dependence between numerical results and domain sizes is greatly decreased, which implies a significant reduction of boundary effects and computational cost. In addition, the method performs well in more complicated cases involving fluid-structure interaction and 3D problems. It will be employed and tested in more complex scientific and industrial applications as future work.
	
	
	
	\bibliographystyle{elsarticle-num} 
	\bibliography{test}

\begin{thebibliography}{10}
\expandafter\ifx\csname url\endcsname\relax
  \def\url#1{\texttt{#1}}\fi
\expandafter\ifx\csname urlprefix\endcsname\relax\def\urlprefix{URL }\fi
\expandafter\ifx\csname href\endcsname\relax
  \def\href#1#2{#2} \def\path#1{#1}\fi

\bibitem{Lucy1977}
L.~Lucy, {A numerical approach to the testing of the fission hypothesis}, The
  Astronomical Journal 82 (1977) 1013--1024.

\bibitem{Gingold1977}
R.~Gingold, J.~Monaghan, {Smoothed particle hydrodynamics: Theory and
  application to non-spherical stars}, Monthly Notices of the Royal
  Astronomical Society 181 (1977) 375--389.

\bibitem{Monaghan1994}
J.~Monaghan, {Simulating free surface flows with SPH}, Journal of Computational
  Physics 110 (1994) 399--406.

\bibitem{Morris1997}
J.~Morris, P.~Fox, Y.~Zhu, {Modeling low Reynolds number incompressible flows
  using SPH}, Journal of Computational Physics 136 (1997) 214--226.

\bibitem{Benz1995}
W.~Benz, E.~Asphaug, {Simulations of brittle solids using smooth particle
  hydrodynamics}, Computer Physics Communications 87 (1995) 253--265.

\bibitem{Gray2001}
J.~Gray, J.~Monaghan, R.~Swift, {SPH elastic dynamics}, Computer Methods in
  Applied Mechanics and Engineering 190 (2001) 6641--6662.

\bibitem{Rafiee2009}
A.~Rafiee, K.~Thiagarajan, {An SPH projection method for simulating
  fluid-hypoelastic structure interaction}, Computer Methods in Applied
  Mechanics and Engineering 198 (2009) 110028.

\bibitem{Zhang2021}
C.~Zhang, M.~Rezavand, X.~Hu, {A multi-resolution SPH Method for
  fluid-structure interactions}, Journal of Computational Physics 429 (2021)
  110028.

\bibitem{Federico2012}
I.~Federico, S.~Marrone, A.~Colagrossi, A.~F., A.~M., Simulating {2D}
  open-channel flows through an {SPH} model, European Journal of Mechanics
  B/Fluids 34 (2012) 35--46.

\bibitem{Tan2015}
S.~Tan, N.~Cheng, Y.~Xie, S.~Shao, {Incompressible SPH simulation of open
  channel flow over smooth bed}, Journal of Hydro-environment Research 9 (2015)
  340--353.

\bibitem{Staroszczyk2014}
R.~Staroszczyk, Incompressible {SPH} model for simulating violent free-surface
  fluid flows, Archives of Hydro-Engineering and Environmental Mechanics 61
  (2014) 61--83.

\bibitem{Marrone2011}
S.~Marrone, M.~Antuono, A.~Colagrossi, G.~Colicchio, D.~Le~Touzé, G.~Graziani,
  Delta-sph model for simulating violent impact flows, Computer Methods in
  Applied Mechanics and Engineering 200 (2011) 1526--1542.

\bibitem{Chen}
R.~Chen, S.~Shao, X.~Liu, X.~Zhou, {Applications of shallow water SPH model in
  mountainous rivers}, Journal of Applied Fluid Mechanics 8 (2015) 863--870.

\bibitem{Vacondio}
R.~Vacondio, B.~Rogers, P.~Stansby, P.~Mignosa, {SPH modeling of shallow flow
  with open boundaries for practical flood simulation}, Journal of Hydraulic
  Engineering 138 (2012) 530--541.

\bibitem{Matthieu}
D.~Matthieu, L.~David, A.~Bertrand, {SPH modeling of shallow-water coastal
  flows}, Journal of Hydraulic Research 48 (2010) 118--125.

\bibitem{Tafuni2018}
A.~Tafuni, J.~Domínguez, R.~Vacondio, A.~Crespo, A versatile algorithm for the
  treatment of open boundary conditions in smoothed particle hydrodynamics
  {GPU} models, Computer Methods in Applied Mechanics and Engineering 342
  (2018) 604--624.

\bibitem{shuoguo}
S.~Zhang, X.~Hu, C.~Mi, Research on braking force of aerodynamic brake panel of
  high-speed train based on {SPH} method, Physics of Gases 5 (2020) 0--0.

\bibitem{Martin2009}
L.~Martin, M.~Basa, N.~Quinlan, Permeable and non‐reflecting boundary
  conditions in {SPH}, International Journal for Numerical Methods in Fluids 61
  (2009) 709--724.

\bibitem{Molteni2013}
D.~Molteni, R.~Grammauta, E.~Vitanza, Simple absorbing layer conditions for
  shallow wave simulations with smoothed particle hydrodynamics, Ocean
  Engineering 62 (2013) 78--90.

\bibitem{Braun2015}
S.~Braun, L.~Wieth, R.~Koch, H.~Bauer, {A framework for permeable boundary
  conditions in SPH: Inlet, Outlet, Periodicity}, 2015.

\bibitem{Ferrand2017}
M.~Ferrand, A.~Joly, C.~Kassiotis, D.~Violeau, A.~Leroy, F.~Morel, B.~Rogers,
  Unsteady open boundaries for {SPH} using semi-analytical conditions and
  riemann solver in {2D}, Computer Physics Communications 210 (2017) 29--44.

\bibitem{Colagrossi2013}
S.~Marrone, A.~Colagrossi, M.~Antuono, G.~Colicchio, G.~Graziani, {An accurate
  SPH modeling of viscous flows around bodies at low and moderate Reynolds
  numbers}, Journal of Computational Physics 245 (2013) 456--475.

\bibitem{Wang2019}
P.~Wang, A.~Zhang, F.~Ming, P.~Sun, H.~Cheng, {A novel nonreflecting boundary
  condition for fluid dynamics solved by smoothed particle hydrodynamics},
  Journal of Fluid Mechanics 860 (2019) 81--114.

\bibitem{Khorasanizade2015}
S.~Khorasanizade, J.~Sousa, {An innovative open boundary treatment for
  incompressible SPH}, International Journal for Numerical methods in Fluids 80
  (2016) 161--180.

\bibitem{Alvarado-Rodriguez2017}
C.~Alvarado-Rodriguez, J.~Klapp, L.~Sigalotti, J.~Dominguez, E.~Sanchez,
  {Nonreflecting outlet boundary conditions for incompressible flows using
  SPH}, Computers and Fluids 159 (2017) 177--188.

\bibitem{Negi2020}
P.~Negi, P.~Ramachandran, A.~Haftu, {An improved non-reflecting outlet boundary
  condition for weakly-compressible SPH}, Computer Methods in Applied Mechanics
  and Engineering 367 (2020) 113119.

\bibitem{Lee2008}
E.~Lee, C.~Moulinec, R.~Xu, D.~Violeau, D.~Laurence, P.~Stansby, Comparisons of
  weakly compressible and truly incompressible algorithms for the {SPH} mesh
  free particle method, Journal of Computational Physics 227 (2008) 8417--8436.

\bibitem{Dilts}
G.~Dilts, {Moving-least-squares-particle hydrodynamics—I. Consistency and
  stability}, International Journal for Numerical Methods in Engineering 44
  (1999) 1115--1155.

\bibitem{Haque}
A.~Haque, G.~Dilts, Three-dimensional boundary detection for particle methods,
  Journal of Computational Physics 226 (2007) 1710--1730.

\bibitem{Lin2019}
Y.~Lin, G.~Liu, G.~Wang, A particle-based free surface detection method and its
  application to the surface tension effects simulation in smoothed particle
  hydrodynamics ({SPH}), Journal of Computational Physics 383 (2019) 196--206.

\bibitem{Marrone2010}
S.~Marrone, A.~Colagrossi, D.~{Le Touzé}, G.~Graziani, {Fast free-surface
  detection and level-set function definition in SPH solvers}, Journal of
  Computational Physics 229 (2010) 3652--3663.

\bibitem{Adami2013}
S.~Adami, X.~Hu, N.~Adams, A transport-velocity formulation for smoothed
  particle hydrodynamics, Journal of Computational Physics 241 (2013) 292--307.

\bibitem{Zhang2021CPC}
C.~Zhang, M.~Rezavand, Y.~Zhu, Y.~Yu, D.~Wu, W.~Zhang, J.~Wang, X.~Hu,
  {SPHinXsys: An open source multi-physics and multi-resolution library based
  on smoothed particle hydrodynamics}, Computer Physics Communications 267
  (2021) 108066.

\bibitem{Hu2006}
X.~Hu, N.~Adams, {A multi phase SPH method for macroscopic and mesoscopic
  flows}, Journal of Computational Physics 213 (2006) 844--861.

\bibitem{Zhang2020}
C.~Zhang, M.~Rezavand, X.~Hu, {Dual-criteria time stepping for weakly
  compressible smoothed particle hydrodynamics}, Journal of Computational
  Physics 404 (2020) 109135.

\bibitem{Zhang2017}
C.~Zhang, X.~Hu, N.~Adams, A generalized transport-velocity formulation for
  smoothed particle hydrodynamics, Journal of Computational Physics 337 (2017)
  216--232.

\bibitem{article2017Chi}
C.~Zhang, X.~Hu, N.~Adams, A weakly compressible {SPH} method based on a
  low-dissipation {R}iemann solver, Journal of Computational Physics 335 (01
  2017).

\bibitem{rezavand2021generalised}
M.~Rezavand, C.~Zhang, X.~Hu, Generalised and efficient wall boundary condition
  treatment in gpu-accelerated smoothed particle hydrodynamics, arXiv preprint
  arXiv:2110.02621 (2021).

\bibitem{Turek2007}
S.~Turek, J.~Hron, Proposal for numerical benchmarking of fluid–structure
  interaction between an elastic object and laminar incompressible flow,
  Vol.~53, 2007, pp. 371--385.

\bibitem{Wendland1995}
H.~Wendland, {Piecewise polynomial, positive definite and compactly supported
  radial functions of minimal degree}, Advances in Computational Mathematics 4
  (1995) 389--396.

\bibitem{Quinlan2006}
N.~J. Quinlan, M.~Basa, M.~Lastiwka, Truncation error in mesh-free particle
  methods, International Journal for Numerical Methods in Engineering 66~(13)
  (2006) 2064--2085.

\bibitem{Springel_2010}
V.~Springel, Smoothed particle hydrodynamics in astrophysics, Annual Review of
  Astronomy and Astrophysics 48~(1) (2010) 391--430.

\bibitem{litvinov2015towards}
S.~Litvinov, X.~Hu, N.~A. Adams, Towards consistence and convergence of
  conservative sph approximations, Journal of Computational Physics 301 (2015)
  394--401.

\bibitem{Ellero2011SPHSO}
M.~Ellero, N.~Adams, Sph simulations of flow around a periodic array of
  cylinders confined in a channel, International Journal for Numerical Methods
  in Engineering 86 (2011) 1027--1040.

\bibitem{MARRONE2013456}
S.~Marrone, A.~Colagrossi, M.~Antuono, G.~Colicchio, G.~Graziani, An accurate
  sph modeling of viscous flows around bodies at low and moderate reynolds
  numbers, Journal of Computational Physics 245 (2013) 456--475.

\bibitem{Federico2010}
I.~Federico, S.~Marrone, A.~Colagrossi, F.~Aristodemo, P.~Veltri, Simulating
  free-surface channel flows through {SPH}, 2010.

\bibitem{Tafuni2016}
A.~Tafuni, J.~Domínguez, R.~Vacondio, I.~Sahin, A.~Crespo, Open boundary
  conditions for large-scale {SPH} simulations, 2016.

\bibitem{White2006}
F.~White, Viscous Fluid Flow, 3rd Edition, McGraw–Hill, New York, 2006.

\bibitem{Brehm2015}
C.~Brehm, C.~Hader, H.~Fasel, A locally stabilized immersed boundary method for
  the compressible navier–stokes equations, Journal of Computational Physics
  295 (2015) 475--504.

\bibitem{Almarouf2017}
M.~Al-Marouf, R.~Samtaney, A versatile embedded boundary adaptive mesh method
  for compressible flow in complex geometry, Journal of Computational Physics
  337 (2017) 339--378.

\bibitem{Liu1998}
C.~Liu, X.~Zheng, C.~Sung, Preconditioned multigrid methods for unsteady
  incompressible flows, Journal of Computational Physics 139 (1998) 35--57.

\bibitem{Le2006}
D.~Le, B.~Khoo, J.~Peraire, An immersed interface method for viscous
  incompressible flows involving rigid and flexible boundaries, Journal of
  Computational Physics 220 (2006) 109--138.

\bibitem{Russell2003}
D.~Russell, Z.~Wang, A cartesian grid method for modeling multiple moving
  object in {2D} incompressible viscous flow, Journal of Computational Physics
  191 (2003) 177--205.

\bibitem{Taira2007}
T.~Kunihiko, C.~Tim, The immersed boundary method: A projection approach,
  Journal of Computational Physics 225 (2007) 2118--2137.

\bibitem{Jin1993}
G.~Jin, M.~Braza, A nonreflecting outlet boundary condition for incompressible
  unsteady navier-stokes calculations, Journal of Computational Physics 107
  (1993) 239--253.

\bibitem{Tritton1959}
D.~J. Tritton, Experiments on the flow past a circular cylinder at low
  {R}eynolds numbers, Journal of Fluid Mechanics 6 (1959) 547–567.

\bibitem{Han2018}
C.~Zhang, M.~Rezavand, X.~Hu, A multi-resolution sph method for fluid-structure
  interactions, Journal of Computational Physics 429 (2021) 110028.

\bibitem{Bhardwaj2012}
R.~Bhardwaj, R.~Mittal, Benchmarking a coupled immersed-boundary-finite-element
  solver for large-scale flow-induced deformation, AIAA Journal 50 (2012)
  1638--1642.

\bibitem{Tian2014}
F.~Tian, H.~Dai, H.~Luo, J.~Doyle, B.~Rousseau, Fluid–structure interaction
  involving large deformations: {3D} simulations and applications to biological
  systems, Journal of Computational Physics 258 (2014) 451–469.

\bibitem{Campregher2009}
R.~Campregher, J.~Militzer, S.~Mansur, A.~Neto, Computations of the flow past a
  still sphere at moderate {R}eynolds numbers using an immersed boundary
  method, Journal of The Brazilian Society of Mechanical Sciences and
  Engineering 31 (2009) 344--352.

\bibitem{magnaudet1995}
J.~Magnaudet, M.~Rivero, J.~Fabre, Accelerated flows past a rigid sphere or a
  spherical bubble. part 1. steady straining flow, Journal of Fluid Mechanics
  284 (1995) 97–135.

\bibitem{Tabata1998}
M.~Tabata, K.~Itakura, A precise computation of drag coefficients of a sphere,
  International Journal of Computational Fluid Dynamics 9 (1998) 303--311.

\bibitem{ROOS1971}
F.~W. Roos, W.~W. Willmarth, Some experimental results on sphere and disk drag,
  AIAA Journal 9~(2) (1971) 285--291.

\end{thebibliography}
	
	
	
	
	
\end{document}